%% file: main.tex
\documentclass[3p,onecolumn]{elsarticle}

\usepackage{paperheader}
\let\oriCiteO\citeA

\definecolor{gray1}{HTML}{E0E0E0}  
\definecolor{gray2}{HTML}{B0B0B0}  
\definecolor{gray3}{HTML}{707070}  
\definecolor{cold1}{HTML}{A8DADC}  
\definecolor{cold2}{HTML}{457B9D}  
\definecolor{cold3}{HTML}{1D3557}  

\RenewDocumentCommand{\citeA}{O{} O{} m}{%
 \renewcommand{\citenumfont}[1]{A##1}%
 \oriCiteO[#1][#2]{#3}%
 \renewcommand{\citenumfont}[1]{##1}%
}

\newcommand{\ie}{i.e.,\xspace}
\newcommand{\eg}{e.g.,\xspace}

\newcommand{\etal}{\textsl{et al.}\xspace}

\begin{document}
\begin{frontmatter}
    
    \markboth{Iuliano and Di Nucci}{Smart Contract Vulnerabilities, Tools, and Benchmarks: an Updated Systematic Literature Review}
    
    \title{Smart Contract Vulnerabilities, Tools, and Benchmarks:\\ an Updated Systematic Literature Review}

    \author[1]{Gerardo Iuliano}
		\ead{geiuliano@unisa.it}
    \author[1]{Dario Di Nucci}
		\ead{ddinucci@unisa.it}
    \affiliation[1]{
    organization={University of Salerno},
    country={Italy}}		
    \input{defs/abstract}

		
\end{frontmatter}









\input{sections/introduction}
\input{sections/background}

\input{sections/methodology}

\input{sections/results}

\input{sections/discussion}

\input{sections/threats}

\input{sections/related_work}
\input{sections/conclusion}

\input{defs/ack}

\bibliographystyle{elsarticle-num}
\bibliography{bibs/references}

\bibliographystyleA{elsarticle-num}
\bibliographyA{bibs/appendix}

\balance
\end{document}

%% file: defs/abstract.tex
\begin{abstract}
Smart contracts are self-executing programs on blockchain platforms like Ethereum, which have revolutionized decentralized finance by enabling trustless transactions and the operation of decentralized applications. Despite their potential, the security of smart contracts remains a critical concern due to their immutability and transparency, which expose them to malicious actors. Numerous solutions for vulnerability detection have been proposed, but it is still unclear which one is the most effective.
This paper presents a systematic literature review that explores vulnerabilities in Ethereum smart contracts, focusing on automated detection tools and benchmark evaluation. We reviewed 3,380 studies from five digital libraries and five major software engineering conferences, applying a structured selection process that resulted in 222 high-quality studies. The key results include a hierarchical taxonomy of 192 vulnerabilities grouped into 13 categories, a comprehensive list of 219 detection tools with corresponding functionalities, methods, and code transformation techniques, a mapping between our taxonomy and the list of tools, and a collection of 133 benchmarks used for tool evaluation. We conclude with a discussion about the insights into the current state of Ethereum smart contract security and directions for future research.
\end{abstract}

%% file: sections/introduction.tex
\section{Introduction}
\label{sec:intro}

Since Bitcoin was introduced in 2009 \cite{nakamoto2008bitcoin}, blockchain and decentralized solutions have gained considerable attention and adoption \cite{alharby2018blockchain}. Although blockchain was initially associated with cryptocurrencies, its potential has expanded, with smart contracts representing a notable and promising application.
Smart contracts (SCs) are self-executing programs that run on blockchain platforms such as Ethereum, where transactions are automatically enforced without intermediaries. Originally, the term \textit{smart contract} referred to the concept of automating traditional legal contracts using software \cite{szabo1997formalizing}. However, they have been adopted in many other domains such as finance, healthcare care and supply chain management by ensuring trustless computation. 
Smart contracts are pivotal in decentralized systems due to their inherent transparency properties, tamper resistance, and immutability. 
In decentralized finance (DeFi), smart contracts form the backbone of financial protocols, allowing decentralized applications (dApps) to facilitate lending, borrowing, trading, and other financial services without traditional intermediaries \cite{chaliasos2024smart}. Ethereum, the most widely used blockchain for DeFi and dApps, supports a vast ecosystem of smart contracts managing valuable assets, including fungible and non-fungible tokens (NFTs) \cite{wang2021non}. This surge in decentralized financial activity has attracted significant attention from developers and attackers, with billions of dollars now at stake.

Despite their potential, smart contracts face many challenges, particularly regarding security \cite{zou2019smart}. Developing secure smart contracts is inherently difficult due to platform-specific limitations, complex business logic, and the immutability of deployed contracts. Once a contract is deployed on-chain, any error, weakness, or vulnerability becomes permanent, making smart contracts appealing targets for malicious actors. The transparency of the blockchain allows attackers to analyze and exploit contracts, often leading to high-stakes breaches and financial losses in DeFi protocols. Furthermore, the interconnected nature of smart contracts, where contracts can interact and trigger each other, adds another layer of complexity, making vulnerabilities harder to detect and mitigate.

Existing taxonomies like DASP top 10 or SWC Registry are open and collaborative projects that join efforts to discover smart contract vulnerabilities within the security community. The main issue is their relevance due to outdatedness; in fact, the latest update of the DASP Top 10 is from 2018, while the SWC Registry is from 2020. The OWASP Smart Contract Top 10 (2025) is the only updated taxonomy to list the top 10 vulnerabilities found in smart contracts. Unfortunately, it comprises only the top 10 vulnerabilities.
In recent years, several studies have revealed new vulnerabilities that affect smart contracts. Fang \etal~\citeA{fang2023beyond} conducted a security analysis of the modifiers used in real-world Ethereum smart contracts and identified insecure modifiers that can be bypassed from one or more unprotected smart contract functions. Huang \etal~\citeA{Huang2024Revealing} identified eight general patterns of library misuse: three arise during library development and five during library utilization, covering the entire library lifecycle. Library misuse refers to incorrect implementation or improper use of libraries, which can result in unexpected results, such as the introduction of vulnerabilities or the inclusion of unsafe libraries. The latest taxonomies do not collect recent updates. Furthermore, we observed several discrepancies between them in terms of categories adopted to classify vulnerabilities and synonyms used to refer to the same vulnerabilities. In particular, focusing on the most recent and comprehensive taxonomies such as those proposed by Rameder \etal~\cite{Rameder2022}, Li \etal~\citeA{li2024static}, and Vidal \etal~\citeA{vidal2024openscv} (OpenSCV), we observed that the categories used to organize vulnerabilities are different between taxonomies. Different taxonomies share a few categories, and each taxonomy uses a different approach to classify and organize vulnerabilities. The intersection of categories in different taxonomies does not provide a comprehensive set of criteria to categorize all vulnerabilities reported in the literature. Each taxonomy has a characteristic that distinguishes it from the other. In addition, synonyms or alternative names of vulnerabilities are used to identify different vulnerabilities. For example, OpenSCV~\citeA{vidal2024openscv} assigns the same synonym to multiple distinct vulnerabilities, creating a core issue when consulting a taxonomy. Especially when performing reverse engineering from alternative names, it becomes unclear which vulnerability is actually being referenced. Moreover, the growing adoption of dApps has led to the emergence of new types of vulnerability, particularly in multiblockchain environments. As systems increasingly integrate and facilitate communication between two or more blockchains, they introduce a new class of risks known in the literature as cross-system vulnerabilities. In particular, several studies~\cite{duan2023attacks, lee2023sok, zhang2022xscope} have identified a range of problems associated with cross-chain technology. A comprehensive taxonomy is essential for vulnerability awareness, but its construction is complicated by the lack of clarity and agreement in existing studies.

Based on the challenges affecting current taxonomies and the emerging vulnerabilities discussed in recent studies, we defined the first research question: \textit{Which vulnerabilities are mentioned? How are vulnerabilities classified?} The result of $RQ_1$ is a structured and hierarchical taxonomy that collates vulnerabilities into 14 different categories. Vulnerabilities are posed on different levels based on their level of abstraction and the details provided about the cause and condition in which they arise. The first level contains 120 vulnerabilities, the second contains 66, while the third contains six.

The research on vulnerabilities goes hand in hand with the research on vulnerability detection. Several studies have proposed novel solutions to detect vulnerabilities automatically, analyze gas and resource consumption, exploit vulnerabilities, provide reactive defenses, and repair smart contracts. The number of different methods adopted when implementing a tool is significantly high (symbolic execution, taint analysis, fuzzing, formal methods, etc.), along with the different approaches used to represent and analyze the source code (Abstract Syntax Tree, Control Flow Graph, Data Flow Graph, Transaction trace, etc). Each time a new tool is proposed, it improves the current state-of-the-art performance or achieves a new goal. For example, Li \etal~\citeA{li2024cobra} proposed a tool capable of finding two previously undisclosed vulnerabilities, \ie \textit{CVE-2023-36979} and \textit{CVE-2023-36980}. Given the high number of vulnerabilities and the different implementable solutions, we explore the landscape of tools proposed to address problems affecting smart contracts. We formulated the second research question: \textit{Which methods do automated tools use to detect vulnerabilities?} As a result, we collected 219 automated analysis tools.

For each tool, we have collected its main functionality, the method, and the code transformation technique adopted. Given that each vulnerability has its unique characteristics and poses distinct challenges for effective detection. To identify which vulnerabilities are most actively studied and targeted, we mapped the vulnerabilities included in our taxonomy to the detection tools designed to address them. The third research question \textit{Which vulnerabilities are addressed by the tools?} shows that the focus is on a narrow set of 79 vulnerabilities, and many of them remain unexplored. We found that some vulnerabilities, such as Price oracle manipulation,  are not detectable by existing static analysis tools and require a deeper understanding of the contract by analyzing natural language from a semantic perspective \citeA{Z.ZhangEtAl2023}. In addition, we found that tools often omit their compatibility with the Solidity version, complicating their effectiveness comparison.

To close the loop, we decided to collect information about the evaluation of tools, and then we formulated the fourth research question: \textit{How are the tools evaluated?} We collected a set of 133 benchmarks used for tool evaluation or to conduct other experiments. About half of them are open source, while the most widely used is SmartBugs-wild, which consists of mainly smart contracts written using Solidity 0.4.x.

Our study updates the systematic literature review conducted by Rameder \etal~\cite{Rameder2022}. They analyzed more than 300 studies published up to 2021, examining the classification of vulnerabilities, detection methods, security analysis tools, and benchmarks for evaluating tools. Their work laid the foundation for systematically understanding the complexities of smart contract vulnerabilities, the tools available to address them, and the open challenges and gaps in automated detection systems. We have overcome previous limitations such as the absence of a structured snowballing process, a comprehensive mapping between tools and vulnerabilities, and the unchecked benchmarks for duplicates. 
We answered four research questions to provide an updated and more complete perspective on Ethereum vulnerabilities, detection tools, and benchmarks. 
We analyzed five digital libraries and five main conferences on software engineering and conducted two snowballing processes to collect a total of 3,380 papers. We then applied a carefully documented process to select and evaluate 332 studies of immediate relevance. Quality appraisal reduced the analysis to 222 studies with sufficient intrinsic and contextual data quality that allowed us to answer our RQs. As a result, we realized a structured taxonomy that organizes hierarchically 192 vulnerabilities in 13 categories, a list of 219 automated detection tools with relative functionalities, methods, and code transformation techniques, a mapping between our taxonomy and the list of tools, and a list of 133 benchmarks used for tool evaluation. 
We provide the following contributions:
\begin{itemize}
    \item A taxonomy of vulnerabilities comprising 192 vulnerabilities hierarchically organized on three levels and into 13 categories, serving as a comprehensive aggregation of the most classifications.
    \item A list of 219 tools used in the literature to address various problems concerning SCs, providing their functionalities along with the method and code transformation technique they apply.
    \item A mapping between our taxonomy and the tools that clarify the research focus and gaps.
    \item A list of 133 benchmarks used for tool evaluation, tool comparison, or performing experiments. 
\end{itemize}

The paper is structured as follows. \Cref{sec:background} describes the background of Ethereum SCs and their vulnerabilities. \Cref{sec:methodology} exposes the research method and design adopted in our study and describes all the steps followed to answer our research questions. The results of each research question are reported in \Cref{sec:results}. \Cref{sec:discussion} is dedicated to discussing our findings compared to the Rameder \etal~\cite{Rameder2022} and lists the research opportunities. \Cref{sec:threats} defines the threats to validity, \Cref{sec:related_work} discusses the related work, highlighting the gaps and our contribution, and finally \Cref{sec:conclusion} concludes the paper.

%% file: sections/background.tex
\section{Background}
\label{sec:background}
This section presents preliminary information on Ethereum, smart contracts, and the vulnerabilities associated with them.

\subsection{Ethereum}
Ethereum is a decentralized open source blockchain platform that facilitates the creation and execution of smart contracts and dApps~\cite{buterin2014sc&dapps}. Proposed by Vitalik Buterin in late 2013~\cite{buterin2013ethereum}, Ethereum was designed to extend blockchain technology beyond simple data storage to include code execution, thus creating a versatile platform for developers. Unlike Bitcoin~\cite{nakamoto2008bitcoin}, which works primarily as a digital currency, Ethereum allows the development and deployment of dApps. At its core, Ethereum features the Ethereum Virtual Machine (EVM), a decentralized virtual machine that executes code on the Ethereum network. This platform supports a near Turing-complete programming language, enabling complex computations and dApps. A key feature of Ethereum is its use of three Merkle Patricia~\cite{merkletree} trees to maintain blockchain state information, which facilitates efficient and verifiable data queries. Ether (ETH), Ethereum's native cryptocurrency, powers transactions and rewards miners, ensuring the network's security and functionality.

\subsection{Smart Contracts}
The idea of smart contracts, which Nick Szabo initially introduced in 1994~\cite{szabo1994smartcontract}, has found a new life on the Ethereum blockchain. Smart contracts in Ethereum are self-executing contracts with terms directly written in the code~\cite{antonopoulos2018masteringSc}. These contracts run on the Ethereum network, a decentralized platform that eliminates the need for intermediaries by automating contract execution. Each smart contract has a unique address and is initiated by a transaction between a sender and the contract. Execution incurs a computational cost measured in gas, which regulates resource usage and rewards miners. Developers write smart contracts in Solidity~\cite{dannen2017introSolidity} or Vyper~\cite{vyper}, two languages explicitly designed for this purpose. Once deployed on the blockchain, these contracts become immutable and tamper-proof~\cite{kaushal2021immutable}. They automatically execute predefined conditions, ensuring trust and reliability. Smart contracts enable seamless collaboration, enforce contract clauses, and are pivotal for developing dApps on the Ethereum platform.

\subsection{Smart Contract Vulnerabilities}
Smart contracts on the Ethereum blockchain offer significant benefits but are susceptible to vulnerabilities, further compounded by blockchain technology's immutable nature~\cite{politou2019blockMutability}, stemming from coding errors and bugs in software development. 
Immutability ensures transparency and trust, but makes code defects permanent. Fixing such defects requires creating new contracts and user migration, which is a complex and risky task, underscoring the importance of exhaustive security measures. 
For upgradeable contracts that are not transparent to the user, the user will know their upgrades directly. For upgradeable contracts that are transparent to the user, they may be upgraded secretly, changing the functionality without the user's knowledge. Huang \etal~\citeA{huang2024sword} studied upgradeable smart contracts in Ethereum, providing information on two common techniques used to upgrade SC: the \textit{proxy pattern} and the \textit{metamorphic contract factory}. 
The \textit{proxy pattern} is a design approach in smart contracts in which a proxy contract forwards all calls to a separate implementation (logic) contract using the delegate call function. Unlike a normal call, the delegate call executes the logic contract's code in the context of the proxy, allowing access to the proxy's storage. This enables upgradable smart contracts, as the logic can be changed without altering the proxy's address or storage.
The \textit{metamorphic contract factory} is a mechanism that creates metamorphic contracts, contracts that can be redeployed with new code at the same address. This is achieved using the \textit{CREATE2} opcode, which allows the address of a contract to be precomputed before deployment. After the original contract self-destructs, the factory can deploy a new version at the same address. However, this process resets the contract's storage, replacing it entirely with the new implementation's code.
Smart contracts can contain mistakes once deployed, potentially leading to unintended behaviors or exploitations.

The \textit{reentrancy} attack~\cite{Rameder2022,samreen2020reentrancy}\citeA{vidal2024openscv, li2024static} stands out among these vulnerabilities, where a contract calls another before updating its state, opening the door to recursive calls that could drain funds or disrupt the intended execution.
Furthermore, smart contracts often interact with external contracts and data sources, introducing additional vulnerabilities if these interactions are not checked appropriately.

\textit{Timestamp Dependency}~\cite{Rameder2022}\citeA{ChuEtAl2023, wu2024comprehensive, vidal2024openscv, li2024static} refers to a vulnerability in smart contracts where the contract's logic relies on the \textit{block.timestamp} (or \textit{now}) value to make critical decisions, such as triggering payouts, closing auctions, or generating randomness. Since miners have limited control over the block timestamp (within a reasonable range), this can be exploited to manipulate the contract's behavior, especially in scenarios involving small time windows or incentives.

\textit{Authorization through tx.origin}~\cite{Rameder2022}\citeA{vidal2024openscv, li2024static} is a security vulnerability that occurs when a smart contract uses \textit{tx.origin} to check the identity of the caller for access control. The \textit{tx.origin} returns the original external account that initiated the transaction, even if the call was forwarded through multiple contracts. This can be exploited in phishing-style attacks, where a malicious contract tricks a user into calling it. Then that contract calls the vulnerable contract using the user's address as \textit{tx.origin}, bypassing the intended access control. Instead, developers should use msg.sender for secure authorization, as it correctly reflects the immediate caller of the function.

\textit{Integer Overflow and Underflow}~\cite{Rameder2022}\citeA{SunEtAl2022a, KushwahaEtAl2022a, AgarwalEtAl2022, MunirAndTaha2023, vidal2024openscv, li2024static} occur when arithmetic operations on unsigned integers exceed their maximum or minimum limits.
An overflow occurs when a value exceeds the maximum limit (e.g., \textit{uint256} going above $2^{256} - 1$), causing it to wrap around to zero.
An underflow happens when a value goes below zero (e.g., subtracting 1 from \textit{uint256(0)}), wrapping around to the maximum value. These bugs can lead to unexpected behavior, such as bypassing balance checks or locking/unlocking tokens incorrectly. Starting with Solidity 0.8.0, these operations automatically revert on overflow or underflow, helping prevent such vulnerabilities by default.

\textit{Vulnerable delegatecall}~\cite{Rameder2022}\citeA{dZaazaaAndElBakkali2023a, dZaazaaAndElBakkali2023b, DiasEtAl2021, vidal2024openscv, li2024static} arises when a smart contract uses \textit{delegatecall} to execute external code without properly validating the target address or ensuring the safety of the called contract. Since \textit{delegatecall} runs the external contract's code in the context of the caller, it has full access to the caller's storage and permissions. If misused, an attacker can trick the contract into executing malicious logic, potentially leading to loss of control over the contract state or funds. To avoid this vulnerability, \textit{delegatecall} should only be used with trusted, well-audited code, and dynamic or user-controlled addresses should never be passed as targets.

\textit{Unprotected selfdestruct}~\cite{Rameder2022}\citeA{Staderini2022, ZhouEtAl2022, vidal2024openscv, li2024static} is a vulnerability that occurs when a smart contract allows anyone, or unauthorized users, to trigger the \textit{selfdestruct} function. This opcode removes the contract from the blockchain and forwards its remaining ether balance to a specified address. If not properly restricted (e.g., with an \textit{onlyOwner} modifier), an attacker could permanently destroy the contract, disrupting its functionality, and potentially stealing funds. To prevent attacks, \textit{selfdestruct} should only be callable by trusted accounts, typically using access control mechanisms like \textit{Ownable}.

The potential of smart contracts to enable dApps is vast, although realizing this potential requires prioritizing security considerations. While blockchain's public nature enhances transparency, it also exposes smart contract codes to potential attackers, who can exploit weaknesses upon scrutiny. Mitigating smart contract vulnerabilities requires rigorous code audit and testing, adherence to best practices, and the implementation of upgradeable mechanisms. By prioritizing security measures, we can harness the full potential of smart contracts while minimizing associated risks.

%% file: sections/methodology.tex
\begin{figure*}[ht]
    \centering
    \includegraphics[width=1\linewidth]{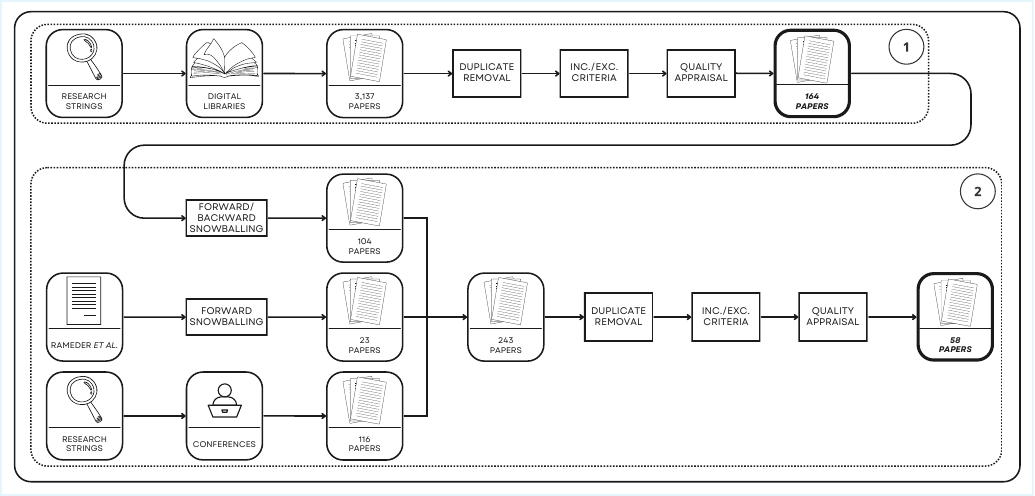}
    \label{fig:research_method}
    \caption{Research Method}
\end{figure*}

\section{Research Method}
\label{sec:methodology}

Nepomuceno and Soares~\cite{NepomucenoAndSoares2019} highlighted the desire among researchers to keep SLRs current, even if others carry out the update process.
However, Mendes \etal~\cite{MendesEtAl2020} pointed out the importance of employing decision support mechanisms to assist the software engineering community in determining the need for SLR updates.
In fact, to decide whether the literature by Rameder \etal~\cite{Rameder2022} needed to be updated, we opted to apply the 3PDF framework introduced by Garner \etal~\cite{GarnerEtAl2016} and approved by Mendes \etal~\cite{MendesEtAl2020}.

The decision framework includes three steps to be applied sequentially. 
The first step consists of answering three questions, and if the answer to at least one of Step 1's questions is 'NO,' there is no arguable need to update an SLR. Of course, the SLR addresses current problems related to Ethereum SCs, detection tools, and benchmarks; it is accessible through the open-access publication on Frontiers in Blockchain and uses a rigorous methodology, ensuring valid and well-conducted results.
The second step invites one to check whether more recent research offers significant contributions that justify updating or supplementing the systematic review. It requires at least one positive answer.
In recent years, important events related to Blockchain technology have occurred. In 2021, there was the NFT boom; in 2022, the transition of Ethereum to PoS was finalized; and in 2023, the increase in blockchain-based projects was 19\% compared to the previous year, and several companies around the world are still quietly investing in blockchain~\footnote{\url{https://www.forbes.com/sites/ninabambysheva/2023/02/07/forbes-blockchain-50-2023/}}. These changes have triggered strong interest in the research world, which has contributed to numerous studies.
The third step requires an assessment of the impact of any research updates on the SLR conclusions. In addition, the final step requires at least one positive answer. The addition of recent studies would bring new perspectives, increase the comprehensiveness of the review, improve credibility, and update the conclusions based on the latest available data.

Finally, the decision framework revealed that the SLR is suitable for upgrading, as outlined in \Cref{tab:3pdf_results}.
We deliberately chose not to modify the search strings and data extraction form of the original study, given the findings of Nepomuceno and Soares~\cite{NepomucenoAndSoares2019}, who observed that such alterations typically result in the creation of a new SLR rather than an update.
By incorporating insights from previous research and using the 3PDF framework, we aimed to ensure a systematic and informed approach to updating the SLR on automated vulnerability analysis of Ethereum smart contracts. Our methodology underscores the importance of maintaining up-to-date literature while respecting established frameworks and guidelines.

\begin{table}[ht]
    \centering
    \footnotesize
    \caption{Application of the 3PDF Framework to Rameder \etal~\cite{Rameder2022}}
    \rowcolors{3}{gray!7}{white}
    \begin{tabular}{lcccccccr}
        \toprule
        \textbf{SLR} & \textbf{1.a} & \textbf{1.b} & \textbf{1.c} & \textbf{2.a} & \textbf{2.b} & \textbf{3.a} & \textbf{3.b} & \textbf{Result}\\
        \midrule
        \cite{Rameder2022} & YES & YES & YES & - & YES & - & YES & Upgradeable\\
        \bottomrule
        \multicolumn{9}{p{13cm}}{
        \textbf{1.a} Does the published SLR still address a current question?
        }\\
        \multicolumn{9}{p{13cm}}{
        \textbf{1.b} Has the SLR had good access or use?
        }\\
        \multicolumn{9}{p{13cm}}{
        \textbf{1.c} Has the SLR used valid methods and was well-conducted?
        }\\
        \multicolumn{9}{p{13cm}}{
        \textbf{2.a} Are there any new relevant methods?
        }\\
        \multicolumn{9}{p{13cm}}{
        \textbf{2.b} Are there any new studies or new information?
        }\\
        \multicolumn{9}{p{13cm}}{
        \textbf{3.a} Will adopting new methods change the findings, conclusions, or credibility?
        }\\
        \multicolumn{9}{p{13cm}}{
        \textbf{3.b} Will the inclusion of new studies/information/data change findings, conclusions, or credibility?
        }\\
        \bottomrule
    \end{tabular}
    \label{tab:3pdf_results}
\end{table}

\subsection{Goal and Research Questions}

The \emph{goal} of this work is to investigate current research studies and the most significant contributions in the field of smart contract vulnerabilities. Concurrently, offering a comprehensive and systematic perspective can contribute to the research community by facilitating assessments and discussions on future directions related to these issues. We also aimed to comprehend how the research area has evolved. Consequently, as mentioned in the previous section, we examined the same research questions (RQs) posed by Rameder \etal~\cite{Rameder2022} except for one. 
In fact, we decided to exclude RQ$_5$ (\ie ``What are the open challenges of automated detection?'') because its systematic and rigorous replication was challenging due to the lack of a data extraction form or supplementary material that clarified how information from the primary studies was collected (\eg thematic analysis or content analysis). Consequently, we could not determine whether Rameder et al. used a systematic process to address that research question. Therefore, to avoid overclaiming systematicity, we analyzed the open challenges related to automated detection of software vulnerabilities, and discussed them in \Cref{sec:discussion}.
In detail, we posed the following \textit{RQs}.

\geriRQ{1}{Which vulnerabilities are mentioned? How are vulnerabilities classified?}

$RQ_1$ supports the understanding of smart contract vulnerabilities and their classification, which is essential for effective security management. A comprehensive taxonomy serves as a point of aggregation, helping researchers, practitioners, and auditors systematically consult, compare, and build on existing results. A taxonomy can improve risk assessment and mitigation, prioritizing efforts and allocating resources more effectively and efficiently.

\geriRQ{2}{Which methods do automated tools use to detect vulnerabilities?}

$RQ_2$ analyzes automated tools and solutions proposed to enhance, support, and facilitate the work of practitioners, researchers, and security auditors. Understanding which methods were used and how they were applied can reveal the capabilities and limitations of existing solutions, thus guiding the development of new techniques and informing the selection of tools.

\geriRQ{3}{Which vulnerabilities are addressed by the tools?}

$RQ_3$ helps assess coverage and gaps in automated vulnerability detection. It enables the selection of complementary tools for comprehensive protection, highlighting areas that require manual review or additional measures.

\geriRQ{4}{How are the tools evaluated?}

$RQ_4$ examines how the performance of vulnerability detection tools is evaluated. Proper evaluation is crucial to ensure that tools perform as expected and to allow effective comparisons between different solutions. A deeper understanding of the benchmarks and evaluation practices used in the literature could also provide insight into current trends and future directions in the field.

\begin{table}[ht]
    \centering
    \footnotesize
    \caption{Query Strings on Different Search Engines}
    \label{tab:query}
    \rowcolors{2}{white}{gray!5}
    \begin{tabular}{lp{10cm}}
        \toprule
        \textbf{Database} & \textbf{Query} \\
        \midrule
        ACM Digital Library & Title: ("smart contract" OR "smart contracts") AND AllField: (vulnerability OR vulnerabilities OR bug OR bugs OR tool OR security OR analysis OR detection OR verification) \\
        Google Scholar & allintitle: ("smart contract" OR "smart contracts" OR Ethereum) AND (vulnerability OR vulnerabilities OR bug OR bugs OR security OR analysis OR detection OR tool OR verification) \\
        IEEE Xplore & "All Metadata": ("smart contract" OR "smart contracts") AND (vulnerability OR vulnerabilities OR bug OR bugs OR tool) AND (security OR analysis OR detection OR verification) \\
        Science Direct & Title, abstract, keywords: "smart contract" AND (vulnerability OR vulnerabilities OR bug OR tool OR security OR analysis OR detection OR verification) \\
        TU Wien CatalogPlus & title contains ("smart contract" OR "smart contracts") AND Subject contains (security OR vulnerability OR vulnerabilities OR bug OR bugs Or analysis OR detection OR tool OR verification) \\
        \bottomrule
    \end{tabular}
\end{table}

\subsection{Identification of Relevant Literature}

\Cref{fig:research_method} depicts the process we followed to update the systematic literature review proposed by Rameder \etal~\cite{Rameder2022}.
Brereton \etal~\cite{Brereton2007} emphasize the importance of searching multiple databases because no single repository contains all pertinent papers. We conducted our search in five digital libraries in Step 1, with the aim of identifying relevant studies in a comprehensive way. We used the same search string (see \Cref{tab:query}) as the original SLR; consequently, the choices and motivations behind each search string are detailed in the original SLR~\cite{Rameder2022}, which also highlighted the need for a structured snowballing process to avoid overlooking essential studies.
Therefore, in addition to analyzing the libraries, we conducted a second search in Step 2. We applied backward and forward snowballing to the papers extracted from Google Scholar that passed the quality assessment stage to ensure a thorough exploration of the literature and capture relevant studies that may have been missed in the initial search.
Furthermore, we applied forward snowballing on Rameder \etal~\cite{Rameder2022} to identify additional relevant studies through references to the initial paper. 
Finally, we analyzed publications from major conferences such as ICSE, FSE, ASE, ISSTA, and ICST. These highly relevant conferences often feature high-quality research on the analyzed topics; therefore, this further step provides valuable information on current trends and developments in smart contract vulnerabilities.

\subsection{Inclusion and Exclusion Criteria}
We adopted the inclusion and exclusion criteria of the original SLR with two modifications. The original SLR, conducted at the end of January 2021, included publications from 2014. Given the nature of our study, we modified this exclusion criterion from ``published before 2014'' to ``published before 2021'', in accordance with the timeframe of our study. The second change is about the fourth exclusion criteria; it passed from ``Not accessible online via the library network of TU Wien'' to ``Not accessible online via the library network of the University of Salerno''. All other criteria remained unchanged and are shown in \Cref{tab:inc_exc_table}.

Exclusion Criterion 1 specifies the removal of papers unrelated to Ethereum. While this criterion excludes studies focusing solely on other blockchains, it does not exclude papers that, although based on other blockchains, relate their findings to Ethereum. This approach enabled us to identify cross-chain vulnerabilities, as cross-system architectures often integrate Ethereum with other blockchain platforms. Notably, Vitalik Buterin~\cite{buterin2016chain} provides a comprehensive analysis of the interoperability problem in blockchain ecosystems, highlighting the relevance of cross-chain considerations.

\begin{table}[ht]
    \centering
    \footnotesize
    \caption{Inclusion and Exclusion Criteria}
    \label{tab:inc_exc_table}
    \rowcolors{2}{white}{gray!5}
    \begin{tabular}{l}
        \toprule
        \textbf{Inclusion Criteria} \\
        \midrule
        1 - SC security bugs or vulnerabilities \\
        2 - SC vulnerability detection, analysis, or security verification tools \\
        3 - Automated detection or verification methods \\
        \toprule
        \textbf{Exclusion Criteria} \\
        \midrule
        1 - Unrelated to Ethereum \\
        2 - Not written in English \\
        3 - A patent, blog entry, or other grey literature \\
        4 - Not accessible online via the library network of the University of Salerno \\
        5 - Published before 2021 \\
        \bottomrule
    \end{tabular}
\end{table}

\subsection{Selection and Classification}
We carefully reviewed and assigned each study to one of these three distinct groups.

\textbf{Systematic Literature Reviews (SLRs)} include comprehensive reviews that systematically collect, critically analyze, and synthesize existing research on smart contract security.
    
\textbf{Surveys and review studies} on smart contract security analysis tools, methods, and vulnerabilities provide overviews and comparative analyses of the state of the art.
    
\textbf{Primary Studies (PSs)} consist of original research on developing and evaluating smart contract security analysis and vulnerability detection tools and methods. Primary Studies provide novel methodologies and empirical findings.

\subsection{Quality Appraisal}
\label{sub:quality_appraisal}
The quality of the papers was evaluated in detail by applying the metrics for intrinsic and contextual data quality described by Strong \etal~\cite{Strong2002}.

\textbf{Intrinsic Data Quality (IDQ)} assesses the accuracy, objectivity, credibility, and reputation of the publication venues. 
We converted other metrics, including the SCImago Journal Ranking, the CORE Journal or Conference Ranking, and the Scopus CiteScore percentile ranking, into scores ranging from 0 to 1. \Cref{tab:idq_score} shows the IDQ score for the ranking of the venues.
In cases where a journal or conference has yet to be ranked for the current year, we consider the most recent ranking available. We select the highest score as the IDQ (Indicator of Journal/Conference Quality) if multiple rankings are available.

\textbf{Contextual Data Quality (CDQ)} assesses the relevance, added value, timeliness, completeness, and amount of data, and thus depends on the context of the evaluation, \eg the purpose of the systematic review, the research questions, and the data to extract and analyze. The CDQ score is the arithmetic mean of the answers to the questions, each a number between 0 and 1. The questionnaire is based on the DARE criteria suggested by Kitchenham and Charters~\cite{Kitchenham2007} and consists of seven questions for SLRs and surveys (see \Cref{tab:cqd_slr}) and two questions for Primary Studies (see \Cref{tab:cqd_ps}).

\textbf{Final Data Quality (FDQ)} combines IDQ and FDQ. We compute the FDQ score as the arithmetic mean of the other two scores; thus, it is a number between 0 and 1.
We map CDQ and FDQ to a three-point Likert scale for readability and ease of classification. We consider a score low if it is below 0.5, high if it is at least 0.8, and medium otherwise. Studies with a CDQ or FDQ score below 0.5 are excluded.

\begin{table}[ht]
    \centering
    \footnotesize
    \caption{IDQ Assessment by Venue Ranking}
    \label{tab:idq_score}
    \rowcolors{2}{gray!5}{white} 
    \begin{tabular}{lllr}
        \toprule
        \textbf{SCImago} & \textbf{CORE} & \textbf{Scopus} & \textbf{IDQ}\\
        \midrule
        Q1 & A*, A & $\geq 75^{th}$ percentile & 1.0\\
        Q2 & B & $\geq 50^{th} \land <75^{th}$ percentile & 0.8\\
        Q3, Q4 & C & $\geq$ 25$^{th} \land <50^{th}$ percentile & 0.5\\
        \multicolumn{3}{l}{not indexed, but pub. by IEEE, ACM, Springer} & 0.4\\
         &  & $<$ 25$^{th}$ percentile & 0.2\\
        \multicolumn{3}{l}{other sources, \eg arXiv, preprints, theses, reports} & 0.2\\
        \bottomrule
    \end{tabular}
\end{table}

\subsection{Data Extraction}

Once we identified the final set of sources, we extracted information relevant to our research questions, following the Nepomuceno and Soares guidelines~\cite{NepomucenoAndSoares2019}. We used the same data extraction forms as Rameder \etal~\cite{Rameder2022}, ensuring consistency in collecting and organizing data on vulnerabilities, analysis tools, datasets, and test sets.
We collected detailed information about vulnerabilities, including their identifier or code, name and alternate names, description, category, and reference paper. 
With respect to tools and frameworks, we collected essential information such as their name, description, and the article that introduced them. We also reported whether the tool is open source, provided links to the tool's page or GitHub repository, and provided the publication date and the last update date. Additionally, we documented the source language. Technical specifications were also recorded, including the input needed by the tools (\ie EVM bytecode or source code), its purpose, the type of analysis conducted, and any code transformation techniques and analysis methods employed. This comprehensive collection allowed us to assess the tools thoroughly.
For datasets and test sets used to compute tool benchmarks, we noted the reference article introducing the dataset, the tools or projects using it, and the repository link. We also recorded the number of contracts within the dataset and the form in which the contracts are represented (\ie Solidity source code, bytecode, address). Furthermore, we checked the analysis results and any identified exploits. This detailed approach ensured that we clearly understood the roles of the datasets and test sets in tool evaluation.
By meticulously collecting and organizing this information, we ensured an accurate and systematic approach to understanding vulnerabilities, analysis tools, and performance benchmarks. This methodical process enhanced the reliability and comprehensiveness of our study.

\input{tables/CDQ_slr_survey}

\input{tables/CDQ_ps}

\subsection{Search Process Execution}

The search was carried out between February and March 2025. We started the research with Step 1, which yielded 3,137 results from the five digital libraries. After removing 392 duplicates, we retained 2,745 unique papers. Then, we applied inclusion and exclusion criteria. \Cref{tab:criteria} shows the occurrences of each criterion. An article was included if it satisfied at least one inclusion criterion and no exclusion criteria. After applying the inclusion and exclusion criteria, 260 papers were retained for the quality appraisal. In total, quality appraisal in Step 1 produced 164 papers, of which 12 were SLRs~\citeA{abubakar2024systematic, faruk2024systematic,
hejazi2025comprehensive, YuliantoEtAl2023, KushwahaEtAl2022a, KushwahaEtAl2022b, PiantadosiEtAl2022, Rameder2021, dZaazaaAndElBakkali2023b, MahugnonAndJules2022, VaccaEtAl2021, kiani2024ethereum}, 22 were surveys ~\citeA{wu2024we,li2024static,vidal2024openscv,Zheng2023rudder,Huang2024Revealing,
hajihosseinkhani2024unveiling,mishra2024smart,huang2024sword,obradovic2024blockchain,
wu2024comprehensive,zhu2024survey, ChuEtAl2023, ChenEtAl2022, Z.ZhangEtAl2023, DaojingHeEtAl2023, QianEtAl2023a, JiangEtAl2023, ChaliasosEtAl2024, ZhouEtAl2022, Staderini2022, dZaazaaAndElBakkali2023a, AgarwalEtAl2022}, and 130 were primary studies~\citeA{iuliano2025automated,chen2024improving,zhang2024towards,sun2024gptscan,luo2024scvhunter,deng2024safeguarding,chen2024verifying,wang2024skyeye,chen2024opentracer,li2024cobra,wu2024advscanner,wang2024contracttinker,eshghie2024highguard,eshghie2024oracle,chen2024demystifying,liu2024funredisp,liu2024funredisp2,chen2024identifying,ye2024midas,chen2023chatgpt,guo2024smart,ye2024funfuzz,song2022esbmc,liu2022finding,li2021gas,zheng2024lent,zhang2024evm,mothukuri2024llmsmartsec,Ozdemir2024,jain2024integrated,chen2024safecheck,he2024parse,liu2024ffgdetector,qian2024mufuzz,ibba2024mindthedapp,ding2024hunting,wang2024smartinv,khan2024involuntary,wang2024scvd,guo2024sernet,li2024femd,dai2024smart,luo2024fellmvp,li2024evofuzzer,wu2023defiranger,khanzadeh2024solosphere,zeng2023solgpt,he2023unknown,chu2024deepfusion,he2024reensat,wang2024contractgnn,cao2023sccheck,dong2024erinys,fang2023beyond,DuanEtAl2023, JieEtAl2023, L.ZhangEtAl2022a, VivarEtAl2021, OtoniEtAl2023, GhalebEtAl2023, JinEtAl2022, SunEtAl2023, SujeethaAndAkila2023, NarayanaAndSathiyamurthy2021, SoudEtAl2023, XieEtAl2023, L.ZhangEtAl2022c, HwangEtAl2022, CaiEtAl2023, LiuEtAl2023a, YangEtAl2024, QianEtAl2023c, CaoEtAl2023a, YuEtAl2021, ChenEtAl2021, QianEtAl2023b, JeonEtAl2023, HuEtAl2023b, EshghieEtAl2021, DeSalveEtAl2024, JianzhongSuEtAl2022, JiEtAl2023, TorresEtAl2022, PasquaEtAl2023, LutzEtAl2021, GhalebEtAl2022, AshizawaEtAl2021, WangEtAl2022, LinoyEtAl2022, Mazurek2021, ZhuEtAl2023, LongHeEtAl2023, GongEtAl2023, MaEtAl2023, HuangEtAl2021, YangEtAl2023, NguyenEtAl2023, Hong-NingEtAl2021, FeiEtAl2023, SunEtAl2022a, ZhengEtAl2022, L.YuEtAl2023, BarboniEtAl2023, LiuEtAl2023b, BoseEtAl2021, CaoEtAl2023b, LiangAndZhai2023, AliEtAl2021, WangEtAl2023, NguyenEtAl2021, J.ZhangEtAl2023, GuoEtAl2024, L.ZhangEtAl2022, diAngeloEtAl2023, LiEtAl2022, LiaoEtAl2023, PraitheeshanEtAl2021, HuEtAl2023a, L.ZhangEtAl2022b, BarboniEtAl2022, DaiEtAl2022, H.ZhangEtAl2023, OlsthoornEtAl2022, ChenEtAl2023, Y.ZhangEtAl2022, R.YuEtAl2023, HeEtAl2024, HajiHosseinKhaniEtAl2024, LiEtAl2023, JiamingYeEtAl2022}.

Afterward, we started Step 2, conducting forward and backward snowballing on the 164 papers obtained in Step 1. During snowballing, pre-2021 papers were not considered due to one of the exclusion criteria. Snowballing contributed 104 papers. We also conducted forward snowballing on Rameder \etal~\cite{Rameder2022}, which helped to find another 23 papers. Finally, we analyzed the publications at major conferences. We manually applied a research string to papers to collect only relevant papers. Conferences contributed another 116 papers. In total, in Step 2, we analyzed 243 papers, 133 of which were removed because they were duplicates. We applied the inclusion and exclusion criteria to the remaining 110 papers and obtained 72 papers. 
Quality appraisal introduced 58 papers, of which seven were surveys~\citeA{
ren2021empirical,zheng2024dappscan,DiasEtAl2021, diAngeloAndSalzer2023, diAngeloEtAl2024, MunirAndTaha2023, WeiEtAl2023} and 51 were primary studies~\citeA{
yang2024uncover,godboley2024poster,liang2024towards,wang2025contractsentry,jj2024enhancing,rodler2023ef,liu2024automated,liu2022invcon,zhong2024prettysmart,liao2024smartaxe,wang2024efficiently,groce2021echidna,ji2021increasing,chen2021sigrec,gao2020checking,wang2024contractcheck,zhang2024scanogenerator,yang2025csafuzzer,gao2024sguard,wang2024dfier,SunEtAl2022b, XuEtAl2021, TorresEtAl2021, RongzeXuEtAl2022, LakadawalaEtAl2024, LiuEtAl2024, HanEtAl2024, ZengEtAl2022, ControEtAl2021, NassirzadehEtAl2021, ShouEtAl2023, FengEtAl2023, NguyenEtAl2022, SongEtAl2023, WuHEtAl2021, PanEtAl2021, RutaoYuEtAl2021, Z.ZhangEtAl2022, MengRenEtAl2021, HuangEtAl2022, LiuZEtAl2021, LiaoEtAl2022, SunbeomSoEtAl2021, ZhouEtAl2021, ZhukovAndKorkhov2023, ChoiEtAl2021, SamreenAndAlalfi2021, TorresEtAl2021b, MiEtAl2021, JiangEtAl2021, XueEtAl2022}.

In this paper, [Ax] citations refer to studies selected in our SLR and listed in the SLR References section, while [x] citations refer to works included in the main References section. 

 \begin{table}[htbp]
    \centering{
    \scriptsize
    \caption{Statistics about the Search Process Execution.}
    \label{tab:criteria}
    \resizebox{\columnwidth}{!}{
    \begin{tabular}{
    l 
    >{\columncolor{gray!7}}r 
    rr 
    >{\columncolor{gray!7}}r 
    >{\columncolor{gray!7}}r
    >{\columncolor{gray!7}}r
    rrrrr 
    >{\columncolor{gray!7}}r 
    r}
    \toprule
        ~ & \cellcolor{white}\textbf{Init Set} & \textbf{Dupl} & \textbf{Unique} & \cellcolor{white}\textbf{Inc 1} & \cellcolor{white}\textbf{Inc 2} & \cellcolor{white}\textbf{Inc 3} & \textbf{Exc 1} & \textbf{Exc 2} & \textbf{Exc 3} & \textbf{Exc 4} & \textbf{Exc 5} & \cellcolor{white}\textbf{Post Crit} & \textbf{QA}\\
    \midrule
        \textbf{Step 1} & 3,137 & 392 & 2,745 & 92 & 158 & 72 & 385 & 25 & 35 & 29 & 4 & 260 & 164 \\ 
        \textbf{Step 2} & 243 & 133 & 110 & 11 & 44 & 25 & 5 & 2 & 0 & 1 & 1 & 72 & 58 \\ 
        \textbf{Total} & 3,380 & 525 & 2,855 & 103 & 202 & 97 & 390 & 27 & 35 & 30 & 5 & 332 & 222 \\
    \bottomrule
    \end{tabular}
    }
    }
\end{table}

We analyzed 3,380 publications, 3,137 in step one and 243 in step two. The first step contributed 164 papers, and the second contributed 58 papers, for a total of 222 papers.
As shown in \Cref{tab:criteria}, the most frequent inclusion criterion was the second (\ie SC vulnerability detection, analysis, or security verification tools), while the most frequent exclusion criterion was the first (\ie Unrelated to Ethereum).

Table~\ref{tab:qa_results} summarizes the quality assessment scores of the papers analyzed in the review. The average IDQ score increased from 0.74 to 0.86 among the papers that passed the QA step, indicating that higher-quality studies were retained. Moreover, the quartile values show that the majority of the included papers exhibit high and consistent quality, with 75\% of the studies scoring above 0.8, confirming the overall robustness and homogeneity in terms of quality of the selected literature.

\begin{table}[ht]
\centering
\myColor{
\caption{Quality Assessment results for evaluated papers.}
\label{tab:qa_results}
\resizebox{0.8\columnwidth}{!}{
\begin{tabular}{lcccc}
\toprule
\textbf{Paper set} & \textbf{IDQ average} & \textbf{1st QR} & \textbf{2nd QR (Median)} & \textbf{3rd QR} \\
\midrule
332 papers (reached QA step) & 0.74 & 0.50 & 1.00 & 1.00 \\
222 papers (passed QA step)  & 0.86 & 0.80 & 1.00 & 1.00 \\
\bottomrule
\end{tabular}
}}
\end{table}

\subsubsection{Search-string sensitivity to evolving terminology}
The SLR we are updating was initially conducted in 2021. In recent years, new terminology has emerged, partly due to updates to the Ethereum Virtual Machine (EVM). Our study adopts the same search strings used in 2021. To ensure these strings remain sufficiently comprehensive, we conducted a sensitivity analysis to assess how the inclusion of newly introduced terms in the search string would affect results. We conducted a sensitivity analysis on Step 2 of our research method. We included 29 new emerging keywords in the research string used to explore the main conferences. Newer terms led to the discovery of two new papers not covered by the original research string. The two additional papers suggest that the difference is marginal ($\approx5\%$), which indicates that our original search query was already reasonably comprehensive. After all, terms like \textit{vulnerability}, \textit{detection}, \textit{security}, \textit{bugs}, are well-established and widely used in the literature, effectively capturing the vast majority of relevant studies despite the emergence of newer terminology. We included the sensitive analysis in the replication package.

\subsection{Data Synthesis and Analysis}
\label{sec:data_analysis}
We analyzed the studies selected above to answer our research questions.
To answer $RQ_1$, we focused on the studies that mention vulnerabilities. In particular, to classify vulnerabilities and organize them hierarchically, we followed a series of steps:

\begin{enumerate}
\item \textbf{Pilot Study.} We randomly selected and analyzed 20 sources to establish an initial taxonomy draft, as suggested by Ralph~\cite{Guidelines4Taxonomy}.

\item \textbf{Pilot Study: Inter-rater Assessment.} We performed an inter-rater assessment of the vulnerabilities to reach unanimity on their classification, clustering, and hierarchical position, leading to several changes.

\item \textbf{Full Study.} The rest of the collected papers were analyzed following the consensus reached in the previous step. The sources were split into two halves, each independently analyzed by one researcher.

\item \textbf{Full Study: Inter-rater Assessment.} After the independent extraction of vulnerabilities, the researchers looked at each other's taxonomy to inter-rater assess them. All discrepancies were solved through discussions and reasoning.
\end{enumerate}

Each researcher followed the following steps:

\begin{enumerate}
\item \textbf{Vulnerability Collection.} We extracted all information related to vulnerabilities, \eg their name, ID, description, cause, classification, and eventually, the alternative names used to refer to them.

\item \textbf{Duplicate Removal.} We focused exclusively on vulnerabilities with the same name or similar description to reduce the number of vulnerabilities but also to ensure that in the literature, two vulnerabilities with the same name describe the same issue.

\item \textbf{Vulnerability Clustering.} We noted numerous vulnerabilities with similar names, which allowed us to divide them into unique clusters and analyze them deeply. 

Clusters of synonyms were created by joining terms that describe the same scenario. We primarily focused on the vulnerability description to understand the type of problem it represents. Then we added the vulnerability name to the cluster of names representing the same scenario. For each cluster, we documented the decision rule used to select a representative name for all its synonyms. Since our work extends and updates a previous taxonomy~\cite{Rameder2022}, we first prioritized names already used in that taxonomy. For newly introduced vulnerabilities, we prioritized names already used in other taxonomies.


\item \textbf{Vulnerability Classification.} To categorize vulnerabilities, we focused on their root causes rather than their effects or attack vectors. Root causes are less ambiguous and help minimize classification ambiguities, providing a more precise, more reproducible taxonomy. We used categories proposed by Rameder \etal~\cite{Rameder2022}, as starting set.
When the categories were insufficient to cover a particular vulnerability, we introduced a new category.

\item \textbf{Hierarchical Organization.} The previous cluster analysis revealed that vulnerabilities are present in the literature with numerous alternative names and are described in various levels of detail. Therefore, we organized the vulnerabilities hierarchically, placing them on different levels: the first level, which is more general and representative, and the subsequent levels, which are more detailed.

\item \textbf{Existing Terminology Usage.} Due to the vast terminology used in the literature to refer to vulnerabilities, our work of classifying and organizing them was based exclusively on existing terminology without introducing new identifiers and names or adapting existing nomenclatures. 
Since our work extends and updates a previous taxonomy~\cite{Rameder2022}, we prioritized the names already employed. We selected the most frequent names in the state of the art for the new vulnerabilities. When the number of occurrences in the literature was equal, we selected the most explanatory name.
\end{enumerate}

The approach we used, particularly Step 4, enables us to adapt the taxonomy to emerging vulnerabilities that were not previously considered in existing taxonomies and are not represented by any categories in the state-of-the-art taxonomies. 
The entire process makes the taxonomy easily upgradable by researchers in future works, as updates to the Ethereum Virtual Machine or Solidity may introduce new vulnerabilities.
When new vulnerabilities are discovered, the proposed taxonomy should be consulted to determine whether an existing category is suitable for classification. If no existing category is sufficiently representative, the taxonomy should be extended to include a new representative category.

\paragraph{Semi-structured Interviews for Taxonomy Validation}

The proposed taxonomy extends the validated taxonomy of Rameder et al.~\cite{Rameder2022} by incorporating additional vulnerabilities from other validated taxonomies, such as OpenSCV~\citeA{vidal2024openscv}.
Vulnerabilities whose classifications remained unchanged from these already validated sources were excluded from our validation process. Accordingly, the validation focused on vulnerabilities originating from non-validated taxonomies, those from validated taxonomies whose classifications had been revised, and newly introduced vulnerabilities positioned in hierarchical relationships to existing ones.
We excluded already-validated vulnerabilities from three taxonomies~\cite{Rameder2022}~\citeA{vidal2024openscv, li2024static}. The remaining vulnerabilities were validated through semi-structured interviews with four security and Solidity experts. 
We divided the vulnerabilities into two clusters, each of which was validated by two experts. 

The semi-structured interviews were conducted in two stages.
In the first stage, we validated the vulnerability categories and their completeness. Given a vulnerability and the list of categories, the participant was asked to select the most appropriate category based on the vulnerability’s root cause. If, according to the participant, no category adequately represented the vulnerability, we discussed it with them to reach a shared agreement.
In the second stage, we validated the hierarchical structure. Given the vulnerability under validation and the list of vulnerabilities in the selected category from the first stage, the participant was asked to identify any hierarchical relationships with other vulnerabilities in the same category. All responses were accompanied by justifications to assess the consistency and validity of the agreements.
The semi-structured interviews allowed us to combine systematic validation with exploratory discussion.
This format enabled participants not only to provide their judgments on specific validation tasks but also to share broader reflections and observations related to the taxonomy.
By offering the flexibility to explore topics beyond the predefined questions, the interviews allowed participants to highlight cross-cutting aspects, identify potential ambiguities, and propose improvements that might not have been expected by the interviewer but were nevertheless relevant to the nature of the validation process.

To answer $RQ_2$, we applied Closed Coding~\cite{saldana2021coding}, adopting the same categories used by Rameder \etal~\cite{Rameder2022} to collect all the information from the tool. The three main areas analyzed are functionality, code transformation technique, and methods, each comprising several categories: 

\begin{description}
\item[\textbf{Functionalities:}] Vulnerability Detection, Program Correctness, Gas or Resource Analysis, Bulk Analysis or Full Automation, Exploit Generation, Reactive Defenses, Manual Analysis or Developer Support.

\item[\textbf{Code Transformation Techniques:}] Control Flow Graph, Data Flow, Transaction Analysis, Traces, Abstract Syntax Tree, Decompilation, Compiler, Intermediate Representation, Specification Language, Disassembly, Finite State Machine.

\item[\textbf{Methods:}] Symbolic Execution, Formal Verification, Constraint Solving, Model Checking, Abstract Interpretation, Fuzzing, Runtime Verification, Concolic Testing, Mutation Testing, Pattern Matching, Code Instrumentation, Machine Learning, Taint Analysis.
\end{description}

We also collected additional information, such as tool name, brief description, link to the repository, date of publication, last update, development languages, input type, and analysis type.
The vulnerabilities identified in $RQ_1$ and the tools identified in $RQ_2$ provided the basis to answer $RQ_3$.  
We identified the vulnerabilities detected by each tool and created a detailed mapping between them.

To ensure that the mapping accurately reflected tools targeting explicit vulnerabilities, we applied an inclusion–exclusion process based on the tool's objectives and scope. Each tool mentioned in the reviewed studies was analyzed according to its primary purpose, detection granularity, and methodological focus. Tools were included only if they were specifically designed to identify or analyze well-defined vulnerability types in smart contracts. Conversely, we excluded tools that serve general compilation or decompilation purposes, those that perform contract-level vulnerability prediction through machine learning without identifying specific weakness types, and mutation-based approaches that do not implement vulnerability-oriented operators. This filtering ensured that only tools with explicit vulnerability-detection capabilities were retained in the mapping.

Finally, to answer $RQ_4$, we adopted the Closed Coding approach~\cite{saldana2021coding}. We collected various benchmarks to evaluate tools or conduct other experiments. We gathered the same information extracted by Rameder \etal, like the smart contract types (\ie source code, bytecode, or addresses), their size in terms of the number of SCs, the associated links, if present, and the analysis results when available. Furthermore, we mapped the benchmarks to the tools to evaluate and compare their performance.

\subsection{Replication Package}
To facilitate the validation and replication of our study, we have made the replication package online~\footnote{https://doi.org/10.6084/m9.figshare.27224493}. It contains the full list of studies that emerged from queries on digital libraries, the results of snowballing, and all conference papers. The inclusion/exclusion criteria and the results of the qualitative analysis are presented for each paper. Finally, a complete and comprehensive taxonomy, a full list of tools, a mapping between tools and vulnerabilities, and benchmarks used for tool evaluation or other experiments are included. We also provide a webpage\footnote{{https://gerardoiuliano.github.io/slr.html}} and a GitHub repository\footnote{{https://github.com/GerardoIuliano/SLR-Ethereum}} for easy consultation of the results.

%% file: tables/CDQ_slr_survey.tex
\begin{table}[h]
    \centering
    \scriptsize
    \caption{CDQ Assessment Criteria of SLRs and Surveys}
    \begin{tabular}{lp{13.8cm}}
    
        \toprule
        
        \multicolumn{2}{p{16cm}}{\textbf{Q1: Is the methodology and literature search of the review systematic and documented? Are all relevant studies at the time likely to be covered?}} \\
        \midrule
        \rowcolor{gray!6}
        \textbf{1.0} & Yes, in comprehensive detail, systematic and good quality. To be (re)classified as SLR. \\
        \textbf{Not Graded} & No, to be (re)classified as survey.\\
        
        \midrule
        
        \multicolumn{2}{p{16cm}}{\textbf{Q2: Does the study focus on work about smart contract vulnerabilities, detection tools, or approaches to automated detection or verification?}} \\
        \midrule
        \rowcolor{gray!6}
        \textbf{1.0} & Yes, the research topic closely related and covers major parts of this work. \\
        \textbf{0.6} & It covers related topics, but important areas differ or are missing.\\
        \rowcolor{gray!6}
        \textbf{0.3} & Some similar areas are covered.\\
        \textbf{0.0} & The review focuses on different research topics, only a very small part is related.\\
        
        \midrule
        
        \multicolumn{2}{p{16cm}}{\textbf{Q3: Does the study cover the evaluation of smart contract vulnerabilities?}} \\
        \midrule
        \rowcolor{gray!6}
        \textbf{1.0} & Yes, the work includes a survey, evaluation and detailed description of vulnerabilities including classifications. \\
        \textbf{0.6} & Descriptions or classifications of certain vulnerabilities are given, but it is not a major focus of the work.\\
        \rowcolor{gray!6}
        \textbf{0.3} &  Some vulnerabilities are superficially listed or mentioned.\\
        \textbf{0.0} &  No, the focus is not on the survey of vulnerabilities.\\
        
        \midrule
        
        \multicolumn{2}{p{16cm}}{\textbf{Q4: How many tools or analysis/verification methods does the review identify, classify, compare or evaluate?}} \\
        \midrule
        \rowcolor{gray!6}
        \textbf{1.0} & more than 12. \\
        \textbf{0.6} & 8 to 12.\\
        \rowcolor{gray!6}
        \textbf{0.3} &  4 to 7.\\
        \textbf{0.0} &  less than 4.\\

        \midrule
        
        \multicolumn{2}{p{16cm}}{\textbf{Q5: Did the author(s) assess and compare the quality/validity of tools or automated detection/verification approaches in detail?}} \\
        \midrule
        \rowcolor{gray!6}
        \textbf{1.0} & Yes, in detail and high quality, including practical experiments or comprehensive/in-depth coverage and classification. \\
        \textbf{0.6} & Yes, compared/evaluated in some detail, but only theoretically.\\
        \rowcolor{gray!6}
        \textbf{0.3} &  Only superficial description.\\
        \textbf{0.0} &  No.\\

        \midrule
        
        \multicolumn{2}{p{16cm}}{\textbf{Q6: Is a set of smart contract benchmarks provided?}} \\
        \midrule
        \rowcolor{gray!6}
        \textbf{1.0} & Yes. \\
        \textbf{Not Graded} & No.\\

        \midrule
        
        \multicolumn{2}{p{16cm}}{\textbf{Q7: How current is the review?}} \\
        \midrule
        \rowcolor{gray!6}
        \textbf{1.0} & 2023/2024. \\
        \textbf{0.5} & 2022. \\
        \rowcolor{gray!6}
        \textbf{0.0} & 2021.\\
        
        \bottomrule
        
    \end{tabular}
    \label{tab:cqd_slr}
\end{table}

%% file: tables/CDQ_ps.tex
\begin{table}[!h]
    \centering
    \scriptsize
    \caption{CDQ Assessment Criteria of Primary Studies (PSs)}
    \begin{tabular}{lp{15cm}}
    
        \toprule
        
        \multicolumn{2}{p{15.8cm}}{\textbf{Q1: Is the PS referenced in a selected SLR or survey? How often is it cited according to Google Scholar?}} \\
        \midrule
        \rowcolor{gray!6}
        \textbf{1.0} &  referenced in selected SLR or Survey, or cited more than 12 times according to Google Scholar. \\
        \textbf{0.6} &  8 to 12 citations. \\
        \rowcolor{gray!6}
        \textbf{0.3} &  4 to 7 citations. \\
        \textbf{0.0} &  less than 4 citations.\\
        
        \midrule
        
        \multicolumn{2}{p{15.8cm}}{\textbf{Q2: Is the PS related to an executable tool or to a framework for verifying security properties or identifying vulnerabilities of Ethereum smart contracts?}} \\
        \midrule
        \rowcolor{gray!6}
        \textbf{1.0} & Yes. \\
        \textbf{0.0} & No. \\
        
        \bottomrule
        
    \end{tabular}
    \label{tab:cqd_ps}
\end{table}

%% file: sections/results.tex
\input{figures/fig_publication_trend}

\section{Analysis of the Results}
\label{sec:results}

Before delving into the results to address the RQs, we provide some meta-information about the studies in our systematic literature review. \Cref{fig:publication_trend} shows the number of papers published yearly. The plot shows a slight growth in the number of papers published recently. However, the graph represents only the papers that have passed the quality assessment. The trend may suggest that several other papers will be published soon, which motivates our work. Furthermore, 2025 was excluded from the plot because we analyzed only two months of data.

\input{sections/results_rq1}
\input{sections/results_rq2}

\input{sections/results_rq3}

\input{sections/results_rq4}

%% file: figures/fig_publication_trend.tex
\begin{figure}[ht]
\centering
\footnotesize

\begin{tikzpicture}
\begin{axis}[
    ybar,
    width=8cm,
    height=4.5cm,
    bar width=0.25,
    enlargelimits=0.25,
    ylabel={Number of publications},
    xtick=data,
    xticklabels={2021, 2022, 2023, 2024},
    legend style={at={(0.5,-0.2)},
      anchor=north,
      legend columns=-1,
      draw=none,
      /tikz/every even column/.append style={column sep=0.8cm}},
    ]

    \addplot+[nodes near coords, nodes near coords style={color=black}, color=cold1] coordinates {(2021,3) (2022,3) (2023,2) (2024,3)};
    \addplot+[nodes near coords, nodes near coords style={color=black}, color=cold2] coordinates {(2021,2) (2022,4) (2023,10) (2024,13)};
    \addplot+[nodes near coords, nodes near coords style={color=black}, color=cold3] coordinates {(2021,32) (2022,33) (2023,42) (2024,73)};
    
    \legend{SLR, Survey, PS}
\end{axis}
\end{tikzpicture}
\caption{Publication trend per year for SLRs, Surveys, and Primary Studies.}
\label{fig:publication_trend}
\end{figure}

%% file: sections/results_rq1.tex
\subsection{RQ1. Which Vulnerabilities are Mentioned? How are Vulnerabilities Classified?}
\label{sec:rq1}

Following the steps described in \Cref{sec:data_analysis}, we answered to $RQ_1$ realizing a structured taxonomy.
Initially, we identified 561 vulnerabilities. The first challenge was to remove all duplicates, which allowed us to reduce this number to 420. 
The second step was to analyze the remaining vulnerabilities to cluster them.
We divided the 420 vulnerabilities into 192 unique clusters. For each cluster, a vulnerability was placed as a cluster representative. In contrast, the remaining 228 vulnerabilities were posed as alternative names. In other words, we have identified 192 vulnerability clusters with alternative names used in the literature to refer to the same phenomenon.
For example, one cluster comprises \textit{Block State Dependency}~\citeA{ChoiEtAl2021} and \textit{Block Info Dependency}~\citeA{ChenEtAl2022}, two vulnerabilities with similar names and expressing the same concept. Specifically, this vulnerability arises from smart contracts that rely on blockchain state attributes, which can be exploited due to the predictable nature of specific blockchain attributes. In the context of Ethereum, miners can manipulate these attributes of the block to delay submission by up to 900 ms~\citeA{dZaazaaAndElBakkali2023b}. 
Another example is the cluster that includes \textit{Timestamp dependency}~\citeA{ChuEtAl2023, ZhouEtAl2022, KushwahaEtAl2022a, KushwahaEtAl2022b, Staderini2022}~\cite{Rameder2022}, \textit{Timestamp restrictions}~\citeA{VivarEtAl2021}, \textit{Time Constraints}~\cite{Rameder2022}, \textit{Time manipulation}~\citeA{AgarwalEtAl2022}~\cite{Rameder2022}, \textit{Block values as a proxy for time}~\citeA{dZaazaaAndElBakkali2023a, dZaazaaAndElBakkali2023b}, \textit{Incorrect use of event blockchain variables for time}~\citeA{vidal2024openscv}. These vulnerabilities express the same concept: the block's timestamp should not be used for precise calculations or critical time dependencies in the contract logic.
As previously explained, many vulnerabilities could overlap across multiple categories; therefore we focused on their root cause rather than their effect or attack vector.
For instance, Unchecked low-level call/send return value (3A) refers to cases where a return value is not correctly handled or checked. While an unhandled exception may cause a Denial of Service, the vulnerability itself stems from the missing or unhandled check of the return value. Therefore, we placed 3A under \textit{Exception \& Error Handling Disorders}.
Similarly, Reentrancy (1A1) is an input- or transaction-driven vulnerability. Once exploited, it can drain all funds (making the contract Greedy, 1J) or exhaust gas through recursive calls (leading to DoS due to block gas limit, 4I). Its effects can vary, but its primary cause is a crafted malicious transaction that triggers the vulnerable external call, along with a malicious contract prepared to exploit it. For these reasons, we selected the \textit{Malicious Environment, Transactions, or Input} category. 
Finally, Timestamp Dependency (2A1) arises from dependency on a block property. If a timestamp is used to lock resource access, it may introduce an access-control weakness; if it is used to generate randomness, it becomes a weak source of randomness. Again, the primary cause, block property dependency, guided our categorization of 2A1 as \textit{Environment/Blockchain Dependency}.

The next step was to hierarchically organize the clusters based on the level of detail in their descriptions. As shown in \Cref{tab:final_tax}, we organized 113 vulnerabilities at the first level, 73 at the second level, and six at the third level. 

The hierarchical organization reveals that the well-known vulnerability, \textit{1A1-Reentrancy}, is actually a specific instance of a broader vulnerability, \textit{1A-Call to the unknown}~\cite{Rameder2022}. In addition, the \textit{Reentrancy} vulnerability exhibits four distinct behaviors (1A1.1, 1A1.2, 1A1.3, and 1A1.4). Similarly, the \textit{3B1-Mishandled out-of-gas exception} is a subcategory of the broader \textit{3B-Unexpected throw or revert} vulnerability, as also exposed in previous research by Rameder \etal~\cite{Rameder2022}.
The sheer variety of names that refer to a single vulnerability category is remarkable. The literature references vulnerabilities such as 1B, 2D, 3C, 5B1, 5B5, 7G, and 7I6 using at least seven names.
Rameder's taxonomy~\cite{Rameder2022} addresses this issue by grouping several vulnerabilities under alternative names or synonyms for a single vulnerability. For instance, \textit{5C1-Missing protection against signature replay attacks}, \textit{5C2-Lack of proper signature verification}, \textit{5C3-Hash collisions with multiple variable-length arguments}, \textit{5C4-Function clashing}, and \textit{5C5-Signature malleability} are all alternative names for \textit{5C-Signature-based vulnerabilities}. Conversely, the taxonomy proposed by Zaazaa and El Bakkali~\citeA{dZaazaaAndElBakkali2023a} classifies each vulnerability independently and further delineates their differences. This divergence in classification was resolved by organizing vulnerabilities into hierarchical levels: \textit{5C-Signature-based vulnerabilities} occupy the first level due to its general nature, whereas the specific vulnerabilities are placed at the second level to distinguish their unique characteristics. Similar hierarchical structuring has been applied to other vulnerabilities, such as \textit{1A1:1A1.1-1A1.2-1A1.3-1A1.4}, \textit{1B:1B1-1B2}, \textit{2A:2A1-2A2-2A3}, \textit{2B:2B1-2B2-2B3-2B4}, \textit{2E:2E1-2E2-2E3}, \textit{2F:2F1-2F2-2F3}, \textit{3A:3A1-3A2-3A3}, \textit{4C:4C1-4C2}, \textit{5B:5B1-5B2-5B3-5B4-5B5-5B5.1-5B5.2-5B6}, and \textit{6A:6A1-6A2}.

\begin{table}[ht]
    \centering
    \myColor{
    \caption{Experts experiences.}
    \label{tab:experts}
    \resizebox{\columnwidth}{!}{
    \rowcolors{2}{gray!10}{white}
    \begin{tabular}{cccc}
        \toprule
        \textbf{Id} & \textbf{Role} & \textbf{Experience} & \textbf{Affiliation} \\
        \midrule
        Expert 1 & Team Leader and Developer & Five years of experience in Solidity & Company\\
        Expert 2 & Developer & Four years of experience in Solidity & Company\\
        Expert 3 & Developer & Two years and half of experience in Solidity& Company\\
        Expert 4 & Post Doc & Six years of experience in Security and two in Solidity& University of Luxembourg\\
        \bottomrule
    \end{tabular}
    }}
    
\end{table}

\subsubsection{Taxonomy Validation Results}
To validate the proposed taxonomy, we analyzed all new vulnerabilities added to the baseline taxonomy~\cite{Rameder2022}.
Vulnerabilities derived from three already validated taxonomies~\cite{Rameder2022}\citeA{vidal2024openscv, li2024static} were excluded from the validation process when their classifications remained unchanged. In total, 36 vulnerabilities required validation.
We conducted semi-structured interviews with four experts in security and the Solidity language. Further information about the experts is reported in \Cref{tab:experts}. The interviews were structured in two stages: the first aimed to validate the vulnerability categories, and the second to validate the hierarchical relations among vulnerabilities. The 36 vulnerabilities were divided into two clusters of 18 vulnerabilities each, with two experts independently validating each cluster.
To assess inter-rater agreement among experts, we calculated Cohen’s Kappa coefficient for both validation stages and then averaged the results.
For the first cluster, the experts achieved Cohen’s Kappa values of 0.87 for category validation and 0.73 for hierarchy validation, yielding an average of 0.80.
For the second cluster, Cohen’s Kappa values were 0.77 for category validation and 1.00 for hierarchy validation, resulting in an average of 0.88.
All experts identified an inconsistency in the taxonomy arising from vulnerabilities classified as Denial-of-Service (DoS). The taxonomy is well-structured and follows a classification strategy focused on the root causes of vulnerabilities. From their perspective, this approach is highly effective for auditors, as it enables them to identify the underlying cause of a vulnerability immediately and, consequently, to clearly determine where to act to remove or mitigate the issue.
However, the DoS category does not align with this root-cause-oriented strategy, since a DoS represents the effect of a malicious operation rather than its cause. A vulnerability that causes a DoS typically originates from problems in other categories within the taxonomy. For example, a DoS can be caused by block gas limit exhaustion, unhandled exceptions, or dependency on an external resource that becomes unavailable.
During the interviews, the experts identified a specific cause for each vulnerability whose effect was a DoS. Based on their experience and to maintain consistency throughout the taxonomy, they suggested removing the DoS category and reallocating its vulnerabilities to the categories that best represent the root cause of Denial of Service.
As a result of the interviews, we validated vulnerabilities and reclassified those initially categorized as DoS. 
Validation with experts enables us to assess the completeness and correctness of the categories, thereby creating a consistent taxonomy that organizes vulnerabilities by root cause.
The interview results are available in the replication package.

\geriSumRQ{1}{
We developed a structured taxonomy for Ethereum smart contract vulnerabilities by clustering 420 unique vulnerabilities into 192 groups. The groups are classified into 13 categories based on impact or effect. Each group consists of a representative vulnerability and alternative names, reflecting the various terminology used in the literature for similar issues. The structure organizes vulnerabilities into three levels based on their specificity. The first level contains 113, the second contains 73, while the third contains six. Taxonomy has been validated by four experts.}

\geriKeyPoint{1}{
Our taxonomy covers a wide range of vulnerabilities, unifying all categories identified across existing taxonomies, none of which individually provide complete coverage. The alternative names used in the literature are consistently assigned to a single vulnerability, ensuring clarity and avoiding ambiguity. Vulnerabilities are categorized based on their root causes.
}


\input{tables/revised_taxonomy}

\subsubsection{How to Consult the Taxonomy}
This guide is designed to help auditors navigate the taxonomy and accurately classify vulnerabilities encountered during smart contract auditing. The process begins with identifying the vulnerability of interest (e.g., Block values as a proxy for time). Next, the synonym clusters in the taxonomy should be checked to verify whether the chosen term appears there. If it does, the corresponding representative name should be used as the canonical reference (e.g., Timestamp Dependency).
Once the representative name is identified, its position in the hierarchy can be explored to determine whether it represents a general vulnerability with specialized children or a specialization of a broader parent vulnerability. This hierarchical exploration helps clarify whether the vulnerability should be interpreted as a general pattern or a specific instance.
In this example, Timestamp Dependency is a child of Block State Dependency, meaning it is a specific type of dependency related to the block’s state. Furthermore, examining its position in the hierarchy may reveal additional specializations that differ from the aspect initially considered, providing a more comprehensive understanding of its possible variants (e.g., Block Number  Dependency).

%% file: tables/revised_taxonomy.tex
{\scriptsize
\begin{longtable}{p{0.5cm} p{7.2cm} p{7.3cm}}
    \caption{\footnotesize A Comprehensive List of Vulnerabilities Hierarchically Organized.}
    \label{tab:final_tax}\\
    \toprule
    \footnotesize\textbf{Id} & \footnotesize\textbf{Name} & \footnotesize\textbf{Alternative Names} \\
    \midrule
    \endfirsthead
    \multicolumn{3}{c}{{\footnotesize Continued from previous page}} \\
    \toprule
    \footnotesize\textbf{Id} & \footnotesize\textbf{Name} & \footnotesize\textbf{Alternative Names} \\ 
    \midrule
    \endhead
    \midrule
    \multicolumn{3}{r}{{\footnotesize Continued on next page}} \\
    \midrule
    \endfoot
    \bottomrule
    \endlastfoot
    \multicolumn{3}{l}{\textbf{1 - Malicious Environment, Transactions, or Input}}\\
    \midrule
        1A
        & Call to the unknown~\citeA{dZaazaaAndElBakkali2023a, dZaazaaAndElBakkali2023b, Staderini2022, KushwahaEtAl2022a}
        & Unexpected fallback function~\cite{Rameder2022} \\ 
        1A1
        & Reentrancy~\citeA{dZaazaaAndElBakkali2023a, dZaazaaAndElBakkali2023b, WeiEtAl2023, ChoiEtAl2021, ChuEtAl2023, ZhouEtAl2022, KushwahaEtAl2022b, Staderini2022, ChaliasosEtAl2024}
        & \href{https://swcregistry.io/docs/SWC-107/}{SWC-107}, \href{https://owasp.org/www-project-smart-contract-top-10/2025/en/src/SC05-reentrancy-attacks.html}{OWASP-SC05}, \href{https://dasp.co/\#item-1}{DASP-1}, Cross function race condition, Recursive call vulnerability, Race to empty~\cite{Rameder2022}\\
        1A1.1
        & Reentrancy with ETH transfer~\citeA{li2024static}
        & Reentrancy-eth~\citeA{AgarwalEtAl2022}~\cite{slither}, Unsafe Credit Transfer~\citeA{vidal2024openscv}\\
        1A1.2
        & Reentrancy without ETH transfer~\citeA{li2024static}
        & Reentrancy-no-eth~\citeA{AgarwalEtAl2022}~\cite{slither}, Unsafe System State Changes~\citeA{vidal2024openscv}\\
        1A1.3
        & Reentrancy with same effect~\citeA{li2024static}
        & Reentrancy benign~\citeA{AgarwalEtAl2022}~\cite{slither}\\
        1A1.4
        & Reentrancy with negative events~\citeA{li2024static}
        & Reentrancy events~\citeA{AgarwalEtAl2022}~\cite{slither}\\
\end{longtable}
}

%% file: sections/results_rq2.tex
\subsection{RQ2. Which Methods Do Automated Tools Use to Detect Vulnerabilities?}
\label{sec:rq2}

To address the second research question, we extracted and analyzed the tools from the selected studies in detail. The criteria used to assess the contextual data quality of the primary studies proved to be highly effective. Through this analysis, we identified a total of 219 distinct tools listed in \Cref{tab:tools}. 

\input{figures/fig_tool_functionalities}

\paragraph{Input and Analysis Types}
The input type most analyzed is the source code. Of 219 tools, 142 analyze only source code, 59 analyze only bytecode, and 12 analyze both types. Six tools~\citeA{ye2024midas,li2021gas,wu2023defiranger,chen2024opentracer,liu2022invcon}\cite{zhang2022xscope} analyze addresses or transaction traces. It is important to note that all tools are implemented and tested on Solidity code, and Vyper has never been considered. 

Static analysis is the most common type of analysis, with 145 tools exclusively performing this type of analysis. In contrast, 39 tools perform only dynamic analysis, and 35 tools perform both types of analysis.

\paragraph{Tool's Functionalities}
\Cref{fig:tool_function} details the functionalities implemented by the tools. 
The most frequent functionality observed is Vulnerability Detection, accounting for 82.5\% of cases. This high percentage is expected because many functionalities indirectly contribute to vulnerability detection. These include, for example, tools that analyze the use of gas and resources or offer developer support.
An analysis of the functionalities revealed the need for fully automated tools, i.e., performing vulnerability detection, generating patches, and ultimately fixing the vulnerability by updating the contract using appropriate techniques or patterns. 
Another limitation lies in the tools used for exploit generation. In particular, the few tools that generate attack vectors or vulnerable transactions do so to prove the actual presence of the vulnerability they identify. An example is SMARTEST~\citeA{SunbeomSoEtAl2021} as discussed by Zhang et al.~\citeA{Z.ZhangEtAl2023}, which aims to detect bugs in smart contracts and generate vulnerable transaction sequences that automatically prove flaws.
A relevant aspect that has emerged is the focus on resource and gas use. Minimizing and optimizing resource use leads to a reduction in fees and, thus, gas. All this can be translated into savings on the user's side. The 16\% of the tools focus on this aspect, showing interest in optimizing resources and gas. 
Although most tools focus on vulnerability detection, only a few are integrated into development environments and some still require full implementation. Therefore, further studies are required to investigate their performance.

\input{figures/fig_toolCodeTransformationCategories}

\paragraph{Code Transformation Techniques}
As shown in \Cref{fig:tool_codeTransformation}, the considered tools use different techniques and structures to represent and analyze the code. 
The most used code structures leveraged by static analysis tools are Control Flow Graphs (CFG), Intermediate Representations, Specification Languages, and Abstract Syntax Trees (AST). 
Techniques such as Decompilation, Finite State Machines (FSM), and Disassembly remain underutilized.
Dynamic analyzers mostly use Data Flow Graphs, Transaction Analysis, and Execution Traces. They use intermediate representations very few times, followed by CFGs and ASTs.
When performing dynamic analysis, contract execution traces, transaction analysis, and data flows are crucial considerations. These techniques can capture the behavior of smart contracts when they are executed.
Finally, the most used structures leveraged by tools performing both types of analysis are CFGs, with nine occurrences; Data Flow, Transaction Analysis, and Execution Traces, with seven occurrences; and ASTs, with five occurrences. 

\input{figures/fig_tool_methods}

\paragraph{Tool's Methods}
As detailed in \Cref{fig:tool_methods}, the tools use several methods to analyze code, with some methods being more suitable than others. 
Static analysis tools mostly use machine learning. Numerous tools extract code metrics to feed ML models, as discussed by
Zaazaa and El Bakkali~\citeA{dZaazaaAndElBakkali2023b}, Quian \etal~\citeA{QianEtAl2023a}, and Zhang \etal~\citeA{Z.ZhangEtAl2023}. Peculiar~\citeA{WuHEtAl2021} analyze Crucial Data Flow Graphs extracted from the source code. Ben et al.~\cite{BenEtAlAFS} use AST Fuse program Slicing (AFS) to fuse code characteristics and feed them to deep learning models. SC-VDM~\cite{SC-VDM} transforms smart contract bytecode into grayscale matrix pictures and then uses Convolutional Neural Networks (CNNs) for vulnerability detection. Finally, Huang et al.~\citeA{HuangEtAl2022} developed a smart contract vulnerability detection model based on multitask learning.
In some cases, Machine Learning is supported by Pattern Matching and Syntactical Analysis. For example, GPSCVulDetector~\citeA{LiuEtAl2023a} combines graph neural networks and pattern matching for comprehensive detection, while SVScanner~\citeA{H.ZhangEtAl2023} extracts semantic features from ASTs of different patterns and uses a text CNN to detect contract bugs. 
Symbolic execution and Taint analysis are also used; for example, SIGUARD~\citeA{J.ZhangEtAl2023}, RNVulDet~\citeA{QianEtAl2023b} and Achecker~\citeA{GhalebEtAl2023} combine both methods, SmartFast~\citeA{LiEtAl2022} and SESCon~\citeA{AliEtAl2021} focus only on taint analysis while ReDetect~\citeA{RutaoYuEtAl2021}, and Ethersolve~\citeA{ControEtAl2021, PasquaEtAl2023} focus on symbolic execution.

Dynamic analysis tools mostly use mutation testing and fuzzing methods. On the one hand, tools such as SuMo~\citeA{BarboniEtAl2022}, AMAT~\citeA{SujeethaAndAkila2023}, and MulESC~\citeA{SunEtAl2022a} perform mutation testing to assess the quality of test suites. On the other hand, EVMFuzzer~\cite{EVMFuzzer}, IR-Fuzz~\citeA{LiuEtAl2023b}, Ity-Fuzz~\citeA{ShouEtAl2023}, and SynTest-Solidity~\citeA{OlsthoornEtAl2022} perform fuzzing. Effuzz~\citeA{JiEtAl2023} and EtherFuzz~\citeA{WangEtAl2022} combine both techniques, selectively mutating the input for the fuzzing phase. 

Some tools perform static and dynamic analyses by combining different methods. Smartscan~\citeA{SamreenAndAlalfi2021} identifies potentially vulnerable patterns and then uses dynamic analysis to confirm their precise exploitability. EtherProv~\citeA{LinoyEtAl2022} leverages Solidity source code static and dynamic analysis data through contract bytecode instrumentation and taint analysis. Finally, VulHunter~\citeA{LiEtAl2023} uses hybrid attention and multi-instance learning mechanisms to detect vulnerabilities, then applies symbolic execution to validate their feasibility.

\input{tables/method_ir}

Table \ref{tab:method_x_ir} presents the mapping between analysis methods, the intermediate representations (IRs) they rely on, the input types they process, and the observed frequencies of their occurrences across the analyzed studies.
The Control Flow Graph (CFG) is the most widely adopted IR, appearing in nearly all analysis methods. Its prevalence reflects its central role in capturing execution structure and control dependencies—fundamental aspects for both static and dynamic analyses.
The use of Data Flow, Transaction Analysis, and Execution Traces spans multiple techniques (e.g., Fuzzing, Taint Analysis, Mutation Testing, Runtime Verification), indicating that dynamic information is crucial for vulnerability detection. This trend highlights a shift toward behavior-aware rather than purely syntactic reasoning.
Most methods support both EVM bytecode and source code, though with different emphases: Symbolic Execution, Taint Analysis, and Runtime Verification primarily target bytecode, while Pattern Matching, Abstract Interpretation, and Machine Learning mainly rely on source code. This duality reflects the heterogeneous availability of smart contracts, as not all on-chain contracts have verified source code.
Researchers often combine multiple IRs depending on the abstraction level required by their technique and the type of input available. This diversity illustrates both the maturity and methodological fragmentation of smart contract security analysis.
Despite this variety, most methods still depend on CFGs and Data Flow/Transaction Traces, suggesting that researchers tend to reuse conventional program analysis abstractions while overlooking richer or hybrid IRs (e.g., SSA form, semantic graphs, control–data dependency networks) that could capture more nuanced behaviors.
Furthermore, Formal Verification and Model Checking appear infrequently, indicating that rigorous methods remain underrepresented compared to heuristic approaches, such as Fuzzing and Machine Learning.
Finally, Machine Learning tools show a strong preference for source code (50 occurrences) over bytecode (17), which limits their real-world applicability since most deployed contracts on-chain lack verified source code.
Overall, the mapping reveals a research landscape dominated by control-flow-based and source-level analyses, showing clear gaps remain in IR diversity, integration across analysis paradigms, and applicability to real-world bytecode.

\paragraph{Open Source Tools}
The tools identified in our study offer valuable resources and knowledge that can benefit both researchers and practitioners. They can serve as benchmarks and can be extended, enhanced, or integrated into development pipelines. However, only 101 of these tools are open source and publicly available. Of those, few are ready for immediate use, while many remain in the form of demos or experimental implementations designed to evaluate detection accuracy and require further development to become fully functional.

\geriSumRQ{2}{
We collected 219 distinct tools. The tools focus significantly on source code analysis (142 tools) and static analysis methods (145 tools). Although vulnerability detection is the main functionality of these tools (82.5\%), there is a significant gap in fully automated solutions that encompass patch and exploit generation. Additionally, only 101 tools are open source, with many still requiring further development to achieve full usability. The study emphasizes the need for improved automation and integration of these tools into an Integrated Development Environment to improve security development practices.
}
\geriKeyPoint{2}{
It is unclear which approach performs best for detecting a particular vulnerability. Moreover, few tools are ready for immediate use or easy integration into a development environment. In addition, many solutions are research-oriented with a limited focus on the practical needs of security auditors.
}

\input{tables/tools2}

%% file: figures/fig_tool_functionalities.tex
\begin{figure*}[ht]
\centering
\scriptsize

\begin{tikzpicture}
\begin{axis}[ 
    height=4.7cm,
    width=9cm,
    xbar, 
    xmin=0,
    xlabel={},
    ytick=data,
    yticklabels={
        Supports manual analysis/ Secure development,
        Provides reactive defenses,
        Generates exploits,
        Allows for bulk analysis/ Full automation,
        Analyzes gas/ Resource consumption,
        Verifies program correctness,
        Detects vulnerabilities
    },
    nodes near coords, 
    nodes near coords align={horizontal},
    bar width=0.3cm, 
    enlarge y limits={abs=0.8},
    ]
    \addplot [fill=cold2] coordinates {
        (29, 0)
        (15, 1)
        (8, 2)
        (8, 3)
        (35, 4)
        (50, 5)
        (181, 6)
    };  
\end{axis}
\end{tikzpicture}
\caption{Number of Tools Providing a Specific Functionality.}
\label{fig:tool_function}

\end{figure*}


%% file: figures/fig_toolCodeTransformationCategories.tex
\begin{figure*}[ht]
\centering
\scriptsize

\begin{tikzpicture}
\begin{axis}[ 
    height=10.2cm,
    width=7.8cm,
    xbar, 
    xmin=0,
    xlabel={},
    ytick=data,
    yticklabels={
        Control Flow Graph (CFG),
        {Data flow, Transaction Analysis and Execution Trace},
        Abstract Syntax Tree (AST),
        Decompilation,
        Intermediate Representation or Specification Languages,
        Disassembly,
        Finite State Machine (FSM),
        Control Flow Graph (CFG),
        {Data flow, Transaction Analysis and Execution Trace},
        Abstract Syntax Tree (AST),
        Decompilation,
        Intermediate Representation or Specification Languages,
        Disassembly,
        Finite State Machine (FSM),
        Control Flow Graph (CFG),
        {Data flow, Transaction Analysis and Execution Trace},
        Abstract Syntax Tree (AST),
        Decompilation,
        Intermediate Representation or Specification Languages,
        Disassembly,
        Finite State Machine (FSM)
    },
    nodes near coords, 
    nodes near coords align={horizontal},
    bar width=0.3cm, 
    enlarge y limits={abs=0.8},
    ]
    \addplot [fill=cold2] coordinates {
        (61, 0)
        (25, 1)
        (43, 2)
        (7, 3)
        (39, 4)
        (5, 5)
        (1, 6)
        
        (3, 7)
        (13, 8)
        (3, 9)
        (0, 10)
        (5, 11)
        (0, 12)
        (0, 13)
        
        (15, 14)
        (13, 15)
        (7, 16)
        (2, 17)
        (3, 18)
        (1, 19)
        (0, 20)
    };  
\end{axis}
\draw[thick] (-6.8, 2.91) -- (6.9, 2.91);
\draw[thick] (-6.8, 5.7) -- (6.9, 5.7);

\node[rotate=90, anchor=west] at (6.7, 0.35) {Static Analyzers}; 
\node[rotate=90, anchor=west] at (6.7, 2.94) {Dynamic Analyzers}; 
\node[rotate=90, anchor=west] at (6.7, 6.1) {Both Analyzers}; 

\end{tikzpicture}
\caption{Number of Tools Employing a Specific Code Transformation Technique.}
\label{fig:tool_codeTransformation}
\end{figure*}


%% file: figures/fig_tool_methods.tex
\begin{figure*}[ht]
\centering
\scriptsize

\begin{tikzpicture}
\begin{axis}[ 
    height=15.7cm,
    width=9cm,
    xbar, 
    xmin=0,
    xlabel={},
    ytick=data,
    yticklabels={
        Symbolic Execution,
        Formal Reasoning and Constraint Solving,
        Model Checking,
        Abstract Implementation,
        Fuzzing,
        Runtime Verification,
        Concolic Testing,
        Mutation Testing,
        Pattern Matching and Syntactical Analysis,
        Code Instrumentation,
        Machine Learning,
        Taint Analysis,
        Symbolic Execution,
        Formal Reasoning and Constraint Solving,
        Model Checking,
        Abstract Implementation,
        Fuzzing,
        Runtime Verification,
        Concolic Testing,
        Mutation Testing,
        Pattern Matching and Syntactical Analysis,
        Code Instrumentation,
        Machine Learning,
        Taint Analysis,
        Symbolic Execution,
        Formal Reasoning and Constraint Solving,
        Model Checking,
        Abstract Implementation,
        Fuzzing,
        Runtime Verification,
        Concolic Testing,
        Mutation Testing,
        Pattern Matching and Syntactical Analysis,
        Code Instrumentation,
        Machine Learning,
        Taint Analysis
    },
    nodes near coords, 
    nodes near coords align={horizontal},
    bar width=0.3cm, 
    enlarge y limits={abs=0.8},
    ]
    \addplot [fill=cold2] coordinates {
        (19, 0)
        (15, 1)
        (5, 2)
        (3, 3)
        (1, 4)
        (0, 5)
        (0, 6)
        (0, 7)
        (17, 8)
        (9, 9)
        (78, 10)
        (10, 11)
        (2, 12)
        (3, 13)
        (2, 14)
        (0, 15)
        (16, 16)
        (4, 17)
        (2, 18)
        (11, 19)
        (2, 20)
        (6, 21)
        (2, 22)
        (6, 23)
        (11, 24)
        (5, 25)
        (2, 26)
        (0, 27)
        (11, 28)
        (4, 29)
        (2, 30)
        (7, 31)
        (4, 32)
        (5, 33)
        (4, 34)
        (10, 35)
    }; 
\end{axis}

\draw[thick] (-5.5, 4.747) -- (8, 4.747);
\draw[thick] (-5.5, 9.376) -- (8, 9.376);

\node[rotate=90, anchor=west] at (8, 1.32) {Static Analyzers}; %
\node[rotate=90, anchor=west] at (8, 5.7) {Dynamic Analyzers}; %
\node[rotate=90, anchor=west] at (8, 10.8) {Both Analyzers}; %

\end{tikzpicture}
\caption{Number of Tools Employing a Specific Method}
\label{fig:tool_methods}
\end{figure*}


%% file: tables/method_ir.tex
\begin{table}[!ht]
\centering
{\color{black}
\caption{{Mapping of methods to intermediate code representations and input types, along with their frequencies.}}
\label{tab:method_x_ir}
\resizebox{1\textwidth}{!}{%
\rowcolors{2}{gray!10}{white}
\begin{tabular}{p{4cm}  p{12cm} p{3.5cm}}
\toprule
\textbf{Method} & \textbf{Intermediate Representation (freq)} & \textbf{Input Type (freq)} \\
\midrule

Symbolic Execution & CFG (14), Data Flow/ Transaction Analysis/Traces (8), Intermediate Representation/ Specification Language (Opcode) (4), AST (2), Disassembly (2) & EVM Bytecode (15), Source Code (12) \\
\
Formal Verification \& Constraint Solving & CFG (6), Data Flow/ Transaction Analysis/Traces (4), AST (3), Intermediate Representation/ Specification Language (Opcode) (2), Decompilation/Compiler (1) & EVM Bytecode (4), Source Code (10) \\
\
Model Checking & CFG (1) & EVM Bytecode (1), Source Code (1) \\
\
Abstract Interpretation & CFG (2), Intermediate Representation/ Specification Language (Opcode) (2), AST (1) & Source Code (2) \\
\
Fuzzing & Data Flow/ Transaction Analysis/Traces (11), CFG (10), AST (4), Intermediate Representation/ Specification Language (Opcode) (2), Decompilation/Compiler (1), Disassembly (1) & EVM Bytecode (9), Source Code (15) \\
\
Runtime Verification & Data Flow/ Transaction Analysis/Traces (5), CFG (2), AST (2) & EVM Bytecode (2), Source Code (4) \\
\
Concoli Testing & Data Flow/ Transaction Analysis/Traces (2), CFG (1), AST (1), Intermediate Representation/ Specification Language (Opcode) (1) & EVM Bytecode (2), Source Code (3) \\
\
Mutation Testing & Data Flow/ Transaction Analysis/Traces (7), Intermediate Representation/ Specification Language (Opcode) (2), AST (2), CFG (2), Decompilation/Compiler (1), Disassembly (1) & EVM Bytecode (4), Source Code (9) \\
\
Pattern Matching / Rule-Based & CFG (6), Intermediate Representation/ Specification Language (Opcode) (6), AST (6), Data Flow/ Transaction Analysis/Traces (4), Disassembly (2), Decompilation/Compiler (1) & EVM Bytecode (4), Source Code (14) \\
\
Code Instrumentation & Data Flow/ Transaction Analysis/Traces (6), AST (5), CFG (5), Intermediate Representation/ Specification Language (Opcode) (2) & EVM Bytecode (5), Source Code (11) \\
\
Machine Learning / Deep Learning / AI & CFG (24), Intermediate Representation/ Specification Language (Opcode) (20), AST (17), Data Flow/ Transaction Analysis/Traces (9), Disassembly (2), Decompilation/Compiler (2) & EVM Bytecode (17), Source Code (50) \\
\
Taint Analysis & CFG (12), Data Flow/ Transaction Analysis/Traces (11), Intermediate Representation/ Specification Language (Opcode) (5), AST (2), Decompilation/Compiler (2), Disassembly (1) & EVM Bytecode (13), Source Code (10) \\
\bottomrule
\end{tabular}}}
\end{table}

%% file: tables/tools2.tex
{\scriptsize
\rowcolors{2}{gray!5}{white}
\begin{longtable}{p{4.5cm} p{0.8cm} >{\centering\arraybackslash}p{0.4cm} p{0.7cm} p{1.1cm} >{\centering\arraybackslash}p{1.2cm} >{\centering\arraybackslash}p{1.7cm} >{\centering\arraybackslash}p{0.9cm} >{\centering\arraybackslash}p{1cm} }
    
    \caption{\footnotesize A Comprehensive List of Tools.}
    \label{tab:tools}\\
    \toprule
    \footnotesize\textbf{Tool} & \footnotesize\textbf{Article} & \footnotesize\textbf{Link} & \footnotesize\textbf{Year} & \footnotesize\textbf{Language} & \footnotesize\textbf{Bytecode} & \footnotesize\textbf{Source Code} & \footnotesize\textbf{Static} & \footnotesize\textbf{Dynamic} \\ 
    \midrule
    \endfirsthead
    \midrule
    \multicolumn{9}{l}{{\footnotesize Continued from previous page}} \\
    \toprule
    \footnotesize\textbf{Tool} & \footnotesize\textbf{Article} & \footnotesize\textbf{Link} & \footnotesize\textbf{Year} & \footnotesize\textbf{Language} & \footnotesize\textbf{Bytecode} & \footnotesize\textbf{Source Code} & \footnotesize\textbf{Static} & \footnotesize\textbf{Dynamic} \\ 
    \midrule
    \endhead
    \midrule
    \multicolumn{9}{r}{{\footnotesize Continued on next page}} \\
    \midrule
    \endfoot
    \bottomrule
    \endlastfoot

    SCStudio & \citeA{MengRenEtAl2021} & \href{https://github.com/FISCO-BCOS/SCStudio}{url} & 2021 & Python & ~ & \ding{51} & \ding{51} & \ding{51} \\
    SmartScan & \citeA{SamreenAndAlalfi2021} & \href{https://smartscan.cs.ryerson.ca:4638/}{url} & 2021 & ~ & ~ & \ding{51} & \ding{51} & \ding{51} \\
    SGUARD & \citeA{NguyenEtAl2021} & \href{https://github.com/reentrancy/sGuard}{url} & 2021 & Node.js & \ding{51} & \ding{51} & \ding{51} & \ding{51} \\
    WANA & \citeA{JiangEtAl2021} & \href{https://github.com/gongbell/wana}{url} & 2020 & Python & \ding{51} & ~ & \ding{51} & \ding{51} \\
    Artemis & \cite{WangEtAl2020Artemis} & ~ & ~ & ~ & \ding{51} & ~ & \ding{51} & ~ \\ 
    SCScan & \cite{9343119} & ~ & ~ & ~ & ~ & \ding{51} & \ding{51} & ~ \\ 
    GasFuzzer & \cite{AshrafEtAl2020} & ~ & ~ & ~ & \ding{51} & ~ & \ding{51} & \ding{51} \\ 
    DEPOSafe & \cite{9376204} & ~ & ~ & ~ & \ding{51} & \ding{51} & \ding{51} & \ding{51} \\
\end{longtable}
}

%% file: sections/results_rq3.tex
\subsection{RQ3. Which Vulnerabilities Do the Tools Address?}
\label{sec:rq3}

Several tools were excluded from this mapping for specific reasons. For example, tools that were not explicitly designed to detect vulnerabilities, such as compilers~\citeA{Mazurek2021} and decompilers~\cite{suiche2017porosity}, were omitted.
We also excluded tools that do not focus on specific vulnerabilities, but instead apply machine learning to classify smart contracts as vulnerable~\citeA{DeSalveEtAl2024, OtoniEtAl2023} or use mutation testing without focusing on vulnerability-specific mutation operators~\citeA{OlsthoornEtAl2022, SujeethaAndAkila2023}. These tools are too generic to be accurately mapped.
Next, we individually analyze each tool to identify the vulnerabilities it detects. During this mapping process, several challenges arose. Many tools claimed to detect broad categories of vulnerabilities, such as \textit{Gas-related vulnerabilities}, \textit{Access Control Vulnerabilities}, and \textit{Arithmetic vulnerabilities}~\citeA{GhalebEtAl2023, XuEtAl2021, ShouEtAl2023, NarayanaAndSathiyamurthy2021, ZhengEtAl2022}. However, in our taxonomy, these labels represent categories that include multiple distinct vulnerabilities. Mapping these tools to all vulnerabilities within these categories would lead to broad generalizations. In other cases, tools declared to detect a vulnerability that, in our taxonomy, has different specializations at lower levels. For example, the tool Siguard~\citeA{J.ZhangEtAl2023} declare to detect signature-based vulnerabilities without specifying what they are in detail. In our taxonomy, vulnerability \textit{5C Signature-based vulnerabilities} has five specializations. In this case, we mapped the tools to the vulnerabilities they claim to detect; however, we considered the mapping partial and should be interpreted with caution, as tools may not extend detection to all specialized sub-vulnerabilities at lower levels. Each specialization may be different and may require a different detection mechanism. Performing a truly exhaustive mapping would require an in-depth analysis of each tool’s detection logic to determine which second-level vulnerabilities it can identify. Therefore, we have a complete mapping of tools that specify vulnerabilities, closely aligned with our taxonomy, without creating ambiguities in categories and hierarchy.
We also focused on tools that facilitate patch generation to address vulnerabilities. For example, tools such as Elysium~\citeA{TorresEtAl2022} that rely on other tools for the detection phase were mapped only to the vulnerabilities for which they can generate patches.
Finally, tools that aggregated several others~\citeA{VivarEtAl2021, diAngeloEtAl2023} or that lacked sufficient documentation~\cite{StegemanSolitor, WuEtAl2020}\citeA{WuHEtAl2021} were also not fully mapped. For these, we mapped only the well-defined vulnerabilities that matched our taxonomy.
The mapping includes 162 tools in total. Of these, 113 were fully mapped, and 49 were partially mapped because the tools identified vulnerabilities that were too generic to map precisely.

\input{tables/topten_2}

\Cref{tab:topten} shows that the most frequently analyzed vulnerability is Reentrancy, with 130 tools, followed by Timestamp Dependency, with 78 tools, and Integer Overflow/Underflow, with 62 tools. 
According to the Systematization of Knowledge by Rezaei et al.~\cite{rezaei2025sokrootcause1}, a qualitative analysis of 50 high-impact smart contract exploits between 2022 and 2025 revealed total losses amounting to 1,093,979,779\$. Among these, the Reentrancy vulnerability alone accounted for losses exceeding 118\$ billion.
External Dependency Vulnerabilities caused losses exceeding 222\$ billion, and Access Control Vulnerabilities caused losses exceeding 417\$ billion. 
A price manipulation vulnerability, not listed among the top 10 detected vulnerabilities, caused a total loss of 279,750,000\$ billion. 
The multi-tier root-cause framework for smart contract exploits proposed in the SoK~\cite{rezaei2025sokrootcause1} highlighted that incidents are rarely single-shot failures. Only 19 of the 50 exploits relied solely on a code defect; the rest involved chains of two or more weaknesses. The most significant financial losses often stem from failure areas that code-centric audits may overlook, e.g. \textit{Flawed Economic Design \& Protocol Logic} and \textit{Protocol Lifecycle \& Governance Failures}. 
When automated tools are insufficient to ensure a satisfactory audit support, several defensive practices that have proven effective at preventing past incidents can be adopted. These include the use of invariants to verify state changes or upgrades, as well as staged roll-outs to gradually deploy and test contract updates.
This knowledge is informative and crucial for understanding the current state of smart contract vulnerabilities.
It is worth noting that the tools identified in our study cannot detect all the vulnerabilities in our taxonomy. Of the 192 vulnerabilities, 113 are not addressed by tools. Furthermore, of the 79 vulnerabilities addressed, only 16 are detected by at least 10 tools. In other words, several vulnerabilities can be detected using only a few available tools.

\input{tables/method_vulnerabilities}

Table~\ref{tab:method_x_vulnerability} provides an overview of the association between detection methods and the vulnerabilities they address, with the frequencies. 
The results show that no single technique comprehensively covers all vulnerabilities; rather, some methods demonstrate greater versatility and are more widely adopted across the literature. 
\textit{AI-based}, \textit{Fuzzing}, and \textit{Symbolic Execution} are the most widely used approaches, each addressing multiple high-frequency vulnerability types. 
AI-based tools are used as generic detection and pattern-recognition strategies, reflecting the growing reliance on data-driven vulnerability analysis. 
Their widespread use suggests a shift from handcrafted rules to adaptive and learning-based automated systems.
Symbolic Execution remains one of the core analysis techniques. It is especially suited to path-dependent vulnerabilities such as reentrancy or arithmetic errors. 
However, its high computational cost and limited scalability explain its selective application compared to newer approaches.
Fuzzing is the third most frequent method. Its focus on runtime behavior makes it particularly effective at detecting vulnerabilities in input validation, exception handling, and transaction state changes. This confirms that fuzzing remains a practical solution for uncovering unexpected execution paths and behavioral flaws.
Rule-based and Pattern-Matching techniques focus on recurring vulnerability patterns, particularly those related to unsafe input handling or data flow (1A1, 2A1, 6A). 
They provide robust coverage of well-known weaknesses and are often embedded in static analysis frameworks.
Formal Verification and Model Checking exhibit lower frequencies but high precision, primarily covering categories such as 1A1, 6A, 2A1, and 3A. 
They are often used in high-quality academic studies to verify contract correctness, rather than for large-scale vulnerability detection. Their use highlights a trade-off between accuracy and scalability.
The results suggest a polarization between traditional and modern approaches. 
Traditional methods, such as symbolic execution, fuzzing, and formal verification, offer rigorous correctness guarantees but are limited in terms of scalability. 
Conversely, modern data-driven and stochastic approaches, particularly AI-based analysis and mutation-based approaches, offer higher automation and broader coverage at the cost of precision. Ressi \etal~\cite{ressi2025vulnerability} analyzed the state of the art in machine-learning vulnerability detection for Ethereum smart contracts, by categorizing existing tools and methodologies, evaluating them, and highlighting their limitations. They discussed best practices to enhance the accuracy, scope, and efficiency of vulnerability detection in smart contracts.

\input{tables/method_x_category}

To close the analysis of RQ$_3$, \Cref{tab:methods_categories} highlights gaps in the literature illustrating the adoption of each detection technique across the 13 vulnerability categories. Detection tools do not cover some categories. In particular, categories such as \textit{4 Resource Consumption \& Gas Related Issues}, \textit{7 Bad Coding Quality \& Programming Language Specifics}, \textit{8 Environment Configuration Issues}, \textit{9 Eliminated/Deprecated syntax}, \textit{10 Library Misuses}, \textit{12 Storage \& Memory}, \textit{13 Cross-chain System Vulnerabilities} are not sufficiently covered by detection tools. The lack of coverage in these categories has practical implications: contracts may remain exposed to vulnerabilities that current tools fail to detect, potentially resulting in economic losses. Some of these categories are inherently more challenging to detect automatically, as they often depend on external context, dynamic interactions, or complex multi-contract architectures. There is a clear need for more comprehensive or hybrid detection approaches that combine static, dynamic, and AI-driven analyses. Concepts such as security smells, anti-patterns, and indicators of technical debt can serve as useful heuristics for identifying vulnerabilities in underexplored categories. Furthermore, creating real-world, cross-chain benchmarks encompassing these neglected categories would support the evaluation and improvement of new detection techniques.

\geriSumRQ{3}{
We mapped 162 tools to our taxonomy. In particular, 113 tools were fully mapped while 49 were only partially mapped due to generic descriptions of the vulnerabilities they identified. The remaining 57 tools were not mapped due to their functionalities or poor documentation. The results indicate that the Reentrancy vulnerability is the most commonly analyzed, detected by 130 tools, followed by Timestamp Dependency (78 tools) and Integer Overflow/Underflow (62 tools). 
}

\geriKeyPoint{3}{
Our study revealed significant coverage gaps: 113 vulnerabilities in our taxonomy remain unaddressed. Many tools lack clear information about which Solidity versions they support, making it difficult to assess their applicability. Tools rarely specify their compatibility with the Solidity pragma. As the language evolves, changes introduced in new versions can significantly affect the performance of the tools. Even if a tool is designed to detect a specific vulnerability, its effectiveness may vary across different Solidity versions. This often leads to inconsistent results and an increased likelihood of false positives. Moreover, entire categories are not covered by detection tools.
}

%% file: tables/topten_2.tex
\begin{table}[!h]
    \centering
    {\color{black}
    \caption{Top 10 detected vulnerabilities, impact, and losses in 2022/25 range. The * indicates the vulnerability contributes to the shown amount.}
    \label{tab:topten}
    \resizebox{0.7\columnwidth}{!}{%
    \rowcolors{2}{gray!5}{white}
    
    \begin{tabular}{clrcr}
        \toprule
        \textbf{Id} & \textbf{Vulnerability} & \textbf{Detection Tools} & \textbf{Impact} & \textbf{Losses~\cite{rezaei2025sokrootcause1}} \\
        \midrule
        1A1 & Reentrancy & 130 & High & 118,501,279\$\\       
        2A1 & Timestamp dependency & 78 & Low & *222,750,000\$ \\       
        6A & Integer overflow/underflow & 62 & Medium & 11,800,000\$ \\       
        3A & Unchecked low-level call/send return values & 51 & Medium & 11 600 000\$ \\       
        1D & Vulnerable delegatecall & 35 & High & 4,800,000\$ \\       
        5A1 & Authorization via transaction origin & 32 & Medium & *417,950,000\$ \\       
        5B1 & Unprotected selfdestruction & 31 & High & *417,950,000\$ \\       
        2B & Transaction-ordering dependency & 30 & Medium & *222,750,000\$ \\       
        1J & Greedy contract & 26 & Medium & 41,850,000\$ \\       
        2A & Block State Dependency & 15 & Low & *222,750,000\$ \\       
        \bottomrule
    \end{tabular}}
    }
\end{table}

%% file: tables/method_vulnerabilities.tex
\begin{table}[h!]
\centering
\caption{Mapping of methods used to detect vulnerability and their frequencies.}
\label{tab:method_x_vulnerability}
\resizebox{\textwidth}{!}{%
\rowcolors{2}{gray!7}{white}
{\color{black}
\begin{tabular}{p{7cm} p{14cm}c}
\toprule
\textbf{Method} & \textbf{Vulnerability Id (frequency)} & Tools\\
\midrule











Machine Learning / Deep Learning / AI & 1A1 (57), 2A1 (44), 6A (22), 3A (17), 5A1 (15), 2B (13), 5B1 (10), 1D (10), 9A (8), 1J (6) & 84 \\
Symbolic Execution & 1A1 (16), 3A (9), 6A (8), 1D (7), 1J (7), 2A (7), 2B (6), 5B1 (5), 5A1 (4), 2A1 (4) & 32 \\
Fuzzing & 1A1 (19), 1D (15), 1J (14), 6A (13), 2A1 (11), 3A (11), 2A2 (10), 3B1 (7), 5B1 (7), 2A (5) & 28 \\
Taint Analysis & 1A1 (14), 6A (10), 3A (8), 1D (7), 2A1 (7), 1J (6), 5B1 (6), 2A (5), 3C (5), 5A1 (4) & 26 \\
Pattern Matching / Rule-Based & 1A1 (12), 2A1 (9), 6A (8), 5B1 (6), 5A1 (5), 3A (5), 2B (5), 8B (4), 1J (4), 2C (4) & 23 \\
Formal Verification \& Constraint Solving & 1A1 (9), 6A (8), 2A1 (7), 3A (5), 1D (4), 1J (4), 2B (3), 5A1 (3), 5B1 (3), 4A (2) & 23 \\
Code Instrumentation & 1A1 (11), 3A (6), 2A1 (4), 6A (4), 5A1 (4), 5B1 (3), 2B (2), 1B (2), 1J (1), 2B1 (1) & 20 \\
Mutation Based & 1A1 (11), 1D (9), 1J (9), 3A (9), 6A (9), 2A1 (6), 2A2 (5), 5B1 (5), 2A (4), 2B (4) & 18 \\
Model Checking & 3A (2), 6A (1), 1A1 (1), 1C (1), 11A (1) & 9 \\
Runtime Verification & 1A1 (5), 5A1 (2), 6A (2), 5B5 (1), 1D (1), 1J (1), 2B (1), 5B1 (1), 5B2 (1), 7I6 (1) & 8 \\
Concolic Testing & 1A1 (4), 6A (2), 3A (2), 5A1 (1), 1D (1), 2A (1), 2B (1), 5B1 (1), 1C (1), 11A (1) & 4 \\
Abstract Interpretation & 1A1 (2), 3A (1), 5B1 (1), 5B5.1 (1), 5B5.2 (1), 6A (1), 7G (1), 8B (1) & 3 \\

\bottomrule
\end{tabular}}}
\end{table}

%% file: tables/method_x_category.tex
\begin{table}[ht]
\centering
\myColor{
\caption{Mapping between detection methods and vulnerability categories. A checkmark indicates that at least one vulnerability in the corresponding category is addressed by the method.}
\label{tab:methods_categories}
\resizebox{0.8\columnwidth}{!}{%
\rowcolors{2}{gray!5}{white}
\begin{tabular}{lccccccccccccc}
\toprule
\textbf{Detection Method} & \textbf{1} & \textbf{2} & \textbf{3} & \textbf{4} & \textbf{5} & \textbf{6} & \textbf{7} & \textbf{8} & \textbf{9} & \textbf{10} & \textbf{11} & \textbf{12} & \textbf{13} \\
\midrule
Symbolic Execution & \ding{51} & \ding{51} & \ding{51} &  & \ding{51} & \ding{51} &  &  &  &  &  &  &  \\
Formal Verification \& Constraint Solving & \ding{51} & \ding{51} & \ding{51} & \ding{51} & \ding{51} & \ding{51} &  &  &  &  &  &  &  \\
Model Checking & \ding{51} &  & \ding{51} &  &  & \ding{51} &  &  &  &  & \ding{51} &  &  \\
Abstract Interpretation & \ding{51} &  & \ding{51} &  & \ding{51} & \ding{51} & \ding{51} & \ding{51} &  &  &  &  &  \\
Fuzzing & \ding{51} & \ding{51} & \ding{51} &  & \ding{51} & \ding{51} &  &  &  &  &  &  &  \\
Runtime Verification & \ding{51} & \ding{51} &  &  & \ding{51} & \ding{51} & \ding{51} &  &  &  & \ding{51} &  &  \\
Concolic Testing & \ding{51} & \ding{51} & \ding{51} &  & \ding{51} & \ding{51} &  &  &  &  & \ding{51} &  &  \\
Mutation Based & \ding{51} & \ding{51} & \ding{51} &  & \ding{51} & \ding{51} &  &  &  &  &  &  &  \\
Pattern Matching / Rule-Based & \ding{51} & \ding{51} & \ding{51} &  & \ding{51} & \ding{51} &  & \ding{51} &  &  &  &  &  \\
Code Instrumentation (ALGO) & \ding{51} & \ding{51} & \ding{51} &  & \ding{51} & \ding{51} &  &  &  &  &  &  &  \\
Machine Learning / Deep Learning / AI & \ding{51} & \ding{51} & \ding{51} &  & \ding{51} & \ding{51} &  &  & \ding{51} &  &  &  &  \\
Taint Analysis & \ding{51} & \ding{51} & \ding{51} &  & \ding{51} & \ding{51} &  &  &  &  &  &  &  \\
\bottomrule
\end{tabular}}
}
\end{table}

%% file: sections/results_rq4.tex
\subsection{RQ4. How are the Tools Evaluated?}
\label{sec:rq4}
\Cref{tab:benchmarks} shows collected benchmarks. The "Discussed in" column contains a paper that cites the benchmark, while the "Publication" column contains a paper that releases the benchmark.
We collected 133 benchmarks, of which 100 were used to evaluate and compare the performance of multiple tools; we could not find information regarding the usage of the remaining 33 benchmarks. 
Most benchmarks, \ie 76, contain only smart contracts written in Solidity, while 18 include only bytecode, and 10 include both. We could not find information on the remaining 29 benchmarks.
Furthermore, some tools did not release the benchmarks used for their evaluation, and some benchmarks are closed to users and require access authorization (Vulpedia's benchmark\footnote{\url{https://drive.google.com/file/d/1kizsz0_8B8nP4UNVr0gYjaj25VVZMO8C/edit}}, SCsVulLyzer benchmark\footnote{\url{https://www.yorku.ca/research/bccc/ucs-technical/cybersecurity-datasets-cds/}}).
We found that 77 of the 133 benchmarks are public and accessible, providing a significant resource for the community. In addition, 27 benchmarks include analyses of tool evaluations or experiment results.
We manually reviewed the 77 publicly accessible benchmarks and categorized them as either real-world (RW) or synthetic (Syn). As a result, 47 are real-world, five are synthetic, and eight are both. The remaining 17 benchmarks lack sufficient documentation, so we were unable to classify them. Most synthetic benchmarks are created using SolidiFi.

\input{tables/pragma_coverage}

We conducted a comprehensive analysis of benchmark coverage and composition, manually examining all publicly accessible datasets to determine which Solidity versions, identified via pragma directives, were used for tool evaluations. In total, we extracted 130,974 smart contracts. After removing duplicates by stripping spaces, tabs, and comments and applying a hash function, we obtained 111,223 unique contracts. Excluding contracts without pragma directives left 99,894 contracts, categorized by version in \Cref{tab:pragma_distribution}. Over 83.7\% of these contracts are written in Solidity 0.4.x, reflecting the heavy reliance on legacy contracts from the early Ethereum ecosystem. Only 10.3\% and 5.5\% use versions 0.5.x and 0.6.x, respectively, showing that more recent releases remain underrepresented. This imbalance may limit the generalizability of tool evaluations, as modern compiler features, syntax changes, and security improvements (e.g., 0.7.x and 0.8.x) are rarely tested. Future benchmarks should prioritize the inclusion of newer versions to ensure realistic and comprehensive evaluation.

\input{tables/tool_x_benchmark_coverage}

Next, we assessed curated smart contracts with labeled or manually injected vulnerabilities. From 21,821 collected contracts, 6,319 duplicates were removed, and an additional 3,807 lacked pragma directives, resulting in 11,695 contracts mapped to our proposed vulnerability taxonomy for uniform labeling. \Cref{tab:benchmark_coverage_transposed} presents vulnerability frequencies across Solidity versions, again highlighting the dominance of legacy 0.4.x contracts. The distribution is highly skewed: only a few vulnerabilities, such as \textit{2A1 Timestamp Dependency}, \textit{1A1 Reentrancy}, \textit{3A Unchecked low-level call}, and \textit{6A Integer over-underflow}, appear more than 1,000 times, while a small subset occurs over 100 times. Most vulnerabilities are represented in fewer than 100 contracts, and some are even fewer than 10 (excluded from the table for clarity), resulting in a long-tail distribution. This suggests that benchmarks focus on well-known, easily reproducible vulnerabilities, leaving many types underrepresented and potentially inflating the performance metrics of tools.
The ongoing evolution of Solidity continues to impact static analyzers. Each new version could introduce syntax changes, security mechanisms, and compiler checks, requiring continuous adaptation of analysis tools. In a complementary study, we investigate how Solidity versions affect the performance of vulnerability detection tools~\cite{iuliano2025msr}.

Considering only vulnerabilities that occur more than 10 times in a benchmark, the number of vulnerabilities sufficiently represented is 25, about 13\% of the taxonomy. Without considering emerging vulnerabilities such as Cross-chain vulnerabilities, Library Misuse vulnerabilities, and Machine-Unauditable vulnerabilities that may not yet be present in a benchmark, the vulnerability representation remains low at 16\%, with 25 out of 156.
The remaining 131 vulnerabilities are either insufficiently represented in benchmarks or unbenchmarked. The 62\% of unbenchmarked vulnerabilities (82 out of 131) are not covered by any tool. We can assume that, on the one hand, the lack of benchmarks prevents the development of a vulnerability-detection tool, which explains the low coverage of existing tools. On the other hand, the remaining 49 vulnerabilities (38\%) are covered by tools; however, their representativeness is too low, or benchmarks are unavailable.

\geriSumRQ{4}{
We collected 133 benchmarks used for evaluating tool performance. The majority (76) contain only Solidity contracts, while some include bytecode or both. In particular, 77 benchmarks are publicly accessible and 27 benchmarks provide tool evaluation results that contribute valuable data to the community.
}
\geriKeyPoint{4}{
The most commonly used dataset in the literature is SmartBugs-wild. The Solmost frequently used Solidity version in tool evaluations is 0.4.x, whereas real-world DApps predominantly use8.x. Existing benchmarks are generally small and have not been updated to the latest version of Solidity, reducing their relevance to current smart contract development practices.
}

\input{tables/dataset_2}

%% file: tables/pragma_coverage.tex
\begin{table}[!ht]
    \centering
    \myColor{
    \caption{Distribution of smart contracts across pragma versions}
    \label{tab:pragma_distribution}
    \resizebox{0.7\columnwidth}{!}{
    \begin{tabular}{l|rrrrrr|r}
        \toprule
        \textbf{Pragma Version} & \textbf{0.3.x} & \textbf{0.4.x} & \textbf{0.5.x} & \textbf{0.6.x} & \textbf{0.7.x} & \textbf{0.8.x} & \textbf{Total} \\
        \midrule
        \textbf{Number of SC} & 8 & 83,631 & 10,338 & 5,580 & 325 & 12 & 99,894\\
        \textbf{Percentage (\%)} & 0.01 & 83.72 & 10.35 & 5.58 & 0.33 & 0.01 & 100\\
        \bottomrule
    \end{tabular}
    }}
\end{table}

%% file: tables/tool_x_benchmark_coverage.tex
\begin{table}[!ht]
    \centering
    \myColor{
    \rowcolors{2}{gray!10}{white}
    \caption{Number of labeled Smart Contracts for each vulnerability across pragma versions}
    \label{tab:benchmark_coverage_transposed}
    \resizebox{\columnwidth}{!}{
    \begin{tabular}{l|rrrrrrrrrrrrrrrrrrrrrrrrr}
        \toprule
        \textbf{Pragma} & \textbf{1A1} & \textbf{1B} & \textbf{1B1} & \textbf{1B2} & \textbf{1D} & \textbf{2A1} & \textbf{2A2} & \textbf{2B} & \textbf{2B1} & \textbf{2C} & \textbf{3A} & \textbf{3B1} & \textbf{3C} & \textbf{1J} & \textbf{4I} & \textbf{4D} & \textbf{5A1} & \textbf{5B5} & \textbf{5B} & \textbf{5B1} & \textbf{5B2} & \textbf{5C3} & \textbf{6A} & \textbf{7I} & \textbf{8A} \\
        \midrule
        0.4.x & 1,390 & 288 & 0 & 0 & 74 & 1,586 & 367 & 10 & 4 & 272 & 1,162 & 8 & 20 & 75 & 6 & 0 & 22 & 0 & 19 & 101 & 46 & 0 & 5,139 & 13 & 1 \\ 
        
        0.5.x & 83 & 0 & 30 & 33 & 5 & 37 & 0 & 74 & 32 & 1 & 112 & 2 & 0 & 48 & 82 & 33 & 74 & 34 & 32 & 25 & 0 & 26 & 64 & 9 & 35 \\ 
        
        0.6.x & 17 & 1 & 13 & 15 & 3 & 0 & 0 & 13 & 0 & 0 & 35 & 0 & 0 & 10 & 0 & 13 & 13 & 12 & 0 & 27 & 0 & 2 & 15 & 0 & 3 \\ 
        
        0.7.x & 0 & 0 & 0 & 0 & 3 & 0 & 0 & 0 & 0 & 0 & 0 & 0 & 0 & 2 & 0 & 0 & 0 & 0 & 0 & 3 & 0 & 0 & 0 & 0 & 0  \\ 
        
        0.8.x & 1 & 0 & 0 & 0 & 2 & 0 & 0 & 0 & 0 & 0 & 1 & 0 & 0 & 0 & 0 & 0 & 0 & 0 & 0 & 0 & 0 & 0 & 0 & 0 & 0  \\ 
        \midrule
        \textbf{Total} & 1,491 & 289 & 43 & 48 & 87 & 1,623 & 367 & 97 & 36 & 273 & 1,310 & 10 & 20 & 135 & 88 & 46 & 109 & 46 & 51 & 156 & 46 & 28 & 5,218 & 22 & 39 \\
        \bottomrule
    \end{tabular}
    }}
\end{table}

%% file: tables/dataset_2.tex
{
\scriptsize
\rowcolors{2}{gray!5}{white}
\myColor{
\begin{longtable}{p{0.7cm}p{0.8cm}p{2.2cm}cccccccc} 
    \caption{
    \footnotesize A List of the Benchmarks Used in the Literature.
    } 
    \label{tab:benchmarks}\\          
    \toprule
    \footnotesize\textbf{In} & \footnotesize\textbf{Papers} & \footnotesize\textbf{Tool/Project} & \footnotesize\textbf{Link} & \footnotesize\textbf{\#Contracts} & \footnotesize\textbf{Results} & \footnotesize\textbf{Source} & \footnotesize\textbf{Byte} & \footnotesize\textbf{Address} & \footnotesize\textbf{RW/Syn}\\           
    \midrule
    \endfirsthead
    \midrule
    \multicolumn{10}{l}{{\footnotesize Continued from previous page}} \\          
    \toprule
    \footnotesize\textbf{In} & \footnotesize\textbf{Papers} & \footnotesize\textbf{Tool/Project} & \footnotesize\textbf{Link} & \footnotesize\textbf{\#Contracts} & \footnotesize\textbf{Results} & \footnotesize\textbf{Source} & \footnotesize\textbf{Byte} & \footnotesize\textbf{Address} & \footnotesize\textbf{RW/Syn}\\           
    \midrule
    \endhead
    \midrule
    \multicolumn{10}{r}{{\footnotesize Continued on next page}} \\          
    \midrule
    \endfoot
    \bottomrule
    \endlastfoot
        \citeA{PiantadosiEtAl2022} & \cite{AshouriEtAl2020} &~&~& 12 &~&~& \ding{51} &~&~\\           
        
        \citeA{PiantadosiEtAl2022} & \cite{AzzopardiEtAl2018} &~&~&~&~&~&~&~&~\\
        
        \citeA{PiantadosiEtAl2022} & \cite{BhargavanEtAl2016} &~&~& 112,802 &~&~& \ding{51} &~&~\\                     
        
        \citeA{PiantadosiEtAl2022} & \cite{ChangEtAl2019} & sCompile &~& 36,099 &~& \ding{51} &~&~&~\\                     
        
        \citeA{PiantadosiEtAl2022} & \cite{ChapmanEtAl2019} & Deviant &~& 3 &~& \ding{51} &~&~&~\\                     
        
        \citeA{PiantadosiEtAl2022} & \cite{ChatterjeeEtAl2019} &~&~& 36,764 &~& \ding{51} &~&~&~\\                     
        
        \citeA{PiantadosiEtAl2022} & \cite{ChenEtAl2017} &~&~& 16,000 &~& \ding{51} &~&~&~\\                     
        
        \citeA{PiantadosiEtAl2022} & \cite{DongEtAl2019} &~&~& 2 &~& \ding{51} &~&~&~\\
\end{longtable}}
}

%% file: sections/discussion.tex
\section{Discussion and Implications}
\label{sec:discussion}

This section discusses the differences between our taxonomy and previous ones, the evolution of the tools during the 2016-2021 period, the updated SLR~\cite{Rameder2022}, and the research opportunity highlighted by our results.

\subsection{Differences and Inconsistencies between Taxonomies}
In the following, we provide a synthesis concerning the differences and inconsistencies found across different taxonomies.

\paragraph{Differences}
As highlighted in the literature review that we updated~\cite{Rameder2022}, we confirm that some vulnerabilities describe specific, limited cases of a broader problem. Therefore, we organized our taxonomy into more levels to facilitate understanding of the high-level problems. This structure also allows us to highlight specific vulnerabilities that could arise from high-level issues.
A significant contribution concerns the vulnerability categories identified and utilized in our taxonomy. We have brought clarity to the taxonomy by using the various categories identified in the main taxonomies found in the literature. In particular, we focused on taxonomies that classify vulnerabilities by their causes or effects, using specific categories. We have observed that the intersection of categories from different taxonomies, such as \citeA{li2024static, vidal2024openscv}\cite{Rameder2022}, does not yield a complete set to categorize all vulnerabilities present in the literature. In our taxonomy, we first included categories common to the main taxonomies, then added those partially covered by a few of them. For example, the taxonomy of Li \etal~\citeA{li2024static} includes the category Storage and Memory, not present in previous taxonomies like OpenSCV~\citeA{vidal2024openscv} or the one of Rameder \etal~\cite{Rameder2022}. We decided to include the category in our taxonomy to better categorize vulnerabilities related to the incorrect use of memory or storage. Another example concerns vulnerabilities in libraries. OpenSCV~\citeA{vidal2024openscv}, for instance, does not delve into the misuse of libraries and the vulnerabilities they introduce, as discussed in a previous study by Huang \etal~\citeA{Huang2024Revealing}. We have dedicated a category to library vulnerabilities also because according to Huang \etal~\citeA{Huang2024Revealing}, vulnerable libraries are still spreading after fixed library versions are released, which means 10A, 10B, and 10C are long-term risks to contract security. Another new category concerns the vulnerabilities considered by Zhang \etal~\citeA{Z.ZhangEtAl2023} as Machine Unauditable Bugs (MUBs), which are not detectable with static analysis and require a deeper semantic understanding of the code. The category also includes the severe Price oracle manipulation vulnerability (11A). Finally, we have introduced a category for vulnerabilities that may occur in cross-chain systems. This last category highlights the main issues that can arise when using two or more blockchains together in a system.


\paragraph{Inconsistences}
The work to develop our taxonomy highlighted significant inconsistencies across existing taxonomies in the literature. These inconsistencies arise from differences in the categories used, depending on what each taxonomy considers relevant, as well as in how vulnerabilities are assigned to categories and how alternative names or synonyms for the same vulnerability are grouped.

For instance, in at least four cases (Unsafe Type Inference, Unexpected Ether Balance, External Library Calling, Insufficient Gas Griefing), OpenSCV~\citeA{vidal2024openscv} assigns the same synonym to multiple distinct vulnerabilities. 
A clear example is the term \textit{External Library Calling}, which appears as an alternative name for \textit{1.7.2 Unsafe External Library Call} and \textit{5.14 Use of Malicious Libraries}, even though they are treated as separate entries in two different categories of the taxonomy.
Similarly, \textit{6.1.3 Improper Locking} and \textit{1.4 Improper Locking During External Calls} are placed under different categories despite sharing the alternative name \textit{Unexpected Ether Balance}.
These examples illustrate a core issue that arises when consulting a taxonomy, particularly when starting from alternative names. Let us say that a vulnerability detector finds the Unexpected Ether Balance vulnerability. For an auditor, it becomes unclear which vulnerability is actually being referenced. This can result in multiple entries, sometimes in different categories, with other root causes and impacts.
Our taxonomy addresses these challenges by aiming to unify and clarify the categorization and naming of vulnerabilities. A key focus of our work has been aggregating alternative names and synonyms used in the literature to refer to the same vulnerability. This approach reduces ambiguity and confusion, thereby improving the overall usability of the taxonomy. Furthermore, each category in our taxonomy represents the root cause of the vulnerability, enabling auditors to identify where the vulnerability may lie quickly.

Another example involves \textit{3.1 Improper Gas Requirements Checking}, which treats as synonyms two vulnerabilities that Rameder \etal~\cite{Rameder2022} classify under entirely different categories. Specifically, \textit{Under-priced opcodes} is categorized under \textit{Resource Consumption \& Gas-Related Issues}, while \textit{Gasless send} is placed under \textit{Exception \& Error Handling Disorders}.
Rameder \etal~\cite{Rameder2022} note \textit{Gasless send} as a potential subcategory of \textit{Unexpected Throw or Revert}. Our taxonomy explicitly includes it as a subcategory, providing a more coherent, structured classification.
Inconsistencies among taxonomies significantly impact the mapping between tools and the vulnerabilities they are designed to detect. This issue becomes particularly evident in primary studies that propose new tools. As discussed in \Cref{sec:rq3}, some tools claim to detect vulnerabilities using names that, in existing taxonomies, refer to entire categories rather than specific issues.
An exception to this is the work by Yang \etal~\citeA{yang2025csafuzzer}, which adopts a clear and structured approach to describing each vulnerability their tool can detect. For example, the vulnerability Block State Dependency is defined using a precise logical formula: (TimestampOp $\vee$ BlockNumberOp) $\wedge$ TransferEther. The formula effectively outlines the specific conditions under which the vulnerability occurs.
To our knowledge, this is the only study that provides such detailed specifications for general vulnerability categories, making it easier to understand what the tool is analyzing.
Thanks to this level of detail, it becomes clear that the vulnerability Blockhash Dependency falls outside the scope of the tool. As a result, performance analysis becomes more straightforward and avoids introducing false negatives. In contrast, tools that do not specify how a vulnerability is detected or rely on vague or broad definitions are more likely to produce a high number of false positives.

\paragraph{Comparison with Standard Taxonomies}
We mapped our taxonomy to the SWC Registry, the OWASP Top 10, and the DASP Top 10. The SWC Registry is fully integrated into our taxonomy. We identified a suitable location for each vulnerability in the SWC Registry, facilitating navigation and enhancing the usability of our taxonomy. We also mapped our taxonomy to the OWASP Top 10 and the DASP Top 10, identifying two issues. \textit{OWASP-SC10 Denial of Service} and \textit{DASP-5 Denial of Service} were not mapped because they are too broad. In fact, the examples of DoS provided by OWASP include \textit{DoS with block gas limit, DoS with failed call, DoS Costly Pattern and Loops}, while the examples of DoS provided by DASP include \textit{DoS with block gas limit, and DoS with unexpected throw}. In our taxonomy, each DoS example appears as a distinct entry (\textit{4I, 4I1, 3E, 3F}). Mapping each DoS entry to OWASP-SC10 and DASP-5 would create ambiguity, as different types of DoS caused by different causes are mapped to the same identifier. Therefore, we did not map the DoS category. Another issue is the OWASP-SC07 Flash Loan Attack. The flash loan is a type of attack used to manipulate the price of an asset. In fact, the example provided by OWASP utilizes the Flash Loan attack to exploit the vulnerability OWASP-SC02 Price Oracle Manipulation. As the name suggests, OWASP-SC07 is a type of attack; therefore, we did not map it to our taxonomy. Finally, DASP-10 is not a vulnerability; it is a catch-all category intended to remind people that there are still classes of vulnerabilities that have not yet been identified.

\paragraph{Cross-Chain System Vulnerabilities and Attacks}
Although our review of the literature focuses mainly on smart contract vulnerabilities in Ethereum, it has also revealed emerging insights into other types of vulnerabilities. In particular, several studies~\cite{duan2023attacks, lee2023sok, zhang2022xscope} have identified a range of problems associated with cross-chain technology. The taxonomies proposed in the literature do not capture this type of vulnerability.
Cross-chain technologies have been developed to overcome the interoperability challenges between heterogeneous blockchain systems. Cross-system architectures may integrate Ethereum with other blockchains. In 2016, BTC-Relay~\cite{btcrelay} introduced a one-way cross-chain mechanism between Bitcoin and Ethereum based on a relay chain. That same year, Vitalik Buterin published \textit{Chain Interoperability}~\cite{buterin2016chain}, providing a comprehensive analysis of the interoperability problem in blockchain ecosystems. Cross-chain smart contracts have shown great potential, particularly in DeFi. However, technical incompatibilities, such as those between Ethereum and Fabric, limit the transfer of information between different blockchains~\cite{buterin2016chain}. As a result, various token types proliferate across platforms, complicating token exchange and transaction data sharing~\cite{he2020joint}. 
Duan \etal~\cite{duan2023attacks} compiled a catalog of 14 vulnerabilities affecting cross-chain systems, along with the corresponding attacks, their consequences, and potential defense strategies. In a separate study, Zhang \etal~\cite{zhang2022xscope} conducted an in-depth analysis of the security of cross-chain bridges, identifying three new categories of security bugs. They also introduced \textsc{Xscope}~\cite{zhang2022xscope}, an automated tool designed to detect security violations in cross-chain bridges and uncover real-world attacks.

\subsection{Tool Evolution compared to the 2016-2021 Period}

The number of tools proposed in the literature is increasing strongly. The demand for secure development has also grown with the increasing interest in DApps and security. \Cref{tab:tool_trend} shows the trend of tools released from January 2021 to February 2025. We compared the tools released between 2016 and 2020 with those released between 2021 and early 2025. In particular, we compared how tools have evolved in terms of functionality, methods, and code transformation techniques. 

\begin{table}[ht]
\centering
\caption{Trend of tools released from Jan 2021 to Feb 2025.}
\label{tab:tool_trend}
\begin{tabular}{lccccc}
\toprule
\textbf{Year} & 2021 & 2022 & 2023 & 2024 & 2025 \\
\textbf{Number of Tools} & 44 & 36 & 48 & 66 & 2 \\
\bottomrule
\end{tabular}
\end{table}

\paragraph{Functionalities} In terms of the functionalities of the developed tools, we observed a significant increase in the use of vulnerability detection tools, rising from 79.5\% to 92.8\%. Resource analysis and gas usage are also up sharply, from 12.1\% to 22.3\%, signaling an increasing focus on issues such as sustainability and security. In contrast, there is a decrease in the search for more comprehensive and articulated solutions such as Full Automation and Bulk Analysis, which drop from 41.3\% to 5.7\%. The focus on exploit generation has also decreased, from 7.1\% to 2.9\%. These data suggest a growing emphasis on security and resource efficiency.

\paragraph{Code Transformation Techniques} The use of code transformation techniques has remained more or less stable. A slight increase was seen in the use of CFG, DFG, Execution Traces, and Transactions. Techniques such as disassembly and FSM are less widely used.

\paragraph{Methods} Methods such as mutation testing, fuzzing, taint analysis, and machine learning are increasingly used. Furthermore, methods such as formal reasoning and constraint solving, model checking, pattern matching, and abstract interpretation are omitted. Remain invariable with the use of code instrumentation. Currently, several experiments combine multiple methods. However, the future of research is shifting towards using LLMs capable of performing refactoring and code repair operations, thereby supporting the removal of vulnerabilities from smart contracts.

\paragraph{LLM Era} The methods implemented by the tools reflect the advanced technology of recent years, namely Artificial Intelligence and Large Language Models (LLMs). Several tools adopt LLMs for different types of tasks, all of which were released in 2024. 
GPTScan~\citeA{sun2024gptscan} is one of the first tools that combines GPT with static analysis to detect vulnerabilities in smart contract logic. 
AdvSCan~\citeA{wu2024advscanner}  implements a novel method that uses LLM and static analysis to automatically generate adversarial smart contracts designed to exploit reentrancy vulnerabilities in victim contracts.
ContractThink~\citeA{wang2024contracttinker} is an LLM-empowered tool for real-world vulnerability repair.
SOChecker~\citeA{chen2024identifying} leverages a fine-tuned Llama2 model to complete the code, followed by the application of symbolic execution methods for vulnerability detection.
LLMSmartSec~\citeA{mothukuri2024llmsmartsec}  is a new approach that accurately identifies and fixes smart contract vulnerabilities.
SolGPT~\citeA{zeng2023solgpt} is designed to address the central issue of detecting and mitigating the vulnerabilities inherent in smart contracts.
FELLMVP~\citeA{luo2024fellmvp} integrates ensemble learning with LLMs to classify vulnerabilities in smart contracts. It comprises a fine-tuned LLM for each vulnerability.
SolOSphere~\citeA{khanzadeh2024solosphere}  offers functionality that consists of parsing and deparsing Solidity code, fetching smart contracts from GitHub, and creating a committed environment for gas analysis. It employs an LLM to generate gas-expensive smart contracts.
SmartInv~\citeA{wang2024smartinv} is an automated framework based on foundation models that infers smart contract invariants and detects bugs at scale. It employs an LLM to generate invariants.
Finally, Chen \etal~\citeA{chen2023chatgpt} evaluated the effectiveness of ChatGPT in detecting vulnerabilities in smart contracts, focusing on its detection capabilities and limitations. The results showed that while ChatGPT could identify a wide range of vulnerabilities and worked efficiently, it struggled with precision and had a higher rate of detection failures. Its overall performance was inferior to the other 14 state-of-the-art tools in most vulnerability categories. The authors also analyzed the causes of false positives, categorizing them into four types: \textit{Protected Mechanism Bias, Lack of understanding of actual meaning, Disturbed by comments}, and \textit{Interference with unexecuted code}.

\paragraph{Machine Unauditable Bugs}
Zhang \etal~\citeA{Z.ZhangEtAl2023} systematically investigated real-world smart contract vulnerabilities and studied how many can be exploited by malicious users and cannot be detected by existing analysis tools.
They further categorized vulnerabilities into three groups. The vulnerabilities in the first group are hard to exploit, doubtful, or not directly related to the functionalities of a given project. The second group of vulnerabilities involves the use of simple and general oracles, which do not require an in-depth understanding of the code semantics. Examples include Reentrancy and Arithmetic Overflow. Such vulnerabilities can be detected by data flow tracing, static symbolic execution, and other static analysis tools. The third group of vulnerabilities requires high-level semantic oracles for detection and is closely related to business logic. Most of these vulnerabilities are not detectable by existing static analysis tools. This group comprises six main types of vulnerability: price manipulation, ID-related violations, erroneous state updates, atomicity violation, privilege escalation, and erroneous accounting. According to their findings, a large portion of exploitable bugs in the wild (i.e., 80\%) are not machine-auditable. 
According to DeFiHackLabs~\cite{notiondefihacks}, price manipulation caused losses in DApps of more than 90 million.
Wu \etal~\citeA{wu2023defiranger} describe two types of widespread price manipulation attacks that attackers can use to manipulate the price inside a Decentralized Exchange. Attacks apply traditional financial manipulation methods, i.e, front-running~\cite{frontrunning2023} and cross-market manipulation~\cite{crossmarket2023}, to DeFi ecosystems.
The work of Wang \etal~\citeA{wang2024smartinv} lists several vulnerabilities that require a deeper understanding of the contract, achieved by analyzing natural language from a semantic perspective, i.e., comments, signatures of significant functions, meaningful variable names, and documentation describing transaction scenarios. Price-oracle manipulation is among these vulnerabilities.

\paragraph{Tool Compatibility and Out-of-date Benchmarks}
The SmartBugs wild dataset~\cite{DurieuxEtAl2020ICSE} contains contracts written in Solidity 0.4.x and a few in 0.5.x. It is the most widely used benchmark for evaluating tools. Tools evaluated on older versions of Solidity without explicitly stating their compatibility with Solidity versions may lead developers and auditors to misuse them. To our knowledge, tools evaluated on older Solidity versions do not necessarily fail on newer versions, but their effectiveness remains uncertain.
Tool compatibility with the latest Solidity versions is a critical issue. However, most tools do not specify which Solidity versions they support, leading users to assume broad compatibility. For example, tools such as EVM-Shield~\citeA{zhang2024evm}, Dfier~\citeA{wang2024dfier}, sGuard+~\citeA{gao2024sguard}, LENT-SSE~\citeA{zheng2024lent}, SliSE~\citeA{wang2024efficiently}, ContractCheck~\citeA{wang2024contractcheck}, and PrettySmart~\citeA{zhong2024prettysmart}, published in 2024, were evaluated using the SmartBugs dataset. Li \etal~\citeA{li2021gas}, Jain \etal~\citeA{jain2024integrated}, and the tool SPCon~\citeA{liu2022finding} also use SmartBugs. This dataset consists mainly of Solidity 0.4.x contracts, with only a few from Solidity 0.5.x. Meanwhile, Solidity's latest major release, 0.8.x, was introduced in December 2020. None of these tools explicitly states their compatibility with specific Solidity versions. According to Wu \etal~\citeA{wu2024comprehensive}, a comprehensive protection system must be established and continuously improved based on the actual version to obtain secure smart contracts.
In the list of benchmarks we collected, we found two datasets of smart contracts written in Solidity version 0.8.x \citeA{liao2024smartaxe, zheng2024dappscan}. Zheng \etal~\citeA{zheng2024dappscan} analyzed 1,199 open-source audit reports from 29 security teams. They identified 9,154 weaknesses and developed two distinct datasets, i.e., DAPPSCAN-source and DAPPSCAN-bytecode. Their work highlighted that, even in the real-world dApps, the most widely used version of Solidity is 0.8.x, whereas in the literature, smart contracts 0.8.x are not widely used to evaluate tool performance or conduct other experiments. 
Furthermore, Li \etal~\citeA{li2024static} observed that benchmarks used in existing studies suffer from small sizes or limited vulnerability types. Industry auditing experts involved in their work confirm that the absence of these vulnerability types in the benchmarks is due to their rare occurrence in practical projects. This observation underscores the benchmark’s alignment with the contemporary landscape of smart contract vulnerabilities.
Finally, our mapping between tools and vulnerabilities revealed that 62 tools detect the integer overflow-underflow vulnerability.
All these tools could produce false positives when executed on smart contracts written in Solidity 0.8.x, because the latest version of Solidity performs automatic overflow/underflow checks at runtime by default, so there is no need to use libraries like SafeMath or good programming practices to avoid overflow. Tools that flag the omission of SafeMath (or similar libraries) for arithmetic operations when detecting overflow vulnerabilities could produce false positives in contracts compiled with Solidity 0.8.x or later unless the tool is version-aware. \Cref{fig:changesAcrossVersions} compares two implementations to handle underflow in different Solidity versions.

\begin{figure}[ht]
\centering
\begin{tcolorbox}[colback=gray!5!white, colframe=white, left=6pt, right=0pt, top=2pt, bottom=2pt]
\begin{minipage}[t]{0.45\textwidth}
\begin{lstlisting}[frame=none, caption={Solidity \textgreater{}= 0.8.0}, label={lst:underflow_08}]
pragma solidity 0.8.0;
contract ExampleDC {
    function f() public pure {
        uint x = 0;
        x--; // Revert on underflow
    }
}
\end{lstlisting}
\end{minipage}
\hfill
\begin{minipage}[t]{0.5\textwidth}
\begin{lstlisting}[frame=none, caption={Solidity \textless{} 0.8.0}, label={lst:underflow_04}]
pragma solidity >=0.4.0 <0.8.0;
import "@openzeppelin/SafeMath.sol";
contract ExampleSM {
    using SafeMath for uint;
    function f() public pure {
        uint x = 0;
        x = x.sub(1); //Revert on underflow
    }
}
\end{lstlisting}
\end{minipage}

\end{tcolorbox}
\caption{Underflow check in different Solidity versions.}
\label{fig:changesAcrossVersions}
\end{figure}

In addition to benchmarks, the literature often leverages platforms that report real-world incidents or host security audits to collect vulnerable smart contracts. For example, Wang et al.~\citeA{wang2024skyeye} evaluated 174 real-world adversarial contracts linked to 159 incidents that collectively resulted in approximately 1.36 billion in financial losses. These incidents were gathered from reputable sources such as Rekt\cite{rekt}, DeFiHackLabs~\cite{notiondefihacks}, SlowMist~\cite{slowmist}, CertiK~\cite{certik}, and BlockSec~\cite{blocksec}. Similarly, Sun et al.~\citeA{sun2024gptscan} assessed \textsc{GPTScan} using different datasets of real-world smart contracts like the Top200 dataset\footnote{\url{https://github.com/MetaTrustLabs/GPTScan-Top200}}, which includes contracts associated with the 200 largest market capitalizations, and the Web3Bugs\footnote{\url{https://github.com/MetaTrustLabs/GPTScan-Web3Bugs}} dataset, which comprises large-scale contract projects that underwent audits on the Code4rena platform~\cite{Code4rena}.

A key aspect of benchmarks is how they are created. Most are sourced from Etherscan~\citeA{SamreenAndAlalfi2021, NguyenEtAl2021, ControEtAl2021}, which allows for the consultation of smart contracts deployed on the blockchain. However, others are purpose-built and contain SCs manually injected with vulnerabilities~\citeA{SunEtAl2022b, DaiEtAl2022, ren2021empirical}. For instance, Sun \etal~\citeA{SunEtAl2022b} manually injected 19 SCs with eight different vulnerabilities. This study emphasizes the need for consistent benchmarks free of false positives, numerous in instances, and broad in vulnerability coverage, which can serve as standards for tool evaluation.
Ren \etal~\citeA{ren2021empirical} provide three datasets, one composed of manually injected vulnerabilities, one composed of manually validated contracts, and the last one unlabeled.
Zheng \etal~\citeA{Zheng2023rudder} proposed a manually validated dataset.
The work of HajiHosseinKhani \etal~\citeA{hajihosseinkhani2024unveiling} describes the limitations of datasets and proposes a unified and labeled benchmark that merges four different benchmarks together. The dataset also contains vulnerability-free smart contracts. 
Liu \etal~\citeA{liu2024ffgdetector} highlighted that there are no official datasets or universally recognized reliable third-party datasets available at present. Their contribution would improve tool assessment and facilitate a fair, unbiased comparison of their performance. 

\subsection{Research Opportunities}
\label{sec:opportunity}

Faruk \etal~\citeA{faruk2024systematic} presents a comprehensive overview of the current state of Web3, examining the challenges and opportunities associated with the development of dApps in this environment. It highlights the difficulties faced by blockchain developers in adopting effective software engineering practices and establishing suitable guidelines for blockchain-based software development. In addition, the paper offers open suggestions for each category of challenges, which can be leveraged to address the identified issues. These results contribute meaningful knowledge to the Web3 community, aiding in the development of better strategies for future challenges.

\paragraph{Vulnerability Layers}
Our taxonomy focuses mainly on the vulnerabilities affecting smart contracts, with a small contribution to the analysis of the vulnerabilities in the system. A blockchain comprises several layers in addition to the application and presentation layers to which the SCs belong. The Consensus Layer, Network Layer, Data Layer, and Infrastructure Layer each present additional vulnerabilities that need more attention.
The shift to Proof of Stake (PoS) has solved many of Ethereum's structural vulnerabilities related to mining and centralization (Selfish mining, 51\% Attack, Replay Attack), making the network more secure, sustainable, and resistant to attacks. However, PoS introduces new challenges in governance, staking, and validator protection that need to be addressed with suitable security measures and incentive structures.

\paragraph{Performance Comparison}
The tools proposed in the current state of the art offer significant benefits, offering researchers and practitioners several approaches and methods to adopt. However, understanding the optimal solution based on specific needs remains unclear due to the challenge of obtaining a comprehensive understanding of the tools' performance. Many tools rely on different benchmarks for performance evaluation, and the execution environments and conditions differ, making it challenging to conduct fair and consistent comparisons. Additional analysis is needed to examine the specific implementations, methods, and techniques to address each vulnerability.

\paragraph{Secure-by-design and Sustainability}
There is a notable gap in tools that support developers during the early stages of implementation. Few tools are designed to be integrated into integrated development environments (IDEs) or development pipelines. For example, SCStudio~\citeA{MengRenEtAl2021} is a Visual Studio Code plug-in for the development of secure smart contracts. Given the immutable nature of blockchain, it is necessary to focus on developing secure smart contracts by design, reducing the need for time-consuming patch generation and code repair systems. This approach improves security and offers significant advantages from a sustainability perspective. The blockchain ecosystem is notorious for its high energy consumption, and replacing faulty smart contracts with vulnerability-free versions aggravates this issue. Deploying and discarding faulty contracts is an unsustainable practice that contributes to unnecessary energy waste. By prioritizing secure development practices, we can reduce environmental impact, improve the overall efficiency of blockchain operations, and enhance their security.

\paragraph{Tool Coverage Gaps}
Our mapping of tools to vulnerabilities offers a clear view of how the literature is evolving. Unsurprisingly, most tools focus on well-known vulnerabilities such as Reentrancy, Timestamp Dependence, and Integer Over/Underflow. However, there is a noticeable lack of interest in addressing many other vulnerabilities. According to our findings, the top 10 identified vulnerabilities are addressed using an average of 49 tools each. In contrast, the next 69 vulnerabilities attract significantly less attention, with an average of only three tools per vulnerability. Furthermore, no tool addresses the remaining 58\% of vulnerabilities. The top 10 vulnerabilities are not guaranteed to have the most significant impact and may deserve further scrutiny. 
This mapping reveals a substantial gap in interest for a wide range of vulnerabilities.

\paragraph{Unexplored Solution}

Further analysis is needed to examine the specific implementations, methods, and techniques to address each vulnerability. This would also help identify unexplored solution combinations that could be further developed. Our mapping provides a foundation for this more in-depth analysis. Future research in this field could explore new solutions and combinations that have not yet been implemented or tested.
Hejazi \etal~\citeA{hejazi2025comprehensive} conducted a systematic evaluation of several detection tools using real-world datasets. The results show that while many tools perform well, they do not accurately cover all vulnerability types, highlighting the need for improved integration of detection methodologies. The findings aim to bridge the gaps in existing methods and guide future improvements to enhance the security of dApps.

\paragraph{Software Engineering Practices}
The detection of vulnerabilities in Solidity smart contracts is a rapidly growing trend; yet, despite numerous proposed solutions, no widely accepted standard has been established. Research to date has been rather narrow in scope, and it remains unclear whether existing techniques are sufficient to ensure comprehensive coverage. Being an emerging area, there is a notable lack of well-established concepts coming from software engineering and security.
In this context, the notion of code smells, particularly security smells, deserves further attention, even within decentralized solutions. Initially introduced by Di Sorbo et al.~\cite{DISORBO2022111193}, Gas-wasting code smells opened new perspectives for analyzing contract code and its gas consumption. Gas-wasting code smells were related to the Design patterns for gas optimization proposed by Marchesi et al.~\cite{marchesi2020design}. A recent study by Jiang et al.~\cite{Jiang} leveraged this concept to connect conventional coding practices with the less intuitive gas computation mechanism, resulting in a new detector that uses code smells as indicators of potential issues.
A code smell is a surface indication in the source code that suggests something may be wrong with its design, structure, or implementation. In the context of smart contracts and security, code smells can serve as precursors to potential vulnerabilities. While a code smell alone may not directly lead to exploitation, it often violates best practices and increases the likelihood of security issues.
Beyond code smells, the identification and analysis of harmful yet recurring patterns (or anti-patterns) could provide researchers with new conceptual tools for developing innovative solutions. To date, the most widely recognized pattern in smart contract development is the Check-Effect-Interaction pattern, which is used to implement secure cryptocurrency transfers.

\paragraph{Consolidated Ground Truth}
We verified the ground truth of the benchmarks whenever possible. Given that some benchmark sizes were expressed using the number of chain blocks or files, we further investigated and reported their size in terms of the number of SCs they contain.
New methods and approaches for creating reliable benchmarks have been proposed in the literature. Some of these involve injecting vulnerabilities into smart contracts in a controlled manner using pattern-based or mutation-based techniques.
SolidiFi~\cite{solidifi}, MuSe~\citeA{iuliano2025automated}, and SCAnoGenerator~\cite{scanogenerator} are three tools specifically designed to evaluate the effectiveness of static analysis tools for Solidity smart contracts by systematically injecting known vulnerabilities into the contract code.
SolidiFi introduces seven predefined vulnerability patterns in source code, while MuSe uses pattern-based mutation operators to inject six types of vulnerability into real-world smart contracts.
SCAnoGenerator facilitates automatic anomaly injection by analyzing the control and data flow of Ethereum smart contracts.
Given the limited diversity and size of existing benchmarks, automated solutions for benchmark generation are essential to support a comprehensive and scalable tool evaluation.

The authors of mutation-based tools for vulnerability injection, such as SolidiFi and MuSe, adopt a pattern-matching approach in which vulnerability patterns are derived from real-world cases. Although these patterns capture realistic conditions, they do not guarantee comprehensive coverage of all possible corner cases that may lead to the same vulnerability. Moreover, we found no empirical evaluations comparing the detection performance of tools on synthetic versus real-world vulnerabilities.
Researchers have widely adopted synthetic benchmarks, yet it remains unclear whether performance on these benchmarks accurately reflects real-world detection capabilities. In our previous work~\citeA{iuliano2025automated}, we observed that Slither failed to detect vulnerabilities injected using the pattern-based mutations of MuSe. This result highlights both the need for improvements in static analyzers and the potential of MuSe-generated benchmarks to enhance the evaluation and development of detection tools.
Given these findings, it is reasonable to assume that benchmarks generated with MuSe could help reveal the limitations of existing tools and thereby guide the refinement of vulnerability detectors. Nonetheless, an empirical comparison between synthetic and real-world benchmarks is required to validate this hypothesis and to better understand the representativeness of synthetic test cases in practical contexts.
Finally, the use cases of smart contracts should also be considered. The most common real-world use cases include token transfers, cross-chain bridges, bidding, voting, lotteries, healthcare applications, investment platforms, price oracles, and others~\cite{gupta2021realworld}. Synthetic benchmarks should reflect these predominant use cases to more accurately represent practical scenarios and ensure that detection tools are evaluated against realistic contract behaviors.

%% file: sections/threats.tex
\section{Threats to Validity}
\label{sec:threats}
This section discusses threats to validity, \ie construct, internal, external, and conclusion validity~\cite{threats}. 

\subsection{Construct Validity}
\paragraph{Search Method} Missing relevant primary studies could significantly impact the quality of SLR. Therefore, we adopted different strategies to collect as many studies as possible. We utilized the five most widely used digital libraries. We applied backward and forward snowballing to the results from the Digital Libraries and forward snowballing to Rameder et al.~\cite{Rameder2022}. Additionally, we analyzed the five main conferences in the field under examination to conduct the research as comprehensively as possible.

\paragraph{Inappropriate or Incomplete Terms in Search Strings} The research strings and terms used may affect the construct validity. We used the same research strings from the updated SLR, which were well-discussed and well-motivated, and customized them for each digital library. To mitigate the risk that newly introduced terms in the search string would affect results, we conducted a sensitivity analysis in Step 2 of our research method. The results suggest that the difference is marginal ($\approx 5\%$), indicating that our original search query was already reasonably comprehensive.

\paragraph{Intrinsic Data Quality metrics} A potential construct validity threat arises from the way we calculate the IDQ of publication venues. Specifically, when multiple rankings (SCImago, CORE, Scopus CiteScore) are available, we select the ranking with the highest score as the IDQ. This approach may inflate the perceived quality of journals or conferences, potentially biasing the assessment.
We adopted this method to ensure a consistent and fair upgrade of the original SLR conducted in 2021, allowing for a meaningful comparison over time, and to adhere to best practices for updating an SLR~\cite{NepomucenoAndSoares2019}.

\subsection{Internal Validity}
\paragraph{Publication Bias} Positive results are more likely to be published than negative results. In addition, some studies tend to report particular success stories. We checked the suitability of each study using inclusion/exclusion criteria to avoid considering only papers reporting positive results.

\paragraph{Subjective Quality Assessment} The paper assessment was based on our judgment, which could have been biased and potentially expressed a threat. Inclusion and exclusion criteria helped us reduce the number of studies to be analyzed. We independently performed the quality assessment using a form to evaluate all studies, utilizing a form featuring multiple well-defined and detailed responses that left little room for subjectivity. We used two question forms: one for SLRs and Surveys and one for PS.

\paragraph{Vulnerability Categorization Ambiguity} Some vulnerabilities could reasonably be assigned to more than one category in our taxonomy. To ensure consistency, we classified each vulnerability based on its primary technical cause rather than its potential runtime consequences or attack vectors. 

\subsection{External Validity}
\paragraph{Restricted Resources} The limited number of researchers who worked on the study could affect the external validity. In the planning phase, we established externally validated roles and tasks to be performed. 

\paragraph{Inaccessible Papers and Databases} A relevant aspect concerns the inaccessibility of papers and databases, which was mainly addressed by utilizing numerous digital libraries, constructing customized research strings for each library, and applying exclusion criterion number four, which excludes papers not accessible online via our university's library network.

\subsection{Conclusion Validity}
\paragraph{Study Misclassification} Quality appraisal was based on the classification of the papers (\ie SLRs, surveys, and primary studies) performed at the beginning of the research. Thus, some studies could have been mistakenly classified and later evaluated. However, some quality assessment questions, particularly those related to primary studies, do not apply to surveys or SLRs, contributing to the correct classification of the analyzed studies.

\paragraph{Bias in Study Selection} Misinterpretations of the title or abstract could affect the selection of the studies. During the data extraction process, we checked the citation whenever it contained relevant information for our study. When an original study cited in the papers was not included among the studies selected for our research, but was listed in the extracted papers, it was recovered due to Question 1 of the CDQ form for primary studies. This practice allowed us to recover studies that had initially been discarded. 

\paragraph{Benchmarks Labeling and Validation Process} The labeling process could have been performed manually or automatically.
Some benchmarks are manually labeled~\citeA{GuoEtAl2024, PanEtAl2021, SoudEtAl2023, DaiEtAl2022, JeonEtAl2023, WeiEtAl2023}, whereas automated labeling is quite challenging. Not all vulnerability patterns are used; some are not well-researched, and it is hard to ensure sufficient patterns, leading to benchmarks featuring false positives that directly impact the performance of the tools.
Finally, the validation processes of the benchmarks are usually not thoroughly described and documented.

%% file: sections/related_work.tex
\section{Related Work}
\label{sec:related_work}
In the recent past, some literature reviews targeted SC vulnerabilities and security.
Oualid Zaazaa and Hanan El Bakkali~\citeA{dZaazaaAndElBakkali2023b} conducted a comprehensive examination of scholarly literature from 2016 to 2021, focusing on papers addressing the detection of vulnerabilities in smart contracts. Their objective is to systematically extract and compile all vulnerabilities documented in the selected articles and to analyze the various techniques researchers use to identify these vulnerabilities, with particular emphasis on machine learning (ML) models. In addition, they aim to collect and consolidate a repository of ML datasets to build robust models capable of addressing the problem.
The findings reveal insufficient coverage of the identified vulnerabilities and highlight the limited number (four) of blockchain platforms studied for smart contract security. The authors urge developers to exercise greater diligence in their development practices to improve blockchain network security and reduce user risks. 
Compared to the study, we delved into tool design, analyzing its functionality, methods, and code transformation techniques that it implements.

Oualid Zaazaa and Hanan El Bakkali performed a detailed examination and classification of vulnerabilities~\citeA{dZaazaaAndElBakkali2023a}. Compared to their work, we organized the vulnerabilities hierarchically and categorized them into 14 categories.

Kushwaha \etal conducted two different studies. On the one hand, they systematically reviewed the security vulnerabilities of the Ethereum smart contracts, including well-known attacks and their preventive methods~\citeA{KushwahaEtAl2022a}. They discussed the vulnerabilities based on their root and subcauses and systematically presented well-known attacks. On the other hand, they systematically reviewed and classified Ethereum smart contract analysis tools~\citeA{KushwahaEtAl2022b}. They analyzed 86 security analysis tools developed for Ethereum blockchain smart contracts, regardless of tool type or analysis approach, highlighted several challenges, and recommended future research.
In our work, we overcome their limitations. We have considered tools that do not have names or have been proposed in pre-print articles and provided a collection of benchmarks for tool evaluation.

Vidal \etal~\cite{VIDAL2024112160} review recent smart contract vulnerability detection research, focusing on detection techniques, targeted vulnerabilities, and evaluation datasets. It maps vulnerabilities to the DASP and SWC classification schemes, identifying research trends and gaps. The study highlights that existing schemes, especially SWC, may not capture newer vulnerabilities, suggesting that updated classification frameworks are needed. The heterogeneity in the capabilities and evaluation settings of the detection tools hinders the comparison, pointing to the importance of standardized benchmarks to evaluate the effectiveness of the tools. 
Compared to this study, we have implemented a structured taxonomy and analyzed more aspects of vulnerability detection tools, such as code transformation techniques and implemented functionalities.

Vidal \etal~\citeA{vidal2024openscv} proposed OpenSCV, a new open and hierarchical taxonomy for smart contract vulnerabilities. OpenSCV is designed to reflect current industry practices, accommodate future changes, and support contributions from the broader community. The taxonomy was developed through an extensive analysis of existing research on vulnerability classification, community-maintained classification schemes, and smart contract vulnerability detection tools. The authors demonstrate that OpenSCV captures the detection capabilities advertised by current tools and emphasizes its relevance to smart contract security research. An expert-based evaluation was performed to validate the taxonomy using a questionnaire involving researchers active in the field. 
Our taxonomy differs from the previous in several aspects. We focused on vulnerabilities in Solidity smart contracts on the Ethereum blockchain. OpenSCV also includes vulnerabilities related to other blockchains, such as Hyperledger Fabric and EOSIO, and other languages, such as Golang. Our taxonomy has a category for fixed vulnerabilities, like vulnerabilities related to the EVM and not to the Solidity language, while OpenSCV does not divide them. The EVM has undergone different upgrades to resolve some vulnerabilities.
We proposed a new category related to library vulnerabilities. We included eight new library vulnerabilities presented by Huang et al.~\citeA{Huang2024Revealing} and new CVEs (CVE-2023-36979 and CVE-2023-36980) presented by Li \etal~\citeA{li2024cobra}.
Our study systematically reviews a wide period with respect to Vidal \etal~\citeA{vidal2024openscv}, including papers up to February 2025, and then new vulnerabilities. 

Kaixuan Li \etal~\citeA{li2024static} focused on the application of static tools for security testing. In particular, they developed a tool-oriented taxonomy aimed at unifying the taxonomies used by detection tools. They developed an up-to-date and fine-grained vulnerability taxonomy for smart contracts, including 45 unique types. Their taxonomy is based on taxonomies, such as the DASP Top 10 (2018) and the SWC Registry (2020), and includes the vulnerability identifiers used by security testing tools. 
Our taxonomy takes into account more preexisting taxonomies and is not tool-oriented. We include vulnerability identifiers that come from primary studies that introduced new tools, existing taxonomies proposed in the academic literature, and open-source taxonomies, like DASP Top 10. In our taxonomy, we also included the updated version of the OWASP Smart Contract Top 10\footnote{\url{https://owasp.org/www-project-smart-contract-top-10/}} (January 2025). 

Hejazi \etal~\citeA{hejazi2025comprehensive} collected 256 tools developed between 2018 and 2024 to analyze vulnerabilities in smart contracts, categorize them by methodology, such as fuzzing, symbolic execution, and others, and further classify them based on their detection capabilities. Their evaluation followed a two-layered approach. First, a theoretical analysis grouped tools by origin (academic or industry) and analysis approach (static, dynamic, or hybrid) using customized evaluation criteria. Second, an experimental assessment tested selected tools on real-world datasets, requiring significant time and computational resources. Their findings revealed that academic tools tend to prioritize documentation, flexibility, and adaptability, while industry tools emphasize speed and seamless integration into production environments. In contrast, the tools collected in our study focus exclusively on detecting vulnerabilities in Solidity smart contracts on the Ethereum blockchain. Our analysis goes beyond categorization by also collecting detailed information on the code transformation techniques, methodologies, and functionalities implemented by each tool. Using our taxonomy, we provide a comprehensive mapping between tools and the vulnerabilities that they are designed to detect. This mapping highlights existing gaps and reveals trends in how specific techniques and methods are applied to different types of vulnerability. When combined with the performance analysis performed by Hejazi \etal~\citeA{hejazi2025comprehensive}, our results offer a more complete understanding of the current landscape and could inspire future improvements in the smart contract security tooling.


Finally, Rameder \etal~\cite{Rameder2022} conducted the SLR we updated. The study assesses the state of the art in automated vulnerability analysis of Ethereum smart contracts, focusing on vulnerability classifications, detection methods, security analysis tools, and benchmarks for tool evaluation. Their initial search across major online libraries identified more than 1,300 publications. After applying a structured protocol to remove duplicates and implement inclusion/exclusion criteria, they retained 303 publications. The authors assess the quality of these studies based on the reputation of the publication venues and predefined contextual criteria. For 165 publications with at least a medium quality score, they extract data on vulnerabilities, methods, and tools. They then synthesize these data, providing a classification of 54 smart contract vulnerabilities, an overview of 140 tools, and a list of benchmarks used for tool evaluation. Finally, they present a detailed discussion of their findings. Our study overcomes previous limitations, such as the absence of a structured snowballing process and a comprehensive mapping between tools and vulnerabilities. The mapping was not completely carried out. The reason was that each tool employs different methods and may classify contracts differently, even when they seemingly address the same vulnerability. Tools also refer to taxonomies that are incompatible with each other. However, our taxonomy serves as a point of aggregation of many state-of-the-art classifications and allows us to map tools and vulnerabilities.

\paragraph{Research Gaps and our Contributions}
Our paper bridges previous research gaps and overcomes their limitations by expanding existing studies. We propose a hierarchical taxonomy of vulnerabilities distributed on three different levels of abstraction based on their detailed description. We categorized them into a comprehensive group of 14 categories that cover all aspects of vulnerabilities. Our taxonomy is the first to collect all the different categories present in the state of the art in a single point of aggregation. We resolved issues related to the ambiguity of synonyms when consulting existing taxonomies. We analyze tool design, including code transformation techniques and functionality.  Additionally, we include unnamed and pre-print tools, establish a list of benchmarks, execute a structured snowballing process, and provide a comprehensive mapping between tools and vulnerabilities.

%% file: sections/conclusion.tex
\section{Conclusion}
\label{sec:conclusion}

This paper presents a systematic review of the literature on smart contract vulnerabilities, detection tools, and benchmarks used for tool evaluation. The review covers various research aspects and aims to assess the current challenges related to smart contract security. 

Our taxonomy merges the vulnerabilities identified in the literature through a rigorous SLR process. It serves as a comprehensive aggregation of the most state-of-the-art classifications. It addresses the existing confusion in the literature by organizing vulnerabilities into a clear hierarchical structure, reducing inconsistencies in nomenclature and making it easier to navigate. This structured categorization enables users to quickly capture the nature of vulnerabilities, facilitating a more immediate understanding of the potential issues that are being addressed. The curated list of tools allowed us to identify which solutions have been explored for vulnerability detection and which areas remain under-investigated. The mapping between these tools and our taxonomy highlights that current tools tend to focus on a narrow subset of vulnerabilities. Moreover, our analysis of existing benchmarks revealed significant issues, such as their limited size and scope, as well as their lack of updates in line with recent versions of Solidity.

Our SLR revealed many aspects to investigate, and as described in Section 5, there are many opportunities. Our research will focus on constructing new benchmarks by mutating smart contracts, we will explore the detection of vulnerabilities, in particular by conducting software component analysis, and we will set up an infrastructure that allows the tools to run within the same environment, under the same conditions, and using the same benchmarks, in order to compare their performance fairly and unbiased. Our work aims to provide the basis for future research, offering a valuable resource for researchers interested in advancing this field, which could greatly benefit from additional empirical studies and investigations.

%% file: defs/ack.tex
\section*{Acknowledgment}
Funded by the European Union – Next Generation EU, Mission 4 Component 1, CUP D53D23008400006, and by the Italian Government (MUR, PRIN 2022) code 202233YFCJ. Gerardo Iuliano is funded by the European Union – Next Generation EU, Mission 4 Component 2, CUP D42B24002220004, by the Italian Government (MUR), and by AstraKode S.r.l.
The authors thank Fabiano Izzo and the AstraKode team for the fruitful discussions that motivated our literature review and allowed us to elaborate on its implications.

%% file: main.bbl
\begin{thebibliography}{100}
\expandafter\ifx\csname url\endcsname\relax
  \def\url#1{\texttt{#1}}\fi
\expandafter\ifx\csname urlprefix\endcsname\relax\def\urlprefix{URL }\fi
\expandafter\ifx\csname href\endcsname\relax
  \def\href#1#2{#2} \def\path#1{#1}\fi

\bibitem{fang2023beyond}
Y.~Fang, D.~Wu, X.~Yi, S.~Wang, Y.~Chen, M.~Chen, Y.~Liu, L.~Jiang, Beyond “protected” and “private”: An empirical security analysis of custom function modifiers in smart contracts, in: Proceedings of the 32nd ACM SIGSOFT International Symposium on Software Testing and Analysis, 2023, pp. 1157--1168.

\bibitem{Huang2024Revealing}
M.~Huang, J.~Chen, Z.~Jiang, Z.~Zheng, \href{https://doi.org/10.1145/3597503.3623335}{Revealing hidden threats: An empirical study of library misuse in smart contracts}, in: Proceedings of the IEEE/ACM 46th International Conference on Software Engineering, ICSE '24, Association for Computing Machinery, New York, NY, USA, 2024.
\newblock \href {https://doi.org/10.1145/3597503.3623335} {\path{doi:10.1145/3597503.3623335}}.
\newline\urlprefix\url{https://doi.org/10.1145/3597503.3623335}

\bibitem{li2024static}
K.~Li, Y.~Xue, S.~Chen, H.~Liu, K.~Sun, M.~Hu, H.~Wang, Y.~Liu, Y.~Chen, Static application security testing (sast) tools for smart contracts: How far are we?, Proceedings of the ACM on Software Engineering 1~(FSE) (2024) 1447--1470.

\bibitem{vidal2024openscv}
F.~R. Vidal, N.~Ivaki, N.~Laranjeiro, Openscv: An open hierarchical taxonomy for smart contract vulnerabilities, Empirical Software Engineering 29~(4) (2024) 101.

\bibitem{li2024cobra}
W.~Li, X.~Li, Z.~Li, Y.~Zhang, Cobra: interaction-aware bytecode-level vulnerability detector for smart contracts, in: Proceedings of the 39th IEEE/ACM International Conference on Automated Software Engineering, 2024, pp. 1358--1369.

\bibitem{Z.ZhangEtAl2023}
Z.~Zhang, B.~Zhang, W.~Xu, Z.~Lin, \href{https://doi.org/10.1109/ICSE48619.2023.00061}{Demystifying exploitable bugs in smart contracts}, in: Proceedings of the 45th International Conference on Software Engineering, ICSE '23, IEEE Press, 2023, p. 615–627.
\newblock \href {https://doi.org/10.1109/ICSE48619.2023.00061} {\path{doi:10.1109/ICSE48619.2023.00061}}.
\newline\urlprefix\url{https://doi.org/10.1109/ICSE48619.2023.00061}

\bibitem{huang2024sword}
Y.~Huang, X.~Wu, Q.~Wang, Z.~Qian, X.~Chen, M.~Tang, Z.~Zheng, The sword of damocles: Upgradeable smart contract in ethereum, in: Proceedings of the 32nd IEEE/ACM International Conference on Program Comprehension, 2024, pp. 333--345.

\bibitem{ChuEtAl2023}
H.~Chu, P.~Zhang, H.~Dong, Y.~Xiao, S.~Ji, W.~Li, A survey on smart contract vulnerabilities: Data sources, detection and repair, Information and Software Technology (2023) 107221.

\bibitem{wu2024comprehensive}
G.~Wu, H.~Wang, X.~Lai, M.~Wang, D.~He, S.~Chan, A comprehensive survey of smart contract security: State of the art and research directions, Journal of Network and Computer Applications (2024) 103882.

\bibitem{SunEtAl2022a}
J.~Sun, S.~Huang, C.~Zheng, T.~Wang, C.~Zong, Z.~Hui, Mutation testing for integer overflow in ethereum smart contracts, Tsinghua science and technology 27~(1) (2022) 27--40.

\bibitem{KushwahaEtAl2022a}
S.~S. Kushwaha, S.~Joshi, D.~Singh, M.~Kaur, H.-N. Lee, Systematic review of security vulnerabilities in ethereum blockchain smart contract, IEEE Access 10 (2022) 6605--6621.
\newblock \href {https://doi.org/10.1109/ACCESS.2021.3140091} {\path{doi:10.1109/ACCESS.2021.3140091}}.

\bibitem{AgarwalEtAl2022}
R.~Agarwal, T.~Thapliyal, S.~K. Shukla, \href{https://doi.org/10.1007/978-3-030-94029-4_6}{Vulnerability and transaction behavior based detection of malicious smart contracts}, in: Cyberspace Safety and Security: 13th International Symposium, CSS 2021, Virtual Event, November 9–11, 2021, Proceedings, Springer-Verlag, Berlin, Heidelberg, 2021, p. 79–96.
\newblock \href {https://doi.org/10.1007/978-3-030-94029-4_6} {\path{doi:10.1007/978-3-030-94029-4_6}}.
\newline\urlprefix\url{https://doi.org/10.1007/978-3-030-94029-4_6}

\bibitem{MunirAndTaha2023}
S.~Munir, W.~Taha, Pre-deployment analysis of smart contracts -- a survey (2023).
\newblock \href {http://arxiv.org/abs/2301.06079} {\path{arXiv:2301.06079}}.

\bibitem{dZaazaaAndElBakkali2023a}
O.~Zaazaa, H.~E. Bakkali, \href{http://dx.doi.org/10.5121/ijcnc.2023.15603}{Unveiling the landscape of smart contract vulnerabilities: A detailed examination and codification of vulnerabilities in prominent blockchains}, International Journal of Computer Networks and Communications 15~(6) (2023) 55–75.
\newblock \href {https://doi.org/10.5121/ijcnc.2023.15603} {\path{doi:10.5121/ijcnc.2023.15603}}.
\newline\urlprefix\url{http://dx.doi.org/10.5121/ijcnc.2023.15603}

\bibitem{dZaazaaAndElBakkali2023b}
O.~Zaazaa, H.~El~Bakkali, A systematic literature review of undiscovered vulnerabilities and tools in smart contract technology, Journal of Intelligent Systems 32~(1) (2023) 20230038.

\bibitem{DiasEtAl2021}
B.~Dia, N.~Ivaki, N.~Laranjeiro, An empirical evaluation of the effectiveness of smart contract verification tools, in: 2021 IEEE 26th Pacific Rim International Symposium on Dependable Computing (PRDC), 2021, pp. 17--26.
\newblock \href {https://doi.org/10.1109/PRDC53464.2021.00013} {\path{doi:10.1109/PRDC53464.2021.00013}}.

\bibitem{Staderini2022}
M.~Staderini, \href{https://flore.unifi.it/retrieve/e398c382-6050-179a-e053-3705fe0a4cff/Towards\_Assessment\_Improvement\_Solidity\_Smart\_Contracts.pdf}{Towards the assessment and the improvement of smart contract security}, Ph.D. thesis, University of Florence, Florence, Italy, phD thesis (09 2022).
\newline\urlprefix\url{https://flore.unifi.it/retrieve/e398c382-6050-179a-e053-3705fe0a4cff/Towards\_Assessment\_Improvement\_Solidity\_Smart\_Contracts.pdf}

\bibitem{ZhouEtAl2022}
H.~Zhou, A.~Milani~Fard, A.~Makanju, The state of ethereum smart contracts security: Vulnerabilities, countermeasures, and tool support, Journal of Cybersecurity and Privacy 2~(2) (2022) 358--378.

\bibitem{abubakar2024systematic}
M.~Abubakar, B.~I. Onyeashie, I.~Wadhaj, P.~Leimich, H.~Ali, W.~J. Buchanan, A systematic review of the blockchain technology security challenges and threats classification, in: 2024 6th International Conference on Blockchain Computing and Applications (BCCA), IEEE, 2024, pp. 684--697.

\bibitem{faruk2024systematic}
M.~J.~H. Faruk, P.~Raya, M.~K. Siam, J.~Q. Cheng, H.~Shahriar, A.~Cuzzocrea, P.~G. Bringas, A systematic literature review of decentralized applications in web3: Identifying challenges and opportunities for blockchain developers, in: 2024 IEEE International Conference on Big Data (BigData), IEEE, 2024, pp. 6240--6249.

\bibitem{hejazi2025comprehensive}
N.~Hejazi, A.~H. Lashkari, A comprehensive survey of smart contracts vulnerability detection tools: Techniques and methodologies, Journal of Network and Computer Applications (2025) 104142.

\bibitem{YuliantoEtAl2023}
S.~Yulianto, H.~L. Hendric Spits~Warnars, H.~Prabowo, Meyliana, A.~N. Hidayanto, Security risks and best practices for blockchain and smart contracts: A systematic literature review, in: 2023 International Conference on Information Management and Technology (ICIMTech), 2023, pp. 1--6.
\newblock \href {https://doi.org/10.1109/ICIMTech59029.2023.10278055} {\path{doi:10.1109/ICIMTech59029.2023.10278055}}.

\bibitem{KushwahaEtAl2022b}
S.~S. Kushwaha, S.~Joshi, D.~Singh, M.~Kaur, H.-N. Lee, Ethereum smart contract analysis tools: A systematic review, IEEE Access 10 (2022) 57037--57062.
\newblock \href {https://doi.org/10.1109/ACCESS.2022.3169902} {\path{doi:10.1109/ACCESS.2022.3169902}}.

\bibitem{PiantadosiEtAl2022}
V.~Piantadosi, G.~Rosa, D.~Placella, S.~Scalabrino, R.~Oliveto, Detecting functional and security‐related issues in smart contracts: A systematic literature review, Software: Practice and Experience 53 (10 2022).
\newblock \href {https://doi.org/10.1002/spe.3156} {\path{doi:10.1002/spe.3156}}.

\bibitem{Rameder2021}
H.~Rameder, Systematic review of ethereum smart contract security vulnerabilities, analysis methods and tools, PhD thesis (2021).

\bibitem{MahugnonAndJules2022}
S.~M. Rosaire, D.~Jules, Smart contracts security threats and solutions, International Journal of Information Technology and Web Engineering (IJITWE) 17~(1) (2022) 1--30.

\bibitem{VaccaEtAl2021}
A.~Vacca, A.~Di~Sorbo, C.~A. Visaggio, G.~Canfora, A systematic literature review of blockchain and smart contract development: Techniques, tools, and open challenges, Journal of Systems and Software 174 (2021) 110891.

\bibitem{kiani2024ethereum}
R.~Kiani, V.~S. Sheng, Ethereum smart contract vulnerability detection and machine learning-driven solutions: A systematic literature review, Electronics 13~(12) (2024) 2295.

\bibitem{wu2024we}
S.~Wu, Z.~Li, L.~Yan, W.~Chen, M.~Jiang, C.~Wang, X.~Luo, H.~Zhou, Are we there yet? unraveling the state-of-the-art smart contract fuzzers, in: Proceedings of the IEEE/ACM 46th International Conference on Software Engineering, 2024, pp. 1--13.

\bibitem{Zheng2023rudder}
Z.~Zheng, N.~Zhang, J.~Su, Z.~Zhong, M.~Ye, J.~Chen, Turn the rudder: A beacon of reentrancy detection for smart contracts on ethereum, in: 2023 IEEE/ACM 45th International Conference on Software Engineering (ICSE), 2023, pp. 295--306.
\newblock \href {https://doi.org/10.1109/ICSE48619.2023.00036} {\path{doi:10.1109/ICSE48619.2023.00036}}.

\bibitem{hajihosseinkhani2024unveiling}
S.~HajiHosseinKhani, A.~H. Lashkari, A.~M. Oskui, Unveiling smart contracts vulnerabilities: Toward profiling smart contracts vulnerabilities using enhanced genetic algorithm and generating benchmark dataset, Blockchain: Research and Applications (2024) 100253.

\bibitem{mishra2024smart}
D.~P. Mishra, S.~Senapati, T.~Dey, R.~K. Lenka, Smart contract: Tools and challenges, in: 2024 1st International Conference on Cognitive, Green and Ubiquitous Computing (IC-CGU), IEEE, 2024, pp. 1--6.

\bibitem{obradovic2024blockchain}
M.~Obradovi{\'c}, P.~Vuleti{\'c}, Blockchain programming languages vulnerabilities and error mitigation strategies, in: 2024 11th International Conference on Electrical, Electronic and Computing Engineering (IcETRAN), IEEE, 2024, pp. 1--6.

\bibitem{zhu2024survey}
H.~Zhu, L.~Yang, L.~Wang, V.~S. Sheng, A survey on security analysis methods of smart contracts, IEEE Transactions on Services Computing (2024).

\bibitem{ChenEtAl2022}
J.~Chen, X.~Xia, D.~Lo, J.~Grundy, X.~Luo, T.~Chen, Defining smart contract defects on ethereum, IEEE transactions on software engineering 48~(1) (2022) 327--345.

\bibitem{DaojingHeEtAl2023}
D.~He, R.~Wu, X.~Li, S.~Chan, M.~Guizani, Detection of vulnerabilities of blockchain smart contracts, IEEE Internet of Things Journal 10~(14) (2023) 12178--12185.
\newblock \href {https://doi.org/10.1109/JIOT.2023.3241544} {\path{doi:10.1109/JIOT.2023.3241544}}.

\bibitem{QianEtAl2023a}
P.~Qian, R.~Cao, Z.~Liu, W.~Li, M.~Li, L.~Zhang, Y.~Xu, J.~Chen, Q.~He, Empirical review of smart contract and defi security: Vulnerability detection and automated repair (2023).
\newblock \href {http://arxiv.org/abs/2309.02391} {\path{arXiv:2309.02391}}.

\bibitem{JiangEtAl2023}
F.~Jiang, K.~Chao, J.~Xiao, Q.~Liu, K.~Gu, J.~Wu, Y.~Cao, Enhancing smart-contract security through machine learning: A survey of approaches and techniques, Electronics 12~(9) (2023) 2046.

\bibitem{ChaliasosEtAl2024}
S.~Chaliasos, M.~A. Charalambous, L.~Zhou, Smart contract and defi security tools: Do they meet the needs of practitioners?, \url{https://synthical.com/article/ed404388-ffad-11ed-9b54-72eb57fa10b3} (3 2023).
\newblock \href {http://arxiv.org/abs/2304.02981} {\path{arXiv:2304.02981}}.

\bibitem{iuliano2025automated}
G.~Iuliano, L.~Allocca, M.~Cicalese, D.~Di~Nucci, Automated vulnerability injection in solidity smart contracts: A mutation-based approach for benchmark development, arXiv preprint arXiv:2504.15948 (2025).

\bibitem{chen2024improving}
Y.~Chen, Z.~Sun, Z.~Gong, D.~Hao, Improving smart contract security with contrastive learning-based vulnerability detection, in: Proceedings of the IEEE/ACM 46th International Conference on Software Engineering, 2024, pp. 1--11.

\bibitem{zhang2024towards}
B.~Zhang, Towards finding accounting errors in smart contracts, in: Proceedings of the IEEE/ACM 46th International Conference on Software Engineering, 2024, pp. 1--13.

\bibitem{sun2024gptscan}
Y.~Sun, D.~Wu, Y.~Xue, H.~Liu, H.~Wang, Z.~Xu, X.~Xie, Y.~Liu, Gptscan: Detecting logic vulnerabilities in smart contracts by combining gpt with program analysis, in: Proceedings of the IEEE/ACM 46th International Conference on Software Engineering, 2024, pp. 1--13.

\bibitem{luo2024scvhunter}
F.~Luo, R.~Luo, T.~Chen, A.~Qiao, Z.~He, S.~Song, Y.~Jiang, S.~Li, Scvhunter: Smart contract vulnerability detection based on heterogeneous graph attention network, in: Proceedings of the IEEE/ACM 46th International Conference on Software Engineering, 2024, pp. 1--13.

\bibitem{deng2024safeguarding}
X.~Deng, S.~M. Beillahi, C.~Minwalla, H.~Du, A.~Veneris, F.~Long, Safeguarding defi smart contracts against oracle deviations, in: Proceedings of the IEEE/ACM 46th International Conference on Software Engineering, 2024, pp. 1--12.

\bibitem{chen2024verifying}
H.~Chen, L.~Lu, B.~Massey, Y.~Wang, B.~T. Loo, Verifying declarative smart contracts, in: Proceedings of the IEEE/ACM 46th International Conference on Software Engineering, 2024, pp. 1--12.

\bibitem{wang2024skyeye}
H.~Wang, Y.~Hu, H.~Wu, D.~Liu, C.~Peng, Y.~Wu, M.~Fan, T.~Liu, Skyeye: Detecting imminent attacks via analyzing adversarial smart contracts, in: Proceedings of the 39th IEEE/ACM International Conference on Automated Software Engineering, 2024, pp. 1570--1582.

\bibitem{chen2024opentracer}
Z.~Chen, Y.~Liu, S.~M. Beillahi, Y.~Li, F.~Long, Opentracer: A dynamic transaction trace analyzer for smart contract invariant generation and beyond, in: Proceedings of the 39th IEEE/ACM International Conference on Automated Software Engineering, 2024, pp. 2399--2402.

\bibitem{wu2024advscanner}
Y.~Wu, X.~Xie, C.~Peng, D.~Liu, H.~Wu, M.~Fan, T.~Liu, H.~Wang, Advscanner: Generating adversarial smart contracts to exploit reentrancy vulnerabilities using llm and static analysis, in: Proceedings of the 39th IEEE/ACM International Conference on Automated Software Engineering, 2024, pp. 1019--1031.

\bibitem{wang2024contracttinker}
C.~Wang, J.~Zhang, J.~Gao, L.~Xia, Z.~Guan, Z.~Chen, Contracttinker: Llm-empowered vulnerability repair for real-world smart contracts, in: Proceedings of the 39th IEEE/ACM International Conference on Automated Software Engineering, 2024, pp. 2350--2353.

\bibitem{eshghie2024highguard}
M.~Eshghie, C.~Artho, H.~Stammler, W.~Ahrendt, T.~Hildebrandt, G.~Schneider, Highguard: Cross-chain business logic monitoring of smart contracts, in: Proceedings of the 39th IEEE/ACM International Conference on Automated Software Engineering, 2024, pp. 2378--2381.

\bibitem{eshghie2024oracle}
M.~Eshghie, C.~Artho, Oracle-guided vulnerability diversity and exploit synthesis of smart contracts using llms, in: Proceedings of the 39th IEEE/ACM International Conference on Automated Software Engineering, 2024, pp. 2240--2248.

\bibitem{chen2024demystifying}
Z.~Chen, Y.~Liu, S.~M. Beillahi, Y.~Li, F.~Long, Demystifying invariant effectiveness for securing smart contracts, Proceedings of the ACM on Software Engineering 1~(FSE) (2024) 1772--1795.

\bibitem{liu2024funredisp}
Y.~Liu, W.~Song, Funredisp: A function redispatch tool to reduce invocation gas fees in solidity smart contracts, in: Proceedings of the 33rd ACM SIGSOFT International Symposium on Software Testing and Analysis, 2024, pp. 1876--1880.

\bibitem{liu2024funredisp2}
Y.~Liu, W.~Song, Funredisp: Reordering function dispatch in smart contract to reduce invocation gas fees, in: Proceedings of the 33rd ACM SIGSOFT International Symposium on Software Testing and Analysis, 2024, pp. 516--527.

\bibitem{chen2024identifying}
J.~Chen, C.~Chen, J.~Hu, J.~Grundy, Y.~Wang, T.~Chen, Z.~Zheng, Identifying smart contract security issues in code snippets from stack overflow, in: Proceedings of the 33rd ACM SIGSOFT International Symposium on Software Testing and Analysis, 2024, pp. 1198--1210.

\bibitem{ye2024midas}
M.~Ye, X.~Lin, Y.~Nan, J.~Wu, Z.~Zheng, Midas: Mining profitable exploits in on-chain smart contracts via feedback-driven fuzzing and differential analysis, in: Proceedings of the 33rd ACM SIGSOFT International Symposium on Software Testing and Analysis, 2024, pp. 794--805.

\bibitem{chen2023chatgpt}
C.~Chen, J.~Su, J.~Chen, Y.~Wang, T.~Bi, J.~Yu, Y.~Wang, X.~Lin, T.~Chen, Z.~Zheng, When chatgpt meets smart contract vulnerability detection: How far are we?, ACM Transactions on Software Engineering and Methodology (2023).

\bibitem{guo2024smart}
H.~Guo, Y.~Chen, X.~Chen, Y.~Huang, Z.~Zheng, Smart contract code repair recommendation based on reinforcement learning and multi-metric optimization, ACM Transactions on Software Engineering and Methodology 33~(4) (2024) 1--31.

\bibitem{ye2024funfuzz}
M.~Ye, Y.~Nan, H.-N. Dai, S.~Yang, X.~Luo, Z.~Zheng, Funfuzz: A function-oriented fuzzer for smart contract vulnerability detection with high effectiveness and efficiency, ACM Transactions on Software Engineering and Methodology 33~(7) (2024) 1--20.

\bibitem{song2022esbmc}
K.~Song, N.~Matulevicius, E.~B. de~Lima~Filho, L.~C. Cordeiro, Esbmc-solidity: an smt-based model checker for solidity smart contracts, in: Proceedings of the ACM/IEEE 44th International Conference on Software Engineering: Companion Proceedings, 2022, pp. 65--69.

\bibitem{liu2022finding}
Y.~Liu, Y.~Li, S.-W. Lin, C.~Artho, Finding permission bugs in smart contracts with role mining, in: Proceedings of the 31st ACM SIGSOFT International Symposium on Software Testing and Analysis, 2022, pp. 716--727.

\bibitem{li2021gas}
C.~Li, Gas estimation and optimization for smart contracts on ethereum, in: 2021 36th IEEE/ACM International Conference on Automated Software Engineering (ASE), IEEE, 2021, pp. 1082--1086.

\bibitem{zheng2024lent}
P.~Zheng, B.~Su, X.~Luo, T.~Chen, N.~Zhang, Z.~Zheng, Lent-sse: Leveraging executed and near transactions for speculative symbolic execution of smart contracts, in: Proceedings of the 33rd ACM SIGSOFT International Symposium on Software Testing and Analysis, 2024, pp. 566--577.

\bibitem{zhang2024evm}
X.~Zhang, W.~Sun, Z.~Xu, H.~Cheng, C.~Cai, H.~Cui, Q.~Li, Evm-shield: In-contract state access control for fast vulnerability detection and prevention, IEEE Transactions on Information Forensics and Security 19 (2024) 2517--2532.

\bibitem{mothukuri2024llmsmartsec}
V.~Mothukuri, R.~M. Parizi, J.~L. Massa, Llmsmartsec: Smart contract security auditing with llm and annotated control flow graph, in: 2024 IEEE International Conference on Blockchain (Blockchain), IEEE, 2024, pp. 434--441.

\bibitem{Ozdemir2024}
F.~{Özdemir Sönmez}, W.~J. Knottenbelt, \href{https://www.sciencedirect.com/science/article/pii/S1877050923021634}{Contractarmor: Attack surface generator for smart contracts}, Procedia Computer Science 231 (2024) 8--15, 14th International Conference on Emerging Ubiquitous Systems and Pervasive Networks / 13th International Conference on Current and Future Trends of Information and Communication Technologies in Healthcare (EUSPN/ICTH 2023).
\newblock \href {https://doi.org/https://doi.org/10.1016/j.procs.2023.12.151} {\path{doi:https://doi.org/10.1016/j.procs.2023.12.151}}.
\newline\urlprefix\url{https://www.sciencedirect.com/science/article/pii/S1877050923021634}

\bibitem{jain2024integrated}
V.~K. Jain, M.~Tripathi, An integrated deep learning model for ethereum smart contract vulnerability detection, International Journal of Information Security 23~(1) (2024) 557--575.

\bibitem{chen2024safecheck}
H.~Chen, X.~Zhao, Y.~Wang, Z.~Zhen, Safecheck: Detecting smart contract vulnerabilities based on static program analysis methods, Security and Privacy 7~(5) (2024) e393.

\bibitem{he2024parse}
L.~He, X.~Zhao, Y.~Wang, Parse: efficient detection of smart contract vulnerabilities via parallel and simplified symbolic execution, in: Proceedings of the 2024 IEEE/ACM 46th International Conference on Software Engineering: Companion Proceedings, 2024, pp. 272--273.

\bibitem{liu2024ffgdetector}
Z.~Liu, J.~Jiao, Ffgdetector: Vulnerability detection in cross-contract feature flow graph using gcn, in: 2024 IEEE International Symposium on Parallel and Distributed Processing with Applications (ISPA), IEEE, 2024, pp. 751--757.

\bibitem{qian2024mufuzz}
P.~Qian, H.~Wu, Z.~Du, T.~Vural, D.~Rong, Z.~Cao, L.~Zhang, Y.~Wang, J.~Chen, Q.~He, Mufuzz: Sequence-aware mutation and seed mask guidance for blockchain smart contract fuzzing, in: 2024 IEEE 40th International Conference on Data Engineering (ICDE), IEEE, 2024, pp. 1972--1985.

\bibitem{ibba2024mindthedapp}
G.~Ibba, S.~Aufiero, S.~Bartolucci, R.~Neykova, M.~Ortu, R.~Tonelli, G.~Destefanis, Mindthedapp: A toolchain for complex network-driven structural analysis of ethereum-based decentralized applications, IEEE Access 12 (2024) 28382--28394.

\bibitem{ding2024hunting}
Y.~Ding, A.~Gervais, R.~Wattenhofer, H.~Sato, Hunting defi vulnerabilities via context-sensitive concolic verification, in: Proceedings of the 2024 IEEE/ACM 46th International Conference on Software Engineering: Companion Proceedings, 2024, pp. 324--325.

\bibitem{wang2024smartinv}
S.~J. Wang, K.~Pei, J.~Yang, Smartinv: Multimodal learning for smart contract invariant inference, in: 2024 IEEE Symposium on Security and Privacy (SP), IEEE, 2024, pp. 2217--2235.

\bibitem{khan2024involuntary}
Z.~A. Khan, A.~S. Namin, Involuntary transfer: A vulnerability pattern in smart contracts, IEEE Access 12 (2024) 62459--62479.

\bibitem{wang2024scvd}
D.~Wang, J.~Chen, S.~Cai, Q.~Feng, Y.~Chen, X.~Hu, Scvd-sa: A smart contract vulnerability detection method based on hybrid deep learning model and self-attention mechanism, in: 2024 IEEE International Conference on Software Analysis, Evolution and Reengineering-Companion (SANER-C), IEEE, 2024, pp. 207--214.

\bibitem{guo2024sernet}
L.~Guo, W.~Yang, L.~Hu, L.~Ma, Z.~Wu, Z.~Lu, Sernet-t: A smart contract multi-label vulnerability detection model based on se attention and one-dimensional residual network, in: 2024 IEEE International Conference on Cognitive Computing and Complex Data (ICCD), IEEE, 2024, pp. 260--265.

\bibitem{li2024femd}
J.~Li, R.~Li, Y.~Lu, B.~Cai, Y.~Cao, H.~Lin, X.~Li, C.~Li, Femd: Feature enhancement-aided multimodal feature fusion approach for smart contract vulnerability detection, in: 2024 IEEE 30th International Conference on Parallel and Distributed Systems (ICPADS), IEEE, 2024, pp. 641--648.

\bibitem{dai2024smart}
C.~Dai, H.~Ding, W.~Ma, Y.~Wu, Smart contract vulnerability detection with self-ensemble pre-trained language models, in: 2024 International Conference on Computer, Information and Telecommunication Systems (CITS), IEEE, 2024, pp. 1--8.

\bibitem{luo2024fellmvp}
Y.~Luo, W.~Xu, K.~Andersson, M.~S. Hossain, D.~Xu, Fellmvp: An ensemble llm framework for classifying smart contract vulnerabilities, in: 2024 IEEE International Conference on Blockchain (Blockchain), IEEE, 2024, pp. 89--96.

\bibitem{li2024evofuzzer}
B.~Li, Z.~Pan, T.~Hu, Evofuzzer: An evolutionary fuzzer for detecting reentrancy vulnerability in smart contracts, IEEE Transactions on Network Science and Engineering (2024).

\bibitem{wu2023defiranger}
S.~Wu, Z.~Yu, D.~Wang, Y.~Zhou, L.~Wu, H.~Wang, X.~Yuan, Defiranger: detecting defi price manipulation attacks, IEEE Transactions on Dependable and Secure Computing 21~(4) (2023) 4147--4161.

\bibitem{khanzadeh2024solosphere}
S.~Khanzadeh, M.~H. Alalfi, Solosphere: A framework for gas optimization in solidity smart contracts, in: 2024 IEEE International Conference on Software Analysis, Evolution and Reengineering-Companion (SANER-C), IEEE, 2024, pp. 35--45.

\bibitem{zeng2023solgpt}
S.~Zeng, H.~Zhang, J.~Wang, K.~Shi, Solgpt: A gpt-based static vulnerability detection model for enhancing smart contract security, in: International Conference on Algorithms and Architectures for Parallel Processing, Springer, 2023, pp. 42--62.

\bibitem{he2023unknown}
D.~He, K.~Ding, S.~Chan, M.~Guizani, Unknown threats detection methods of smart contracts, IEEE Internet of Things Journal 11~(3) (2023) 4430--4441.

\bibitem{chu2024deepfusion}
H.~Chu, P.~Zhang, H.~Dong, Y.~Xiao, S.~Ji, Deepfusion: Smart contract vulnerability detection via deep learning and data fusion, IEEE Transactions on Reliability (2024).

\bibitem{he2024reensat}
L.~He, X.~Zhao, Y.~Wang, Reensat: Reentrancy vulnerability detection in smart contracts using semantic-enhanced sat evaluation, IEEE Transactions on Reliability (2024).

\bibitem{wang2024contractgnn}
Y.~Wang, X.~Zhao, L.~He, Z.~Zhen, H.~Chen, Contractgnn: Ethereum smart contract vulnerability detection based on vulnerability sub-graphs and graph neural networks, IEEE Transactions on Network Science and Engineering (2024).

\bibitem{cao2023sccheck}
Y.~Cao, F.~Jiang, J.~Xiao, S.~Chen, X.~Shao, C.~Wu, Sccheck: A novel graph-driven and attention-enabled smart contract vulnerability detection framework for web 3.0 ecosystem, IEEE Transactions on Network Science and Engineering 11~(5) (2023) 4007--4019.

\bibitem{dong2024erinys}
C.~Dong, H.~Huang, Y.~Shang, Erinys: Efficient fuzzing by function invoke sequence generation for smart contracts, in: Proceedings of the 2024 8th International Conference on Big Data and Internet of Things, 2024, pp. 236--241.

\bibitem{DuanEtAl2023}
L.~Duan, L.~Yang, C.~Liu, W.~Ni, W.~Wang, A new smart contract anomaly detection method by fusing opcode and source code features for blockchain services, IEEE Transactions on Network and Service Management 20~(4) (2023) 4354--4368.
\newblock \href {https://doi.org/10.1109/TNSM.2023.3278311} {\path{doi:10.1109/TNSM.2023.3278311}}.

\bibitem{JieEtAl2023}
W.~Jie, Q.~Chen, J.~Wang, A.~S. {Voundi Koe}, J.~Li, P.~Huang, Y.~Wu, Y.~Wang, \href{https://www.sciencedirect.com/science/article/pii/S0020025523004565}{A novel extended multimodal ai framework towards vulnerability detection in smart contracts}, Information Sciences 636 (2023) 118907.
\newblock \href {https://doi.org/https://doi.org/10.1016/j.ins.2023.03.132} {\path{doi:https://doi.org/10.1016/j.ins.2023.03.132}}.
\newline\urlprefix\url{https://www.sciencedirect.com/science/article/pii/S0020025523004565}

\bibitem{L.ZhangEtAl2022a}
L.~Zhang, J.~Wang, W.~Wang, Z.~Jin, C.~Zhao, Z.~Cai, H.~Chen, A novel smart contract vulnerability detection method based on information graph and ensemble learning, Sensors 22~(9) (2022) 3581.

\bibitem{VivarEtAl2021}
A.~L. Vivar, A.~L.~S. Orozco, L.~J.~G. Villalba, A security framework for ethereum smart contracts, Computer Communications 172 (2021) 119--129.

\bibitem{OtoniEtAl2023}
R.~Otoni, M.~Marescotti, L.~Alt, P.~Eugster, A.~Hyv\"{a}rinen, N.~Sharygina, \href{https://doi.org/10.1145/3564699}{A solicitous approach to smart contract verification}, ACM Trans. Priv. Secur. 26~(2) (mar 2023).
\newblock \href {https://doi.org/10.1145/3564699} {\path{doi:10.1145/3564699}}.
\newline\urlprefix\url{https://doi.org/10.1145/3564699}

\bibitem{GhalebEtAl2023}
A.~Ghaleb, J.~Rubin, K.~Pattabiraman, \href{https://doi.org/10.1109/ICSE48619.2023.00087}{Achecker: Statically detecting smart contract access control vulnerabilities}, in: Proceedings of the 45th International Conference on Software Engineering, ICSE '23, IEEE Press, 2023, p. 945–956.
\newblock \href {https://doi.org/10.1109/ICSE48619.2023.00087} {\path{doi:10.1109/ICSE48619.2023.00087}}.
\newline\urlprefix\url{https://doi.org/10.1109/ICSE48619.2023.00087}

\bibitem{JinEtAl2022}
H.~Jin, Z.~Wang, M.~Wen, W.~Dai, Y.~Zhu, D.~Zou, Aroc: An automatic repair framework for on-chain smart contracts, IEEE Transactions on Software Engineering 48~(11) (2022) 4611--4629.
\newblock \href {https://doi.org/10.1109/TSE.2021.3123170} {\path{doi:10.1109/TSE.2021.3123170}}.

\bibitem{SunEtAl2023}
X.~Sun, L.~Tu, J.~Zhang, J.~Cai, B.~Li, Y.~Wang, Assbert: Active and semi-supervised bert for smart contract vulnerability detection, Journal of Information Security and Applications 73 (2023) 103423.

\bibitem{SujeethaAndAkila2023}
R.~SUJEETHA, K.~AKILA, Automated mutation analysis for smart contract using ama tool with enhanced ga and machine learning approach, Journal of Theoretical and Applied Information Technology 101~(21) (2023).

\bibitem{NarayanaAndSathiyamurthy2021}
K.~L. Narayana, K.~Sathiyamurthy, Automation and smart materials in detecting smart contracts vulnerabilities in blockchain using deep learning, Materials Today: Proceedings 81 (2023) 653--659.

\bibitem{SoudEtAl2023}
M.~Soud, I.~Qasse, G.~Liebel, M.~Hamdaqa, Automesc: Automatic framework for mining and classifying ethereum smart contract vulnerabilities and their fixes, in: 2023 49th Euromicro Conference on Software Engineering and Advanced Applications (SEAA), 2023, pp. 410--417.
\newblock \href {https://doi.org/10.1109/SEAA60479.2023.00068} {\path{doi:10.1109/SEAA60479.2023.00068}}.

\bibitem{XieEtAl2023}
X.~Xie, H.~Wang, Z.~Jian, Y.~Fang, Z.~Wang, T.~Li, Block-gram: Mining knowledgeable features for efficiently smart contract vulnerability detection, Digital Communications and Networks (2023).

\bibitem{L.ZhangEtAl2022c}
L.~Zhang, W.~Chen, W.~Wang, Z.~Jin, C.~Zhao, Z.~Cai, H.~Chen, Cbgru: A detection method of smart contract vulnerability based on a hybrid model, Sensors (Basel, Switzerland) 22~(9) (2022) 3577.

\bibitem{HwangEtAl2022}
S.-J. Hwang, S.-H. Choi, J.~Shin, Y.-H. Choi, Codenet: Code-targeted convolutional neural network architecture for smart contract vulnerability detection, IEEE Access 10 (2022) 32595--32607.
\newblock \href {https://doi.org/10.1109/ACCESS.2022.3162065} {\path{doi:10.1109/ACCESS.2022.3162065}}.

\bibitem{CaiEtAl2023}
J.~Cai, B.~Li, J.~Zhang, X.~Sun, B.~Chen, Combine sliced joint graph with graph neural networks for smart contract vulnerability detection, Journal of Systems and Software 195 (2023) 111550.

\bibitem{LiuEtAl2023a}
Z.~Liu, P.~Qian, X.~Wang, Y.~Zhuang, L.~Qiu, X.~Wang, Combining graph neural networks with expert knowledge for smart contract vulnerability detection, IEEE Transactions on Knowledge and Data Engineering 35~(2) (2023) 1296--1310.
\newblock \href {https://doi.org/10.1109/TKDE.2021.3095196} {\path{doi:10.1109/TKDE.2021.3095196}}.

\bibitem{YangEtAl2024}
H.~Yang, X.~Gu, X.~Chen, L.~Zheng, Z.~Cui, Crossfuzz: Cross-contract fuzzing for smart contract vulnerability detection, Science of Computer Programming 234 (2024) 103076.

\bibitem{QianEtAl2023c}
P.~Qian, Z.~Liu, Y.~Yin, Q.~He, \href{https://doi.org/10.1145/3543507.3583367}{Cross-modality mutual learning for enhancing smart contract vulnerability detection on bytecode}, in: Proceedings of the ACM Web Conference 2023, WWW '23, Association for Computing Machinery, New York, NY, USA, 2023, p. 2220–2229.
\newblock \href {https://doi.org/10.1145/3543507.3583367} {\path{doi:10.1145/3543507.3583367}}.
\newline\urlprefix\url{https://doi.org/10.1145/3543507.3583367}

\bibitem{CaoEtAl2023a}
Y.~Cao, F.~Jiang, J.~Xiao, S.~Chen, W.~Yang, Y.~Yi, Data flow-driven and attention mechanism-enabled smart contract vulnerability detection for secure and green blockchain-based service networks, in: ICC 2023 - IEEE International Conference on Communications, 2023, pp. 5135--5140.
\newblock \href {https://doi.org/10.1109/ICC45041.2023.10279381} {\path{doi:10.1109/ICC45041.2023.10279381}}.

\bibitem{YuEtAl2021}
X.~Yu, H.~Zhao, B.~Hou, Z.~Ying, B.~Wu, Deescvhunter: A deep learning-based framework for smart contract vulnerability detection, in: 2021 International Joint Conference on Neural Networks (IJCNN), 2021, pp. 1--8.
\newblock \href {https://doi.org/10.1109/IJCNN52387.2021.9534324} {\path{doi:10.1109/IJCNN52387.2021.9534324}}.

\bibitem{ChenEtAl2021}
J.~Chen, X.~Xia, D.~Lo, J.~Grundy, X.~Luo, T.~Chen, Defectchecker: Automated smart contract defect detection by analyzing evm bytecode, IEEE Transactions on Software Engineering 48~(7) (2022) 2189--2207.
\newblock \href {https://doi.org/10.1109/TSE.2021.3054928} {\path{doi:10.1109/TSE.2021.3054928}}.

\bibitem{QianEtAl2023b}
P.~Qian, J.~He, L.~Lu, S.~Wu, Z.~Lu, L.~Wu, Y.~Zhou, Q.~He, Demystifying random number in ethereum smart contract: Taxonomy, vulnerability identification, and attack detection, IEEE Transactions on Software Engineering 49~(7) (2023) 3793--3810.
\newblock \href {https://doi.org/10.1109/TSE.2023.3271417} {\path{doi:10.1109/TSE.2023.3271417}}.

\bibitem{JeonEtAl2023}
S.~Jeon, G.~Lee, H.~Kim, S.~Woo, Design and evaluation of highly accurate smart contract code vulnerability detection framework, Data Mining and Knowledge Discovery (10 2023).
\newblock \href {https://doi.org/10.1007/s10618-023-00981-1} {\path{doi:10.1007/s10618-023-00981-1}}.

\bibitem{HuEtAl2023b}
T.~Hu, B.~Li, Z.~Pan, C.~Qian, Detect defects of solidity smart contract based on the knowledge graph, IEEE transactions on reliability 73~(1) (2024) 1--17.

\bibitem{EshghieEtAl2021}
M.~Eshghie, C.~Artho, D.~Gurov, \href{https://doi.org/10.1145/3463274.3463348}{Dynamic vulnerability detection on smart contracts using machine learning}, in: Proceedings of the 25th International Conference on Evaluation and Assessment in Software Engineering, EASE '21, Association for Computing Machinery, New York, NY, USA, 2021, p. 305–312.
\newblock \href {https://doi.org/10.1145/3463274.3463348} {\path{doi:10.1145/3463274.3463348}}.
\newline\urlprefix\url{https://doi.org/10.1145/3463274.3463348}

\bibitem{DeSalveEtAl2024}
A.~De~Salve, A.~Brighente, M.~Conti, Edit: A data inspection tool for smart contracts temporal behavior modeling and prediction, Future Generation Computer Systems 154 (2024) 413--425.

\bibitem{JianzhongSuEtAl2022}
J.~Su, H.-N. Dai, L.~Zhao, Z.~Zheng, X.~Luo, \href{https://doi.org/10.1145/3551349.3560429}{Effectively generating vulnerable transaction sequences in smart contracts with reinforcement learning-guided fuzzing}, in: Proceedings of the 37th IEEE/ACM International Conference on Automated Software Engineering, ASE '22, Association for Computing Machinery, New York, NY, USA, 2023.
\newblock \href {https://doi.org/10.1145/3551349.3560429} {\path{doi:10.1145/3551349.3560429}}.
\newline\urlprefix\url{https://doi.org/10.1145/3551349.3560429}

\bibitem{JiEtAl2023}
S.~Ji, J.~Wu, J.~Qiu, J.~Dong, Effuzz: Efficient fuzzing by directed search for smart contracts, Information and Software Technology 159 (2023) 107213.

\bibitem{TorresEtAl2022}
C.~Ferreira~Torres, H.~Jonker, R.~State, \href{https://doi.org/10.1145/3545948.3545975}{Elysium: Context-aware bytecode-level patching to automatically heal vulnerable smart contracts}, in: Proceedings of the 25th International Symposium on Research in Attacks, Intrusions and Defenses, RAID '22, Association for Computing Machinery, New York, NY, USA, 2022, p. 115–128.
\newblock \href {https://doi.org/10.1145/3545948.3545975} {\path{doi:10.1145/3545948.3545975}}.
\newline\urlprefix\url{https://doi.org/10.1145/3545948.3545975}

\bibitem{PasquaEtAl2023}
M.~Pasqua, A.~Benini, F.~Contro, M.~Crosara, M.~Dalla~Preda, M.~Ceccato, Enhancing ethereum smart-contracts static analysis by computing a precise control-flow graph of ethereum bytecode, Journal of Systems and Software 200 (2023) 111653.

\bibitem{LutzEtAl2021}
O.~Lutz, H.~Chen, H.~Fereidooni, C.~Sendner, A.~Dmitrienko, A.~R. Sadeghi, F.~Koushanfar, Escort: Ethereum smart contracts vulnerability detection using deep neural network and transfer learning (2021).
\newblock \href {http://arxiv.org/abs/2103.12607} {\path{arXiv:2103.12607}}.

\bibitem{GhalebEtAl2022}
A.~Ghaleb, J.~Rubin, K.~Pattabiraman, \href{https://doi.org/10.1145/3533767.3534378}{etainter: detecting gas-related vulnerabilities in smart contracts}, in: Proceedings of the 31st ACM SIGSOFT International Symposium on Software Testing and Analysis, ISSTA 2022, Association for Computing Machinery, New York, NY, USA, 2022, p. 728–739.
\newblock \href {https://doi.org/10.1145/3533767.3534378} {\path{doi:10.1145/3533767.3534378}}.
\newline\urlprefix\url{https://doi.org/10.1145/3533767.3534378}

\bibitem{AshizawaEtAl2021}
N.~Ashizawa, N.~Yanai, J.~P. Cruz, S.~Okamura, \href{https://doi.org/10.1145/3457337.3457841}{Eth2vec: Learning contract-wide code representations for vulnerability detection on ethereum smart contracts}, in: Proceedings of the 3rd ACM International Symposium on Blockchain and Secure Critical Infrastructure, BSCI '21, Association for Computing Machinery, New York, NY, USA, 2021, p. 47–59.
\newblock \href {https://doi.org/10.1145/3457337.3457841} {\path{doi:10.1145/3457337.3457841}}.
\newline\urlprefix\url{https://doi.org/10.1145/3457337.3457841}

\bibitem{WangEtAl2022}
X.~Wang, J.~Sun, C.~Hu, P.~Yu, B.~Zhang, D.~Hou, Etherfuzz: Mutation fuzzing smart contracts for tod vulnerability detection, Wireless communications and mobile computing 2022 (2022) 1--8.

\bibitem{LinoyEtAl2022}
S.~Linoy, S.~Ray, N.~Stakhanova, Etherprov: Provenance-aware detection, analysis, and mitigation of ethereum smart contract security issues, in: 2021 IEEE International Conference on Blockchain (Blockchain), 2021, pp. 1--10.
\newblock \href {https://doi.org/10.1109/Blockchain53845.2021.00014} {\path{doi:10.1109/Blockchain53845.2021.00014}}.

\bibitem{Mazurek2021}
L.~Mazurek, Ethver: Formal verification of randomized ethereum smart contracts, in: Financial Cryptography and Data Security. FC 2021 International Workshops, Springer Berlin Heidelberg, Berlin, Heidelberg, pp. 364--380.

\bibitem{ZhuEtAl2023}
H.~Zhu, K.~Yang, L.~Wang, Z.~Xu, V.~S. Sheng, Grabit: A sequential model-based framework for smart contract vulnerability detection, in: 2023 IEEE 34th International Symposium on Software Reliability Engineering (ISSRE), IEEE, 2023, pp. 568--577.

\bibitem{LongHeEtAl2023}
L.~He, X.~Zhao, Y.~Wang, J.~Yang, X.~Sun, GraphSA: Smart Contract Vulnerability Detection Combining Graph Neural Networks and Static Analysis, Vol. 372 of FAIA, 2023, Ch. GraphSA: Smart Contract Vulnerability Detection Combining Graph Neural Networks and Static Analysis, pp. 1020--1027.
\newblock \href {https://doi.org/10.3233/FAIA230374} {\path{doi:10.3233/FAIA230374}}.

\bibitem{GongEtAl2023}
G.~Peng, Y.~Wenzhong, W.~Liejun, W.~Fuyuan, H.~KeZiErBieKe, L.~Yuanyuan, \href{http://www.techscience.com/cmc/v76n2/53991}{Gratdet: Smart contract vulnerability detector based on graph representation and transformer}, Computers, Materials \& Continua 76~(2) (2023) 1439--1462.
\newblock \href {https://doi.org/10.32604/cmc.2023.038878} {\path{doi:10.32604/cmc.2023.038878}}.
\newline\urlprefix\url{http://www.techscience.com/cmc/v76n2/53991}

\bibitem{MaEtAl2023}
C.~Ma, S.~Liu, G.~Xu, Hgat: smart contract vulnerability detection method based on hierarchical graph attention network, Journal of cloud computing : advances, systems and applications 12~(1) (2023) 93--13.

\bibitem{HuangEtAl2021}
J.~Huang, S.~Han, W.~You, W.~Shi, B.~Liang, J.~Wu, Y.~Wu, Hunting vulnerable smart contracts via graph embedding based bytecode matching, IEEE transactions on information forensics and security 16 (2021) 2144--2156.

\bibitem{YangEtAl2023}
Z.~Yang, W.~Zhu, M.~Yu, Improvement and optimization of vulnerability detection methods for ethernet smart contracts, IEEE Access 11 (2023) 78207--78223.
\newblock \href {https://doi.org/10.1109/ACCESS.2023.3298672} {\path{doi:10.1109/ACCESS.2023.3298672}}.

\bibitem{NguyenEtAl2023}
H.~H. Nguyen, N.-M. Nguyen, C.~Xie, Z.~Ahmadi, D.~Kudendo, T.-N. Doan, L.~Jiang, Mando-hgt: Heterogeneous graph transformers for smart contract vulnerability detection, in: 2023 IEEE/ACM 20th International Conference on Mining Software Repositories (MSR), IEEE, 2023, pp. 334--346.

\bibitem{Hong-NingEtAl2021}
Z.~Wang, B.~Wen, Z.~Luo, S.~Liu, M-a-r: A dynamic symbol execution detection method for smart contract reentry vulnerability, in: Blockchain and Trustworthy Systems, Communications in Computer and Information Science, Springer Singapore, Singapore, 2021, pp. 418--429.

\bibitem{FeiEtAl2023}
J.~Fei, X.~Chen, X.~Zhao, Msmart: Smart contract vulnerability analysis and improved strategies based on smartcheck, Applied Sciences 13~(3) (2023) 1733.

\bibitem{ZhengEtAl2022}
P.~Zheng, Z.~Zheng, X.~Luo, \href{https://doi.org/10.1145/3533767.3534395}{Park: accelerating smart contract vulnerability detection via parallel-fork symbolic execution}, in: Proceedings of the 31st ACM SIGSOFT International Symposium on Software Testing and Analysis, ISSTA 2022, Association for Computing Machinery, New York, NY, USA, 2022, p. 740–751.
\newblock \href {https://doi.org/10.1145/3533767.3534395} {\path{doi:10.1145/3533767.3534395}}.
\newline\urlprefix\url{https://doi.org/10.1145/3533767.3534395}

\bibitem{L.YuEtAl2023}
L.~Yu, J.~Lu, X.~Liu, L.~Yang, F.~Zhang, J.~Ma, Pscvfinder: A prompt-tuning based framework for smart contract vulnerability detection, in: 2023 IEEE 34th International Symposium on Software Reliability Engineering (ISSRE), IEEE, 2023, pp. 556--567.

\bibitem{BarboniEtAl2023}
M.~Barboni, A.~Morichetta, A.~Polini, F.~Casoni, \href{https://doi.org/10.1007/s11219-023-09637-1}{Resumo: a regression strategy and tool for mutation testing of solidity smart contracts}, Software Quality Journal 32~(1) (2023) 225–253.
\newblock \href {https://doi.org/10.1007/s11219-023-09637-1} {\path{doi:10.1007/s11219-023-09637-1}}.
\newline\urlprefix\url{https://doi.org/10.1007/s11219-023-09637-1}

\bibitem{LiuEtAl2023b}
Z.~Liu, P.~Qian, J.~Yang, L.~Liu, X.~Xu, Q.~He, X.~Zhang, Rethinking smart contract fuzzing: Fuzzing with invocation ordering and important branch revisiting, IEEE transactions on information forensics and security 18 (2023) 1--1.

\bibitem{BoseEtAl2021}
P.~Bose, D.~Das, Y.~Chen, Y.~Feng, C.~Kruegel, G.~Vigna, Sailfish: Vetting smart contract state-inconsistency bugs in seconds, in: 2022 IEEE Symposium on Security and Privacy (SP), IEEE, 2022, pp. 161--178.

\bibitem{CaoEtAl2023b}
Y.~Cao, F.~Jiang, J.~Xiao, S.~Chen, X.~Shao, C.~Wu, Sccheck: A novel graph-driven and attention-enabled smart contract vulnerability detection framework for web 3.0 ecosystem, IEEE Transactions on Network Science and Engineering (2023) 1--12\href {https://doi.org/10.1109/TNSE.2023.3324942} {\path{doi:10.1109/TNSE.2023.3324942}}.

\bibitem{LiangAndZhai2023}
J.~Liang, Y.~Zhai, Scgru: A model for ethereum smart contract vulnerability detection combining cnn and bigru-attention, in: 2023 8th International Conference on Signal and Image Processing (ICSIP), 2023, pp. 831--837.
\newblock \href {https://doi.org/10.1109/ICSIP57908.2023.10270857} {\path{doi:10.1109/ICSIP57908.2023.10270857}}.

\bibitem{AliEtAl2021}
A.~Ali, Z.~U. Abideen, K.~Ullah, Sescon: Secure ethereum smart contracts by vulnerable patterns’ detection, Security and communication networks 2021 (2021) 1--14.

\bibitem{WangEtAl2023}
H.~Wang, Z.~Wang, S.~Liu, J.~Sun, Y.~Zhao, Y.~Wan, T.~D. Nguyen, sfuzz2.0: Storage‐access pattern guided smart contract fuzzing, Journal of software : evolution and process 36~(4) (2023).

\bibitem{NguyenEtAl2021}
T.~D. Nguyen, L.~H. Pham, J.~Sun, Sguard: Towards fixing vulnerable smart contracts automatically, in: 2021 IEEE Symposium on Security and Privacy (SP), 2021, pp. 1215--1229.
\newblock \href {https://doi.org/10.1109/SP40001.2021.00057} {\path{doi:10.1109/SP40001.2021.00057}}.

\bibitem{J.ZhangEtAl2023}
J.~Zhang, Y.~Li, J.~Gao, Z.~Guan, Z.~Chen, \href{https://doi.org/10.1109/ICSE-Companion58688.2023.00019}{Siguard: Detecting signature-related vulnerabilities in smart contracts}, in: Proceedings of the 45th International Conference on Software Engineering: Companion Proceedings, ICSE '23, IEEE Press, 2023, p. 31–35.
\newblock \href {https://doi.org/10.1109/ICSE-Companion58688.2023.00019} {\path{doi:10.1109/ICSE-Companion58688.2023.00019}}.
\newline\urlprefix\url{https://doi.org/10.1109/ICSE-Companion58688.2023.00019}

\bibitem{GuoEtAl2024}
J.~Guo, L.~Lu, J.~Li, Smart contract vulnerability detection based on multi-scale encoders, Electronics 13~(3) (2024) 489.

\bibitem{L.ZhangEtAl2022}
L.~Zhang, J.~Wang, W.~Wang, Z.~Jin, Y.~Su, H.~Chen, Smart contract vulnerability detection combined with multi-objective detection, Computer Networks 217 (2022) 109289.

\bibitem{diAngeloEtAl2023}
M.~di~Angelo, T.~Durieux, J.~F. Ferreira, G.~Salzer, Smartbugs 2.0: An execution framework for weakness detection in ethereum smart contracts, in: 2023 38th IEEE/ACM International Conference on Automated Software Engineering (ASE), IEEE, 2023, pp. 2102--2105.

\bibitem{LiEtAl2022}
L.~Zhaoxuan, S.~Lu, R.~Zhang, R.~Xue, W.~Ma, R.~Liang, Z.~Zhao, S.~Gao, Smartfast: an accurate and robust formal analysis tool for ethereum smart contracts, Empirical Software Engineering 27 (10 2022).
\newblock \href {https://doi.org/10.1007/s10664-022-10218-2} {\path{doi:10.1007/s10664-022-10218-2}}.

\bibitem{LiaoEtAl2023}
Z.~Liao, S.~Hao, Y.~Nan, Z.~Zheng, \href{https://doi.org/10.1145/3597926.3598111}{Smartstate: Detecting state-reverting vulnerabilities in smart contracts via fine-grained state-dependency analysis}, in: Proceedings of the 32nd ACM SIGSOFT International Symposium on Software Testing and Analysis, ISSTA 2023, Association for Computing Machinery, New York, NY, USA, 2023, p. 980–991.
\newblock \href {https://doi.org/10.1145/3597926.3598111} {\path{doi:10.1145/3597926.3598111}}.
\newline\urlprefix\url{https://doi.org/10.1145/3597926.3598111}

\bibitem{PraitheeshanEtAl2021}
P.~Praitheeshan, L.~Pan, X.~Zheng, A.~Jolfaei, R.~Doss, Solguard: Preventing external call issues in smart contract-based multi-agent robotic systems, Information Sciences 579 (2021) 150--166.

\bibitem{HuEtAl2023a}
T.~Hu, J.~Li, X.~Xu, B.~Li, Solitester: Detecting exploitable external-risky vulnerability in smart contracts using contract account triggering method, Journal of Software: Evolution and Process  e2633.

\bibitem{L.ZhangEtAl2022b}
L.~Zhang, Y.~Li, T.~Jin, W.~Wang, Z.~Jin, C.~Zhao, Z.~Cai, H.~Chen, Spcbig-ec: A robust serial hybrid model for smart contract vulnerability detection, Sensors (Basel, Switzerland) 22~(12) (2022) 4621.

\bibitem{BarboniEtAl2022}
M.~Barboni, A.~Morichetta, A.~Polini, \href{https://www.sciencedirect.com/science/article/pii/S0164121222001418}{Sumo: A mutation testing approach and tool for the ethereum blockchain}, Journal of Systems and Software 193 (2022) 111445.
\newblock \href {https://doi.org/https://doi.org/10.1016/j.jss.2022.111445} {\path{doi:https://doi.org/10.1016/j.jss.2022.111445}}.
\newline\urlprefix\url{https://www.sciencedirect.com/science/article/pii/S0164121222001418}

\bibitem{DaiEtAl2022}
M.~Dai, Z.~Yang, J.~Guo, Superdetector: A framework for performance detection on vulnerabilities of smart contracts, Journal of physics. Conference series 2289~(1) (2022) 12010.

\bibitem{H.ZhangEtAl2023}
H.~Zhang, W.~Zhang, Y.~Feng, Y.~Liu, Svscanner: Detecting smart contract vulnerabilities via deep semantic extraction, Journal of Information Security and Applications 75 (2023) 103484.

\bibitem{OlsthoornEtAl2022}
M.~Olsthoorn, D.~Stallenberg, A.~van Deursen, A.~Panichella, \href{https://doi.org/10.1145/3510454.3516869}{Syntest-solidity: automated test case generation and fuzzing for smart contracts}, in: Proceedings of the ACM/IEEE 44th International Conference on Software Engineering: Companion Proceedings, ICSE '22, Association for Computing Machinery, New York, NY, USA, 2022, p. 202–206.
\newblock \href {https://doi.org/10.1145/3510454.3516869} {\path{doi:10.1145/3510454.3516869}}.
\newline\urlprefix\url{https://doi.org/10.1145/3510454.3516869}

\bibitem{ChenEtAl2023}
Q.~Chen, T.~Zhou, K.~Liu, L.~Li, C.~Ge, Z.~Liu, J.~Klein, T.~F. Bissyandé, Tips: towards automating patch suggestion for vulnerable smart contracts, Automated software engineering 30~(2) (2023) 31.

\bibitem{Y.ZhangEtAl2022}
Y.~Zhang, D.~Liu, Toward vulnerability detection for ethereum smart contracts using graph-matching network, Future Internet 14~(11) (2022) 326.

\bibitem{R.YuEtAl2023}
R.~Yu, Y.~Zhang, Y.~Wang, C.~Liu, Txmirror: When the dynamic evm stack meets transactions for smart contract vulnerability detection, Symmetry (Basel) 15~(7) (2023) 1345.

\bibitem{HeEtAl2024}
D.~He, K.~Ding, S.~Chan, M.~Guizani, Unknown threats detection methods of smart contracts, IEEE internet of things journal 11~(3) (2024) 1--1.

\bibitem{HajiHosseinKhaniEtAl2024}
S.~HajiHosseinKhani, A.~H. Lashkari, A.~M. Oskui, Unveiling vulnerable smart contracts: Toward profiling vulnerable smart contracts using genetic algorithm and generating benchmark dataset, Blockchain: Research and Applications 5~(1) (2024) 100171.

\bibitem{LiEtAl2023}
Z.~Li, S.~Lu, R.~Zhang, Z.~Zhao, R.~Liang, R.~Xue, W.~Li, F.~Zhang, S.~Gao, Vulhunter: Hunting vulnerable smart contracts at evm bytecode-level via multiple instance learning, IEEE Transactions on Software Engineering 49~(11) (2023) 4886--4916.
\newblock \href {https://doi.org/10.1109/TSE.2023.3317209} {\path{doi:10.1109/TSE.2023.3317209}}.

\bibitem{JiamingYeEtAl2022}
J.~Ye, M.~Ma, Y.~Lin, L.~Ma, Y.~Xue, J.~Zhao, \href{https://www.sciencedirect.com/science/article/pii/S0164121222001236}{Vulpedia: Detecting vulnerable ethereum smart contracts via abstracted vulnerability signatures}, Journal of Systems and Software 192 (2022) 111410.
\newblock \href {https://doi.org/https://doi.org/10.1016/j.jss.2022.111410} {\path{doi:https://doi.org/10.1016/j.jss.2022.111410}}.
\newline\urlprefix\url{https://www.sciencedirect.com/science/article/pii/S0164121222001236}

\bibitem{ren2021empirical}
M.~Ren, Z.~Yin, F.~Ma, Z.~Xu, Y.~Jiang, C.~Sun, H.~Li, Y.~Cai, Empirical evaluation of smart contract testing: What is the best choice?, in: Proceedings of the 30th ACM SIGSOFT international symposium on software testing and analysis, 2021, pp. 566--579.

\bibitem{zheng2024dappscan}
Z.~Zheng, J.~Su, J.~Chen, D.~Lo, Z.~Zhong, M.~Ye, Dappscan: building large-scale datasets for smart contract weaknesses in dapp projects, IEEE Transactions on Software Engineering (2024).

\bibitem{diAngeloAndSalzer2023}
M.~di~Angelo, G.~Salzer, Consolidation of ground truth sets for weakness detection in smart contracts, in: A.~Essex, S.~Matsuo, O.~Kulyk, L.~Gudgeon, A.~Klages-Mundt, D.~Perez, S.~Werner, A.~Bracciali, G.~Goodell (Eds.), Financial Cryptography and Data Security. FC 2023 International Workshops, Springer Nature Switzerland, Cham, 2024, pp. 439--455.

\bibitem{diAngeloEtAl2024}
M.~Di~Angelo, T.~Durieux, J.~Ferreira, G.~Salzer, \href{https://doi.org/10.1007/s10664-023-10414-8}{Evolution of automated weakness detection in ethereum bytecode: a comprehensive study}, Empirical Software Engineering 29 (11 2023).
\newblock \href {https://doi.org/10.48550/arXiv.2303.10517} {\path{doi:10.48550/arXiv.2303.10517}}.
\newline\urlprefix\url{https://doi.org/10.1007/s10664-023-10414-8}

\bibitem{WeiEtAl2023}
Z.~Wei, J.~Sun, Z.~Zhang, X.~Zhang, X.~Yang, L.~Zhu, Survey on quality assurance of smart contracts (2023).
\newblock \href {http://arxiv.org/abs/2311.00270} {\path{arXiv:2311.00270}}.

\bibitem{yang2024uncover}
S.~Yang, J.~Chen, M.~Huang, Z.~Zheng, Y.~Huang, Uncover the premeditated attacks: detecting exploitable reentrancy vulnerabilities by identifying attacker contracts, in: Proceedings of the IEEE/ACM 46th International Conference on Software Engineering, 2024, pp. 1--12.

\bibitem{godboley2024poster}
S.~Godboley, P.~R. Krishna, Poster: Verisol-mce: Verification-based condition coverage analysis of smart contracts using model checker engines, in: 2024 IEEE Conference on Software Testing, Verification and Validation (ICST), IEEE, 2024, pp. 434--437.

\bibitem{liang2024towards}
R.~Liang, J.~Chen, C.~Wu, K.~He, Y.~Wu, W.~Sun, R.~Du, Q.~Zhao, Y.~Liu, Towards effective detection of ponzi schemes on ethereum with contract runtime behavior graph, ACM Transactions on Software Engineering and Methodology (2024).

\bibitem{wang2025contractsentry}
S.~Wang, X.~Zhao, Contractsentry: a static analysis tool for smart contract vulnerability detection, Automated Software Engineering 32~(1) (2025) 1.

\bibitem{jj2024enhancing}
L.~JJ, K.~Singh, Enhancing oyente: four new vulnerability detections for improved smart contract security analysis, International Journal of Information Technology 16~(6) (2024) 3389--3399.

\bibitem{rodler2023ef}
M.~Rodler, D.~Paa{\ss}en, W.~Li, L.~Bernhard, T.~Holz, G.~Karame, L.~Davi, Ef{\lightning}cf: High performance smart contract fuzzing for exploit generation, in: 2023 IEEE 8th European Symposium on Security and Privacy (EuroS\&P), IEEE, 2023, pp. 449--471.

\bibitem{liu2024automated}
Y.~Liu, C.~Zhang, et~al., Automated invariant generation for solidity smart contracts, arXiv preprint arXiv:2401.00650 (2024).

\bibitem{liu2022invcon}
Y.~Liu, Y.~Li, Invcon: A dynamic invariant detector for ethereum smart contracts, in: Proceedings of the 37th IEEE/ACM International Conference on Automated Software Engineering, 2022, pp. 1--4.

\bibitem{zhong2024prettysmart}
Z.~Zhong, Z.~Zheng, H.-N. Dai, Q.~Xue, J.~Chen, Y.~Nan, Prettysmart: Detecting permission re-delegation vulnerability for token behaviors in smart contracts, in: Proceedings of the IEEE/ACM 46th International Conference on Software Engineering, 2024, pp. 1--12.

\bibitem{liao2024smartaxe}
Z.~Liao, Y.~Nan, H.~Liang, S.~Hao, J.~Zhai, J.~Wu, Z.~Zheng, Smartaxe: Detecting cross-chain vulnerabilities in bridge smart contracts via fine-grained static analysis, Proceedings of the ACM on Software Engineering 1~(FSE) (2024) 249--270.

\bibitem{wang2024efficiently}
Z.~Wang, J.~Chen, Y.~Wang, Y.~Zhang, W.~Zhang, Z.~Zheng, Efficiently detecting reentrancy vulnerabilities in complex smart contracts, Proceedings of the ACM on Software Engineering 1~(FSE) (2024) 161--181.

\bibitem{groce2021echidna}
A.~Groce, G.~Grieco, echidna-parade: A tool for diverse multicore smart contract fuzzing, in: Proceedings of the 30th ACM SIGSOFT International Symposium on Software Testing and Analysis, 2021, pp. 658--661.

\bibitem{ji2021increasing}
S.~Ji, J.~Dong, J.~Qiu, B.~Gu, Y.~Wang, T.~Wang, Increasing fuzz testing coverage for smart contracts with dynamic taint analysis, in: 2021 IEEE 21st International Conference on Software Quality, Reliability and Security (QRS), IEEE, 2021, pp. 243--247.

\bibitem{chen2021sigrec}
T.~Chen, Z.~Li, X.~Luo, X.~Wang, T.~Wang, Z.~He, K.~Fang, Y.~Zhang, H.~Zhu, H.~Li, et~al., Sigrec: Automatic recovery of function signatures in smart contracts, IEEE Transactions on Software Engineering 48~(8) (2021) 3066--3086.

\bibitem{gao2020checking}
Z.~Gao, L.~Jiang, X.~Xia, D.~Lo, J.~Grundy, Checking smart contracts with structural code embedding, IEEE Transactions on Software Engineering 47~(12) (2020) 2874--2891.

\bibitem{wang2024contractcheck}
X.~Wang, S.~Tian, W.~Cui, Contractcheck: Checking ethereum smart contracts in fine-grained level, IEEE Transactions on Software Engineering (2024).

\bibitem{zhang2024scanogenerator}
P.~Zhang, B.~Wang, X.~Luo, H.~Dong, Scanogenerator: Automatic anomaly injection for ethereum smart contracts, IEEE Transactions on Software Engineering (2024).

\bibitem{yang2025csafuzzer}
J.~Yang, X.~Zhao, H.~Zhang, L.~He, S.~Wang, N.~Gou, Csafuzzer: Fuzzing smart contracts combining with static analysis, Empirical Software Engineering 30~(3) (2025) 62.

\bibitem{gao2024sguard}
C.~Gao, W.~Yang, J.~Ye, Y.~Xue, J.~Sun, sguard+: Machine learning guided rule-based automated vulnerability repair on smart contracts, ACM Transactions on Software Engineering and Methodology 33~(5) (2024) 1--55.

\bibitem{wang2024dfier}
Z.~Wang, W.~Dai, M.~Li, K.-K.~R. Choo, D.~Zou, Dfier: A directed vulnerability verifier for ethereum smart contracts, Journal of Network and Computer Applications 231 (2024) 103984.

\bibitem{SunEtAl2022b}
J.~Sun, S.~Huang, X.~Wang, M.~Wang, J.~Du, A detection method for scarcity defect of blockchain digital asset based on invariant analysis, in: 2022 IEEE 22nd International Conference on Software Quality, Reliability and Security (QRS), 2022, pp. 73--84.
\newblock \href {https://doi.org/10.1109/QRS57517.2022.00018} {\path{doi:10.1109/QRS57517.2022.00018}}.

\bibitem{XuEtAl2021}
Y.~Xu, G.~Hu, L.~You, C.~Cao, A novel machine learning-based analysis model for smart contract vulnerability, Security and Communication Networks 2021 (2021) 1--12.

\bibitem{TorresEtAl2021}
C.~F. Torres, A.~K. Iannillo, A.~Gervais, R.~State, Confuzzius: A data dependency-aware hybrid fuzzer for smart contracts, in: 2021 IEEE European Symposium on Security and Privacy, 2021, pp. 103--119.
\newblock \href {https://doi.org/10.1109/EuroSP51992.2021.00018} {\path{doi:10.1109/EuroSP51992.2021.00018}}.

\bibitem{RongzeXuEtAl2022}
R.~Xu, Z.~Tang, G.~Ye, H.~Wang, X.~Ke, D.~Fang, Z.~Wang, \href{https://www.sciencedirect.com/science/article/pii/S221421262200148X}{Detecting code vulnerabilities by learning from large-scale open source repositories}, Journal of Information Security and Applications 69 (2022) 103293.
\newblock \href {https://doi.org/https://doi.org/10.1016/j.jisa.2022.103293} {\path{doi:https://doi.org/10.1016/j.jisa.2022.103293}}.
\newline\urlprefix\url{https://www.sciencedirect.com/science/article/pii/S221421262200148X}

\bibitem{LakadawalaEtAl2024}
H.~Lakadawala, K.~Dzigbede, Y.~Chen, Detecting reentrancy vulnerability in smart contracts using graph convolution networks, in: 2024 IEEE 21st Consumer Communications \& Networking Conference (CCNC), 2024, pp. 188--193.
\newblock \href {https://doi.org/10.1109/CCNC51664.2024.10454763} {\path{doi:10.1109/CCNC51664.2024.10454763}}.

\bibitem{LiuEtAl2024}
Y.~Liu, C.~Wang, Y.~Ma, \href{https://doi.org/10.1007/s10515-024-00418-z}{Dl4sc: a novel deep learning-based vulnerability detection framework for smart contracts}, Automated Software Engineering 31~(1) (2024) 24.
\newblock \href {https://doi.org/10.1007/s10515-024-00418-z} {\path{doi:10.1007/s10515-024-00418-z}}.
\newline\urlprefix\url{https://doi.org/10.1007/s10515-024-00418-z}

\bibitem{HanEtAl2024}
Q.~Han, L.~Wang, H.~Zhang, L.~Shi, D.~Wang, \href{https://doi.org/10.1007/s11227-024-05954-9}{{Ethchecker: a context-guided fuzzing for smart contracts}}, The Journal of Supercomputing 80~(10) (2024) 13949--13975.
\newblock \href {https://doi.org/10.1007/s11227-024-05954-9} {\path{doi:10.1007/s11227-024-05954-9}}.
\newline\urlprefix\url{https://doi.org/10.1007/s11227-024-05954-9}

\bibitem{ZengEtAl2022}
Q.~Zeng, J.~He, G.~Zhao, S.~Li, J.~Yang, H.~Tang, H.~Luo, Ethergis: A vulnerability detection framework for ethereum smart contracts based on graph learning features, in: 2022 IEEE 46th Annual Computers, Software, and Applications Conference (COMPSAC), 2022, pp. 1742--1749.
\newblock \href {https://doi.org/10.1109/COMPSAC54236.2022.00277} {\path{doi:10.1109/COMPSAC54236.2022.00277}}.

\bibitem{ControEtAl2021}
F.~Contro, M.~Crosara, M.~Ceccato, M.~D. Preda, Ethersolve: Computing an accurate control-flow graph from ethereum bytecode, in: 2021 IEEE/ACM 29th International Conference on Program Comprehension (ICPC), 2021, pp. 127--137.
\newblock \href {https://doi.org/10.1109/ICPC52881.2021.00021} {\path{doi:10.1109/ICPC52881.2021.00021}}.

\bibitem{NassirzadehEtAl2021}
B.~Nassirzadeh, H.~Sun, S.~Banescu, V.~Ganesh, Gas gauge: A security analysis tool for smart contract out-of-gas vulnerabilities, in: P.~Pardalos, I.~Kotsireas, Y.~Guo, W.~Knottenbelt (Eds.), Mathematical Research for Blockchain Economy, Springer International Publishing, Cham, 2023, pp. 143--167.

\bibitem{ShouEtAl2023}
C.~Shou, S.~Tan, K.~Sen, \href{https://doi.org/10.1145/3597926.3598059}{Ityfuzz: Snapshot-based fuzzer for smart contract}, in: Proceedings of the 32nd ACM SIGSOFT International Symposium on Software Testing and Analysis, ISSTA 2023, Association for Computing Machinery, New York, NY, USA, 2023, p. 322–333.
\newblock \href {https://doi.org/10.1145/3597926.3598059} {\path{doi:10.1145/3597926.3598059}}.
\newline\urlprefix\url{https://doi.org/10.1145/3597926.3598059}

\bibitem{FengEtAl2023}
H.~Feng, X.~Ren, Q.~Wei, Y.~Lei, R.~Kacker, D.~R. Kuhn, D.~E. Simos, Magicmirror: Towards high-coverage fuzzing of smart contracts, in: 2023 IEEE Conference on Software Testing, Verification and Validation (ICST), 2023, pp. 141--152.
\newblock \href {https://doi.org/10.1109/ICST57152.2023.00022} {\path{doi:10.1109/ICST57152.2023.00022}}.

\bibitem{NguyenEtAl2022}
H.~H. Nguyen, N.-M. Nguyen, H.-P. Doan, Z.~Ahmadi, T.-N. Doan, L.~Jiang, \href{https://doi.org/10.1145/3540250.3558927}{Mando-guru: vulnerability detection for smart contract source code by heterogeneous graph embeddings}, in: Proceedings of the 30th ACM Joint European Software Engineering Conference and Symposium on the Foundations of Software Engineering, ESEC/FSE 2022, Association for Computing Machinery, New York, NY, USA, 2022, p. 1736–1740.
\newblock \href {https://doi.org/10.1145/3540250.3558927} {\path{doi:10.1145/3540250.3558927}}.
\newline\urlprefix\url{https://doi.org/10.1145/3540250.3558927}

\bibitem{SongEtAl2023}
S.~Song, X.~Yu, Y.~Ma, J.~Li, J.~Yu, Multi-model smart contract vulnerability detection based on bigru, in: B.~Luo, L.~Cheng, Z.-G. Wu, H.~Li, C.~Li (Eds.), Neural Information Processing, Springer Nature Singapore, Singapore, 2024, pp. 3--14.

\bibitem{WuHEtAl2021}
H.~Wu, Y.~Qin, B.~Lin, X.~Mao, Y.~Lei, Z.~Zhang, S.~Wang, H.~Zhang, Peculiar: Smart contract vulnerability detection based on crucial data flow graph and pre-training techniques, 2021.
\newblock \href {https://doi.org/10.1109/ISSRE52982.2021.00047} {\path{doi:10.1109/ISSRE52982.2021.00047}}.

\bibitem{PanEtAl2021}
Z.~Pan, T.~Hu, C.~Qian, B.~Li, Redefender: A tool for detecting reentrancy vulnerabilities in smart contracts effectively, in: 2021 IEEE 21st International Conference on Software Quality, Reliability and Security (QRS), 2021, pp. 915--925.
\newblock \href {https://doi.org/10.1109/QRS54544.2021.00101} {\path{doi:10.1109/QRS54544.2021.00101}}.

\bibitem{RutaoYuEtAl2021}
R.~Yu, J.~Shu, D.~Yan, X.~Jia, Redetect: Reentrancy vulnerability detection in smart contracts with high accuracy, in: 2021 17th International Conference on Mobility, Sensing and Networking (MSN), 2021, pp. 412--419.
\newblock \href {https://doi.org/10.1109/MSN53354.2021.00069} {\path{doi:10.1109/MSN53354.2021.00069}}.

\bibitem{Z.ZhangEtAl2022}
Z.~Zhang, Y.~Lei, M.~Yan, Y.~Yu, J.~Chen, S.~Wang, X.~Mao, \href{https://doi.org/10.1145/3551349.3560428}{Reentrancy vulnerability detection and localization: A deep learning based two-phase approach}, in: Proceedings of the 37th IEEE/ACM International Conference on Automated Software Engineering, ASE '22, Association for Computing Machinery, New York, NY, USA, 2023, pp. 1--13.
\newblock \href {https://doi.org/10.1145/3551349.3560428} {\path{doi:10.1145/3551349.3560428}}.
\newline\urlprefix\url{https://doi.org/10.1145/3551349.3560428}

\bibitem{MengRenEtAl2021}
M.~Ren, F.~Ma, Z.~Yin, H.~Li, Y.~Fu, T.~Chen, Y.~Jiang, \href{https://doi.org/10.1145/3460319.3469078}{Scstudio: a secure and efficient integrated development environment for smart contracts}, in: Proceedings of the 30th ACM SIGSOFT International Symposium on Software Testing and Analysis, ISSTA 2021, Association for Computing Machinery, New York, NY, USA, 2021, p. 666–669.
\newblock \href {https://doi.org/10.1145/3460319.3469078} {\path{doi:10.1145/3460319.3469078}}.
\newline\urlprefix\url{https://doi.org/10.1145/3460319.3469078}

\bibitem{HuangEtAl2022}
J.~Huang, K.~Zhou, A.~Xiong, D.~Li, Smart contract vulnerability detection model based on multi-task learning, Sensors 22~(5) (2022) 1829.

\bibitem{LiuZEtAl2021}
Z.~Liu, P.~Qian, X.~Wang, L.~Zhu, Q.~He, S.~Ji, \href{https://doi.org/10.24963/ijcai.2021/379}{Smart contract vulnerability detection: From pure neural network to interpretable graph feature and expert pattern fusion}, in: Z.-H. Zhou (Ed.), Proceedings of the Thirtieth International Joint Conference on Artificial Intelligence, {IJCAI-21}, International Joint Conferences on Artificial Intelligence Organization, 2021, pp. 2751--2759, main Track.
\newblock \href {https://doi.org/10.24963/ijcai.2021/379} {\path{doi:10.24963/ijcai.2021/379}}.
\newline\urlprefix\url{https://doi.org/10.24963/ijcai.2021/379}

\bibitem{LiaoEtAl2022}
Z.~Liao, Z.~Zheng, X.~Chen, Y.~Nan, \href{https://doi.org/10.1145/3533767.3534222}{Smartdagger: a bytecode-based static analysis approach for detecting cross-contract vulnerability}, in: Proceedings of the 31st ACM SIGSOFT International Symposium on Software Testing and Analysis, ISSTA 2022, Association for Computing Machinery, New York, NY, USA, 2022, p. 752–764.
\newblock \href {https://doi.org/10.1145/3533767.3534222} {\path{doi:10.1145/3533767.3534222}}.
\newline\urlprefix\url{https://doi.org/10.1145/3533767.3534222}

\bibitem{SunbeomSoEtAl2021}
S.~So, S.~Hong, H.~Oh, \href{https://www.usenix.org/conference/usenixsecurity21/presentation/so}{{SmarTest}: Effectively hunting vulnerable transaction sequences in smart contracts through language {Model-Guided} symbolic execution}, in: 30th USENIX Security Symposium (USENIX Security 21), USENIX Association, 2021, pp. 1361--1378.
\newline\urlprefix\url{https://www.usenix.org/conference/usenixsecurity21/presentation/so}

\bibitem{ZhouEtAl2021}
T.~Zhou, K.~Liu, L.~Li, Z.~Liu, J.~Klein, T.~F. Bissyandé, Smartgift: Learning to generate practical inputs for testing smart contracts, in: 2021 IEEE International Conference on Software Maintenance and Evolution (ICSME), 2021, pp. 23--34.
\newblock \href {https://doi.org/10.1109/ICSME52107.2021.00009} {\path{doi:10.1109/ICSME52107.2021.00009}}.

\bibitem{ZhukovAndKorkhov2023}
A.~Zhukov, V.~Korkhov, Smartgraph: Static analysis tool for solidity smart contracts, in: O.~Gervasi, B.~Murgante, A.~M. A.~C. Rocha, C.~Garau, F.~Scorza, Y.~Karaca, C.~M. Torre (Eds.), Computational Science and Its Applications -- ICCSA 2023 Workshops, Springer Nature Switzerland, Cham, 2023, pp. 584--598.

\bibitem{ChoiEtAl2021}
J.~Choi, D.~Kim, S.~Kim, G.~Grieco, A.~Groce, S.~K. Cha, Smartian: Enhancing smart contract fuzzing with static and dynamic data-flow analyses, in: 2021 36th IEEE/ACM International Conference on Automated Software Engineering (ASE), 2021, pp. 227--239.
\newblock \href {https://doi.org/10.1109/ASE51524.2021.9678888} {\path{doi:10.1109/ASE51524.2021.9678888}}.

\bibitem{SamreenAndAlalfi2021}
N.~F. Samreen, M.~H. Alalfi, Smartscan: an approach to detect denial of service vulnerability in ethereum smart contracts, in: 2021 IEEE/ACM 4th International Workshop on Emerging Trends in Software Engineering for Blockchain (WETSEB), IEEE, 2021, pp. 17--26.

\bibitem{TorresEtAl2021b}
C.~Ferreira~Torres, A.~K. Iannillo, A.~Gervais, R.~State, The eye of horus: Spotting and analyzing attacks on ethereum smart contracts, in: N.~Borisov, C.~Diaz (Eds.), Financial Cryptography and Data Security, Springer Berlin Heidelberg, Berlin, Heidelberg, 2021, pp. 33--52.

\bibitem{MiEtAl2021}
F.~Mi, Z.~Wang, C.~Zhao, J.~Guo, F.~Ahmed, L.~Khan, Vscl: Automating vulnerability detection in smart contracts with deep learning, in: 2021 IEEE International Conference on Blockchain and Cryptocurrency (ICBC), 2021, pp. 1--9.
\newblock \href {https://doi.org/10.1109/ICBC51069.2021.9461050} {\path{doi:10.1109/ICBC51069.2021.9461050}}.

\bibitem{JiangEtAl2021}
B.~Jiang, Y.~Chen, D.~Wang, I.~Ashraf, W.~Chan, Wana: Symbolic execution of wasm bytecode for extensible smart contract vulnerability detection, in: 2021 IEEE 21st International Conference on Software Quality, Reliability and Security (QRS), 2021, pp. 926--937.
\newblock \href {https://doi.org/10.1109/QRS54544.2021.00102} {\path{doi:10.1109/QRS54544.2021.00102}}.

\bibitem{XueEtAl2022}
Y.~Xue, J.~Ye, W.~Zhang, J.~Sun, L.~Ma, H.~Wang, J.~Zhao, xfuzz: Machine learning guided cross-contract fuzzing, IEEE Transactions on Dependable and Secure Computing 21~(2) (2024) 515--529.
\newblock \href {https://doi.org/10.1109/TDSC.2022.3182373} {\path{doi:10.1109/TDSC.2022.3182373}}.

\bibitem{fu2019evmfuzzer}
Y.~Fu, M.~Ren, F.~Ma, H.~Shi, X.~Yang, Y.~Jiang, H.~Li, X.~Shi, Evmfuzzer: detect evm vulnerabilities via fuzz testing, in: Proceedings of the 2019 27th ACM joint meeting on european software engineering conference and symposium on the foundations of software engineering, 2019, pp. 1110--1114.

\bibitem{zhang2020txspector}
M.~Zhang, X.~Zhang, Y.~Zhang, Z.~Lin, $\{$TXSPECTOR$\}$: Uncovering attacks in ethereum from transactions, in: 29th USENIX Security Symposium (USENIX Security 20), 2020, pp. 2775--2792.

\bibitem{stegeman2018solitor}
L.~Stegeman, Solitor: runtime verification of smart contracts on the ethereum network, Master's thesis, University of Twente (2018).

\bibitem{wu2020kaya}
Z.~Wu, J.~Zhang, J.~Gao, Y.~Li, Q.~Li, Z.~Guan, Z.~Chen, Kaya: A testing framework for blockchain-based decentralized applications, in: 2020 IEEE International Conference on Software Maintenance and Evolution (ICSME), IEEE, 2020, pp. 826--829.

\bibitem{sigmaprime_solidity_security}
{Sigma Prime}, Solidity security: Comprehensive guide to smart contract security, \url{https://blog.sigmaprime.io/solidity-security.html}, accessed: [Date of access] (2023).

\end{thebibliography}


\begin{thebibliography}{100}
\expandafter\ifx\csname url\endcsname\relax
  \def\url#1{\texttt{#1}}\fi
\expandafter\ifx\csname urlprefix\endcsname\relax\def\urlprefix{URL }\fi
\expandafter\ifx\csname href\endcsname\relax
  \def\href#1#2{#2} \def\path#1{#1}\fi

\bibitem{nakamoto2008bitcoin}
S.~Nakamoto, Bitcoin: A peer-to-peer electronic cash system (2008).

\bibitem{alharby2018blockchain}
M.~Alharby, A.~Aldweesh, A.~Van~Moorsel, Blockchain-based smart contracts: A systematic mapping study of academic research (2018), in: 2018 International Conference on Cloud Computing, Big Data and Blockchain (ICCBB), IEEE, 2018, pp. 1--6.

\bibitem{szabo1997formalizing}
N.~Szabo, Formalizing and securing relationships on public networks, First monday (1997).

\bibitem{chaliasos2024smart}
S.~Chaliasos, M.~A. Charalambous, L.~Zhou, R.~Galanopoulou, A.~Gervais, D.~Mitropoulos, B.~Livshits, Smart contract and defi security tools: Do they meet the needs of practitioners?, in: Proceedings of the 46th IEEE/ACM International Conference on Software Engineering, 2024, pp. 1--13.

\bibitem{wang2021non}
Q.~Wang, R.~Li, Q.~Wang, S.~Chen, Non-fungible token (nft): Overview, evaluation, opportunities and challenges, arXiv preprint arXiv:2105.07447 (2021).

\bibitem{zou2019smart}
W.~Zou, D.~Lo, P.~S. Kochhar, X.-B.~D. Le, X.~Xia, Y.~Feng, Z.~Chen, B.~Xu, Smart contract development: Challenges and opportunities, IEEE transactions on software engineering 47~(10) (2019) 2084--2106.

\bibitem{Rameder2022}
H.~Rameder, M.~Di~Angelo, G.~Salzer, Review of automated vulnerability analysis of smart contracts on ethereum, Frontiers in Blockchain 5 (03 2022).
\newblock \href {https://doi.org/10.3389/fbloc.2022.814977} {\path{doi:10.3389/fbloc.2022.814977}}.

\bibitem{duan2023attacks}
L.~Duan, Y.~Sun, W.~Ni, W.~Ding, J.~Liu, W.~Wang, Attacks against cross-chain systems and defense approaches: A contemporary survey, IEEE/CAA Journal of Automatica Sinica 10~(8) (2023) 1647--1667.

\bibitem{lee2023sok}
S.-S. Lee, A.~Murashkin, M.~Derka, J.~Gorzny, Sok: Not quite water under the bridge: Review of cross-chain bridge hacks, in: 2023 IEEE International Conference on Blockchain and Cryptocurrency (ICBC), IEEE, 2023, pp. 1--14.

\bibitem{zhang2022xscope}
J.~Zhang, J.~Gao, Y.~Li, Z.~Chen, Z.~Guan, Z.~Chen, Xscope: Hunting for cross-chain bridge attacks, in: Proceedings of the 37th IEEE/ACM International Conference on Automated Software Engineering, 2022, pp. 1--4.

\bibitem{buterin2014sc&dapps}
V.~Buterin, et~al., A next-generation smart contract and decentralized application platform, white paper 3~(37) (2014) 2--1.

\bibitem{buterin2013ethereum}
V.~Buterin, et~al., Ethereum white paper, GitHub repository 1 (2013) 22--23.

\bibitem{merkletree}
H.~Liu, X.~Luo, H.~Liu, X.~Xia, Merkle tree: A fundamental component of blockchains, in: 2021 International Conference on Electronic Information Engineering and Computer Science (EIECS), 2021, pp. 556--561.
\newblock \href {https://doi.org/10.1109/EIECS53707.2021.9588047} {\path{doi:10.1109/EIECS53707.2021.9588047}}.

\bibitem{szabo1994smartcontract}
N.~Szabo, Smart contracts http://www. fon. hum. uva. nl/rob/courses/informationinspeech/cdrom/literature, LOTwinterschool2006/szabo. best. vwh. net/smart. contracts. html (1994).

\bibitem{antonopoulos2018masteringSc}
A.~M. Antonopoulos, G.~Wood, Mastering ethereum: building smart contracts and dapps, O'reilly Media, 2018.

\bibitem{dannen2017introSolidity}
C.~Dannen, Introducing Ethereum and solidity, Vol.~1, Springer, 2017.

\bibitem{vyper}
M.~Kaleem, A.~Mavridou, A.~Laszka, Vyper: A security comparison with solidity based on common vulnerabilities, in: 2020 2nd Conference on Blockchain Research \& Applications for Innovative Networks and Services (BRAINS), 2020, pp. 107--111.
\newblock \href {https://doi.org/10.1109/BRAINS49436.2020.9223278} {\path{doi:10.1109/BRAINS49436.2020.9223278}}.

\bibitem{kaushal2021immutable}
R.~K. Kaushal, N.~Kumar, S.~N. Panda, V.~Kukreja, Immutable smart contracts on blockchain technology: Its benefits and barriers, in: 2021 9th International Conference on Reliability, Infocom Technologies and Optimization (Trends and Future Directions)(ICRITO), IEEE, 2021, pp. 1--5.

\bibitem{politou2019blockMutability}
E.~Politou, F.~Casino, E.~Alepis, C.~Patsakis, Blockchain mutability: Challenges and proposed solutions, IEEE Transactions on Emerging Topics in Computing 9~(4) (2019) 1972--1986.

\bibitem{samreen2020reentrancy}
N.~F. Samreen, M.~H. Alalfi, Reentrancy vulnerability identification in ethereum smart contracts, in: 2020 IEEE International Workshop on Blockchain Oriented Software Engineering (IWBOSE), IEEE, 2020, pp. 22--29.

\bibitem{NepomucenoAndSoares2019}
V.~Nepomuceno, S.~Soares, \href{https://www.sciencedirect.com/science/article/pii/S0950584919300072}{On the need to update systematic literature reviews}, Information and Software Technology 109 (2019) 40--42.
\newblock \href {https://doi.org/https://doi.org/10.1016/j.infsof.2019.01.005} {\path{doi:https://doi.org/10.1016/j.infsof.2019.01.005}}.
\newline\urlprefix\url{https://www.sciencedirect.com/science/article/pii/S0950584919300072}

\bibitem{MendesEtAl2020}
E.~Mendes, C.~Wohlin, K.~Felizardo, M.~Kalinowski, When to update systematic literature reviews in software engineering (04 2020).

\bibitem{GarnerEtAl2016}
P.~Garner, S.~Hopewell, J.~Chandler, H.~MacLehose, H.~Schünemann, E.~Akl, J.~Beyene, S.~Chang, R.~Churchill, K.~Dearness, G.~Guyatt, C.~Lefebvre, B.~Liles, R.~Marshall, L.~García, C.~Mavergames, M.~Nasser, A.~Qaseem, M.~Sampson, E.~Wilson, When and how to update systematic reviews: Consensus and checklist, BMJ 354 (2016) i3507.
\newblock \href {https://doi.org/10.1136/bmj.i3507} {\path{doi:10.1136/bmj.i3507}}.

\bibitem{Brereton2007}
P.~Brereton, B.~A. Kitchenham, D.~Budgen, M.~Turner, M.~Khalil, \href{https://www.sciencedirect.com/science/article/pii/S016412120600197X}{Lessons from applying the systematic literature review process within the software engineering domain}, Journal of Systems and Software 80~(4) (2007) 571--583, software Performance.
\newblock \href {https://doi.org/https://doi.org/10.1016/j.jss.2006.07.009} {\path{doi:https://doi.org/10.1016/j.jss.2006.07.009}}.
\newline\urlprefix\url{https://www.sciencedirect.com/science/article/pii/S016412120600197X}

\bibitem{buterin2016chain}
V.~Buterin, Chain interoperability, R3 research paper 9 (2016) 1--25.

\bibitem{Strong2002}
D.~Strong, Y.~Lee, R.~Wang, Data quality in context, Communications of the ACM 40 (08 2002).
\newblock \href {https://doi.org/10.1145/253769.253804} {\path{doi:10.1145/253769.253804}}.

\bibitem{Kitchenham2007}
B.~Kitchenham, S.~Charters, Guidelines for performing systematic literature reviews in software engineering 2 (01 2007).

\bibitem{Guidelines4Taxonomy}
P.~Ralph, Toward methodological guidelines for process theories and taxonomies in software engineering, IEEE Transactions on Software Engineering 45~(7) (2019) 712--735.
\newblock \href {https://doi.org/10.1109/TSE.2018.2796554} {\path{doi:10.1109/TSE.2018.2796554}}.

\bibitem{saldana2021coding}
J.~Salda{\~n}a, The coding manual for qualitative researchers (2021).

\bibitem{slither}
J.~Feist, G.~Grieco, A.~Groce, Slither: a static analysis framework for smart contracts, in: 2019 IEEE/ACM 2nd International Workshop on Emerging Trends in Software Engineering for Blockchain (WETSEB), IEEE, 2019, pp. 8--15.

\bibitem{kaur2020scalability}
G.~Kaur, C.~Gandhi, Scalability in blockchain: Challenges and solutions, in: Handbook of Research on Blockchain Technology, Elsevier, 2020, pp. 373--406.

\bibitem{suhail2022blockchain}
S.~Suhail, R.~Hussain, R.~Jurdak, A.~Oracevic, K.~Salah, C.~S. Hong, R.~Matulevi{\v{c}}ius, Blockchain-based digital twins: Research trends, issues, and future challenges, ACM Computing Surveys (CSUR) 54~(11s) (2022) 1--34.

\bibitem{yu2019challenge}
G.~Yu, T.-Z. Nie, X.-H. Li, Y.-F. Zhang, D.-R. Shen, Y.-B. Bao, The challenge and prospect of distributed data management techniques in blockchain systems, Chinese Journal of Computers 42 (2019) 1--27.

\bibitem{xu2022banks}
J.~Xu, N.~Vadgama, From banks to defi: the evolution of the lending market, Enabling the Internet of Value: How Blockchain Connects Global Businesses (2022) 53--66.

\bibitem{rizvi2020identifying}
S.~Rizvi, R.~Orr, A.~Cox, P.~Ashokkumar, M.~R. Rizvi, Identifying the attack surface for iot network, Internet of Things 9 (2020) 100162.

\bibitem{sun2020collaborative}
W.~Sun, L.~Wang, P.~Wang, Y.~Zhang, Collaborative blockchain for space-air-ground integrated networks, IEEE Wireless Communications 27~(6) (2020) 82--89.

\bibitem{lv2022attack}
Z.~Lv, D.~Wu, W.~Yang, L.~Duan, Attack and protection schemes on fabric isomorphic crosschain systems, International Journal of Distributed Sensor Networks 18~(1) (2022) 15501477211059945.

\bibitem{tian2021enabling}
H.~Tian, K.~Xue, X.~Luo, S.~Li, J.~Xu, J.~Liu, J.~Zhao, D.~S. Wei, Enabling cross-chain transactions: A decentralized cryptocurrency exchange protocol, IEEE Transactions on Information Forensics and Security 16 (2021) 3928--3941.

\bibitem{chepurnoy2019multi}
A.~Chepurnoy, A.~Saxena, Multi-stage contracts in the utxo model, in: Data Privacy Management, Cryptocurrencies and Blockchain Technology: ESORICS 2019 International Workshops, DPM 2019 and CBT 2019, Luxembourg, September 26--27, 2019, Proceedings 14, Springer, 2019, pp. 244--254.

\bibitem{malavolta2018anonymous}
G.~Malavolta, P.~Moreno-Sanchez, C.~Schneidewind, A.~Kate, M.~Maffei, Anonymous multi-hop locks for blockchain scalability and interoperability, Cryptology ePrint Archive (2018).

\bibitem{xue2021hedging}
Y.~Xue, M.~Herlihy, Hedging against sore loser attacks in cross-chain transactions, in: Proceedings of the 2021 ACM Symposium on Principles of Distributed Computing, 2021, pp. 155--164.

\bibitem{careem2020reputation}
M.~A.~A. Careem, A.~Dutta, Reputation based routing in manet using blockchain, in: 2020 International Conference on COMmunication Systems \& NETworkS (COMSNETS), IEEE, 2020, pp. 1--6.

\bibitem{sonnino2020replay}
A.~Sonnino, S.~Bano, M.~Al-Bassam, G.~Danezis, Replay attacks and defenses against cross-shard consensus in sharded distributed ledgers, in: 2020 IEEE European Symposium on Security and Privacy (EuroS\&P), IEEE, 2020, pp. 294--308.

\bibitem{schrepel2019collusion}
T.~Schrepel, Collusion by blockchain and smart contracts, Harv. JL \& Tech. 33 (2019) 117.

\bibitem{BenEtAlAFS}
B.~Wang, H.~Chu, P.~Zhang, H.~Dong, Smart contract vulnerability detection using code representation fusion, in: 2021 28th Asia-Pacific Software Engineering Conference (APSEC), 2021, pp. 564--565.
\newblock \href {https://doi.org/10.1109/APSEC53868.2021.00069} {\path{doi:10.1109/APSEC53868.2021.00069}}.

\bibitem{SC-VDM}
K.~Zhou, J.~Cheng, H.~Li, Y.~Yuan, L.~Liu, X.~Li, Sc-vdm: A lightweight smart contract vulnerability detection model, in: Y.~Tan, Y.~Shi, A.~Zomaya, H.~Yan, J.~Cai (Eds.), Data Mining and Big Data, Springer Singapore, Singapore, 2021, pp. 138--149.

\bibitem{EVMFuzzer}
Y.~Fu, M.~Ren, F.~Ma, H.~Shi, X.~Yang, Y.~Jiang, H.~Li, X.~Shi, \href{https://doi.org/10.1145/3338906.3341175}{Evmfuzzer: detect evm vulnerabilities via fuzz testing}, in: Proceedings of the 2019 27th ACM Joint Meeting on European Software Engineering Conference and Symposium on the Foundations of Software Engineering, ESEC/FSE 2019, Association for Computing Machinery, New York, NY, USA, 2019, p. 1110–1114.
\newblock \href {https://doi.org/10.1145/3338906.3341175} {\path{doi:10.1145/3338906.3341175}}.
\newline\urlprefix\url{https://doi.org/10.1145/3338906.3341175}

\bibitem{WangEtAl2020Artemis}
A.~Wang, H.~Wang, B.~Jiang, W.~K. Chan, Artemis: An improved smart contract verification tool for vulnerability detection, in: 2020 7th International Conference on Dependable Systems and Their Applications (DSA), 2020, pp. 173--181.
\newblock \href {https://doi.org/10.1109/DSA51864.2020.00031} {\path{doi:10.1109/DSA51864.2020.00031}}.

\bibitem{9343119}
X.~Hao, W.~Ren, W.~Zheng, T.~Zhu, Scscan: A svm-based scanning system for vulnerabilities in blockchain smart contracts, in: 2020 IEEE 19th International Conference on Trust, Security and Privacy in Computing and Communications (TrustCom), 2020, pp. 1598--1605.
\newblock \href {https://doi.org/10.1109/TrustCom50675.2020.00221} {\path{doi:10.1109/TrustCom50675.2020.00221}}.

\bibitem{AshrafEtAl2020}
I.~Ashraf, X.~Ma, B.~Jiang, W.~Chan, Gasfuzzer: Fuzzing ethereum smart contract binaries to expose gas-oriented exception security vulnerabilities, IEEE Access PP (2020) 1--1.
\newblock \href {https://doi.org/10.1109/ACCESS.2020.2995183} {\path{doi:10.1109/ACCESS.2020.2995183}}.

\bibitem{9376204}
R.~Ji, N.~He, L.~Wu, H.~Wang, G.~Bai, Y.~Guo, Deposafe: Demystifying the fake deposit vulnerability in ethereum smart contracts, in: 2020 25th International Conference on Engineering of Complex Computer Systems (ICECCS), 2020, pp. 125--134.
\newblock \href {https://doi.org/10.1109/ICECCS51672.2020.00022} {\path{doi:10.1109/ICECCS51672.2020.00022}}.

\bibitem{9425913}
S.~Lee, E.-S. Cho, Lightweight extension of an execution environment for safer function calls in solidity/ethereum virtual machine smart contracts, in: 2021 IEEE International Conference on Software Analysis, Evolution and Reengineering (SANER), 2021, pp. 689--695.
\newblock \href {https://doi.org/10.1109/SANER50967.2021.00087} {\path{doi:10.1109/SANER50967.2021.00087}}.

\bibitem{8935603}
Z.~Tian, Smart contract defect detection based on parallel symbolic execution, in: 2019 3rd International Conference on Circuits, System and Simulation (ICCSS), 2019, pp. 127--132.
\newblock \href {https://doi.org/10.1109/CIRSYSSIM.2019.8935603} {\path{doi:10.1109/CIRSYSSIM.2019.8935603}}.

\bibitem{9730357}
Y.~Tang, Z.~Li, Y.~Bai, Rethinking of reentrancy on the ethereum, in: 2021 IEEE Intl Conf on Dependable, Autonomic and Secure Computing, Intl Conf on Pervasive Intelligence and Computing, Intl Conf on Cloud and Big Data Computing, Intl Conf on Cyber Science and Technology Congress (DASC/PiCom/CBDCom/CyberSciTech), 2021, pp. 68--75.
\newblock \href {https://doi.org/10.1109/DASC-PICom-CBDCom-CyberSciTech52372.2021.00025} {\path{doi:10.1109/DASC-PICom-CBDCom-CyberSciTech52372.2021.00025}}.

\bibitem{9724805}
S.~Ji, J.~Dong, J.~Qiu, B.~Gu, Y.~Wang, T.~Wang, Increasing fuzz testing coverage for smart contracts with dynamic taint analysis, in: 2021 IEEE 21st International Conference on Software Quality, Reliability and Security (QRS), 2021, pp. 243--247.
\newblock \href {https://doi.org/10.1109/QRS54544.2021.00035} {\path{doi:10.1109/QRS54544.2021.00035}}.

\bibitem{8870156}
T.~Chen, Z.~Li, Y.~Zhang, X.~Luo, T.~Wang, T.~Hu, X.~Xiao, D.~Wang, J.~Huang, X.~Zhang, A large-scale empirical study on control flow identification of smart contracts, in: 2019 ACM/IEEE International Symposium on Empirical Software Engineering and Measurement (ESEM), 2019, pp. 1--11.
\newblock \href {https://doi.org/10.1109/ESEM.2019.8870156} {\path{doi:10.1109/ESEM.2019.8870156}}.

\bibitem{8949045}
P.~Momeni, Y.~Wang, R.~Samavi, Machine learning model for smart contracts security analysis, in: 2019 17th International Conference on Privacy, Security and Trust (PST), 2019, pp. 1--6.
\newblock \href {https://doi.org/10.1109/PST47121.2019.8949045} {\path{doi:10.1109/PST47121.2019.8949045}}.

\bibitem{3057565}
J.~Correas, P.~Gordillo, G.~Román~Díez, Static profiling and optimization of ethereum smart contracts using resource analysis, IEEE Access PP (2021) 1--1.
\newblock \href {https://doi.org/10.1109/ACCESS.2021.3057565} {\path{doi:10.1109/ACCESS.2021.3057565}}.

\bibitem{Sun_2021}
Y.~Sun, L.~Gu, \href{https://dx.doi.org/10.1088/1742-6596/1820/1/012004}{Attention-based machine learning model for smart contract vulnerability detection}, Journal of Physics: Conference Series 1820~(1) (2021) 012004.
\newblock \href {https://doi.org/10.1088/1742-6596/1820/1/012004} {\path{doi:10.1088/1742-6596/1820/1/012004}}.
\newline\urlprefix\url{https://dx.doi.org/10.1088/1742-6596/1820/1/012004}

\bibitem{Wang2021SmartCV}
B.~Wang, H.~Chu, P.~Zhang, H.~Dong, \href{https://api.semanticscholar.org/CorpusID:246945372}{Smart contract vulnerability detection using code representation fusion}, 2021 28th Asia-Pacific Software Engineering Conference (APSEC) (2021) 564--565.
\newline\urlprefix\url{https://api.semanticscholar.org/CorpusID:246945372}

\bibitem{9680540}
W.~Jie, A.~S.~V. Koe, P.~Huang, S.~Zhang, Full-stack hierarchical fusion of static features for smart contracts vulnerability detection, in: 2021 IEEE International Conference on Blockchain (Blockchain), 2021, pp. 95--102.
\newblock \href {https://doi.org/10.1109/Blockchain53845.2021.00091} {\path{doi:10.1109/Blockchain53845.2021.00091}}.

\bibitem{JianjunEtAl2021}
J.~Huang, S.~Han, W.~You, W.~Shi, B.~Liang, J.~Wu, Y.~Wua, Hunting vulnerable smart contracts via graph embedding based bytecode matching, IEEE Transactions on Information Forensics and Security PP (2021) 1--1.
\newblock \href {https://doi.org/10.1109/TIFS.2021.3050051} {\path{doi:10.1109/TIFS.2021.3050051}}.

\bibitem{3444370}
Z.~Li, W.~Guo, Q.~Xu, Y.~Xu, H.~Wang, M.~Xian, \href{https://doi.org/10.1145/3444370.3444617}{Research on blockchain smart contracts vulnerability and a code audit tool based on matching rules}, in: Proceedings of the 2020 International Conference on Cyberspace Innovation of Advanced Technologies, CIAT 2020, Association for Computing Machinery, New York, NY, USA, 2021, p. 484–489.
\newblock \href {https://doi.org/10.1145/3444370.3444617} {\path{doi:10.1145/3444370.3444617}}.
\newline\urlprefix\url{https://doi.org/10.1145/3444370.3444617}

\bibitem{101007}
S.~Du, H.~Huang, A general framework of smart contract vulnerability mining based on control flow graph matching, in: X.~Sun, J.~Wang, E.~Bertino (Eds.), Artificial Intelligence and Security, Springer Singapore, Singapore, 2020, pp. 166--175.

\bibitem{e3714a3b32d246f0b781dd55b7d6eec7}
R.~Ma, Z.~Jian, G.~Chen, K.~Ma, Y.~Chen, Rejection: A ast-based reentrancy vulnerability detection method, in: W.~Han, L.~Zhu, F.~Yan (Eds.), Trusted Computing and Information Security - 13th Chinese Conference, CTCIS 2019, Revised Selected Papers, Communications in Computer and Information Science, Springer, Germany, 2020, pp. 58--71, publisher Copyright: {\textcopyright} Springer Nature Singapore Pte Ltd 2020.; 13th Chinese Conference on Trusted Computing and Information Security, CTCIS 2019 ; Conference date: 24-10-2019 Through 27-10-2019.
\newblock \href {https://doi.org/10.1007/978-981-15-3418-8_5} {\path{doi:10.1007/978-981-15-3418-8_5}}.

\bibitem{8867880}
M.~Fu, L.~Wu, Z.~Hong, F.~Zhu, H.~Sun, W.~Feng, A critical-path-coverage-based vulnerability detection method for smart contracts, IEEE Access 7 (2019) 147327--147344.
\newblock \href {https://doi.org/10.1109/ACCESS.2019.2947146} {\path{doi:10.1109/ACCESS.2019.2947146}}.

\bibitem{101007978}
Y.~Fan, S.~Shang, X.~Ding, Smart contract vulnerability detection based on dual attention graph convolutional network, in: H.~Gao, X.~Wang (Eds.), Collaborative Computing: Networking, Applications and Worksharing, Springer International Publishing, Cham, 2021, pp. 335--351.

\bibitem{9202747}
Z.~Yang, J.~Keung, M.~Zhang, Y.~Xiao, Y.~Huang, T.~Hui, Smart contracts vulnerability auditing with multi-semantics, in: 2020 IEEE 44th Annual Computers, Software, and Applications Conference (COMPSAC), 2020, pp. 892--901.
\newblock \href {https://doi.org/10.1109/COMPSAC48688.2020.0-153} {\path{doi:10.1109/COMPSAC48688.2020.0-153}}.

\bibitem{Rodler2020EVMPatchTA}
M.~Rodler, W.~Li, G.~O. Karame, L.~Davi, \href{https://api.semanticscholar.org/CorpusID:222090393}{Evmpatch: Timely and automated patching of ethereum smart contracts}, in: USENIX Security Symposium, 2020.
\newline\urlprefix\url{https://api.semanticscholar.org/CorpusID:222090393}

\bibitem{181106632}
W.~J. Tann, X.~J. Han, S.~S. Gupta, Y.~Ong, \href{http://arxiv.org/abs/1811.06632}{Towards safer smart contracts: {A} sequence learning approach to detecting vulnerabilities}, CoRR abs/1811.06632 (2018).
\newblock \href {http://arxiv.org/abs/1811.06632} {\path{arXiv:1811.06632}}.
\newline\urlprefix\url{http://arxiv.org/abs/1811.06632}

\bibitem{8970384}
P.~Qian, Z.~Liu, Q.~He, R.~Zimmermann, X.~Wang, Towards automated reentrancy detection for smart contracts based on sequential models, IEEE Access 8 (2020) 19685--19695.
\newblock \href {https://doi.org/10.1109/ACCESS.2020.2969429} {\path{doi:10.1109/ACCESS.2020.2969429}}.

\bibitem{s23167246}
W.~Deng, H.~Wei, T.~Huang, C.~Cao, Y.~Peng, X.~Hu, Smart contract vulnerability detection based on deep learning and multimodal decision fusion, Sensors 23 (2023) 7246.
\newblock \href {https://doi.org/10.3390/s23167246} {\path{doi:10.3390/s23167246}}.

\bibitem{giesen2022}
J.-R. Giesen, S.~Andreina, M.~Rodler, G.~O. Karame, L.~Davi, \href{https://arxiv.org/abs/2203.00364}{Practical mitigation of smart contract bugs} (2022).
\newblock \href {http://arxiv.org/abs/2203.00364} {\path{arXiv:2203.00364}}.
\newline\urlprefix\url{https://arxiv.org/abs/2203.00364}

\bibitem{3551349}
P.~Tolmach, Y.~Li, S.-W. Lin, \href{https://doi.org/10.1145/3551349.3559560}{Property-based automated repair of defi protocols}, in: Proceedings of the 37th IEEE/ACM International Conference on Automated Software Engineering, ASE '22, Association for Computing Machinery, New York, NY, USA, 2023.
\newblock \href {https://doi.org/10.1145/3551349.3559560} {\path{doi:10.1145/3551349.3559560}}.
\newline\urlprefix\url{https://doi.org/10.1145/3551349.3559560}

\bibitem{suiche2017porosity}
M.~Suiche, Porosity: A decompiler for blockchain-based smart contracts bytecode, DEF con 25~(11) (2017).

\bibitem{pakala2024}
P.-A. Losi, \href{https://www.palkeo.com/en/projets/ethereum/pakala.html}{Pakala: A toolkit for analyzing ethereum smart contracts}, accessed: 2024-10-08 (2024).
\newline\urlprefix\url{https://www.palkeo.com/en/projets/ethereum/pakala.html}

\bibitem{StegemanSolitor}
L.~{Stegeman}, \href{http://essay.utwente.nl/76902/}{Solitor: runtime verification of smart contracts on the ethereum network} (November 2018).
\newline\urlprefix\url{http://essay.utwente.nl/76902/}

\bibitem{WuEtAl2020}
Z.~Wu, J.~Zhang, J.~Gao, Y.~Li, Q.~Li, Z.~Guan, Z.~Chen, Kaya: A testing framework for blockchain-based decentralized applications, in: 2020 IEEE International Conference on Software Maintenance and Evolution (ICSME), 2020, pp. 826--829.
\newblock \href {https://doi.org/10.1109/ICSME46990.2020.00103} {\path{doi:10.1109/ICSME46990.2020.00103}}.

\bibitem{rezaei2025sokrootcause1}
H.~Rezaei, M.~Eshghie, K.~Anderesson, F.~Palmieri, \href{https://arxiv.org/abs/2507.20175}{Sok: Root cause of \$1 billion loss in smart contract real-world attacks via a systematic literature review of vulnerabilities} (2025).
\newblock \href {http://arxiv.org/abs/2507.20175} {\path{arXiv:2507.20175}}.
\newline\urlprefix\url{https://arxiv.org/abs/2507.20175}

\bibitem{ressi2025vulnerability}
D.~Ressi, A.~Span{\`o}, L.~Benetollo, M.~Bugliesi, C.~Piazza, S.~Rossi, Vulnerability detection in solidity smart contracts via machine learning: A qualitative analysis, Blockchain: Research and Applications (2025) 100390.

\bibitem{iuliano2025msr}
G.~Iuliano, D.~Corradini, M.~Pasqua, M.~Ceccato, D.~D. Nucci, \href{https://arxiv.org/abs/2504.05515}{How do solidity versions affect vulnerability detection tools? an empirical study} (2025).
\newblock \href {http://arxiv.org/abs/2504.05515} {\path{arXiv:2504.05515}}.
\newline\urlprefix\url{https://arxiv.org/abs/2504.05515}

\bibitem{AshouriEtAl2020}
M.~Ashouri, {Etherolic}: A practical security analyzer for smart contracts, in: Proceedings of the 35th Annual ACM Symposium on Applied Computing, ACM, 2020, pp. 353--356.

\bibitem{AzzopardiEtAl2018}
S.~Azzopardi, J.~Ellul, G.~J. Pace, Monitoring smart contracts: contractlarva and open challenges beyond, in: Proceedings of the International Conference on Runtime Verification, 2018, pp. 113--137.

\bibitem{BhargavanEtAl2016}
K.~Bhargavan, A.~Delignat-Lavaud, C.~Fournet, et~al., Formal verification of smart contracts, in: Proceedings of the 2016 ACM Workshop on Programming Languages and Analysis for Security, 2016, pp. 91--96.

\bibitem{ChangEtAl2019}
J.~Chang, B.~Gao, H.~Xiao, J.~Sun, Y.~Cai, Z.~Yang, {sCompile}: Critical path identification and analysis for smart contracts, in: Proceedings of the International Conference on Formal Engineering Methods, 2019, pp. 286--304.

\bibitem{ChapmanEtAl2019}
P.~Chapman, D.~Xu, L.~Deng, Y.~Xiong, {Deviant}: A mutation testing tool for solidity smart contracts, in: Proceedings of the 2019 IEEE International Conference on Blockchain (Blockchain), 2019, pp. 319--324.

\bibitem{ChatterjeeEtAl2019}
K.~Chatterjee, A.~K. Goharshady, E.~K. Goharshady, The treewidth of smart contracts, in: Proceedings of the 34th ACM/SIGAPP Symposium on Applied Computing, 2019, pp. 400--408.

\bibitem{ChenEtAl2017}
T.~Chen, X.~Li, Y.~Wang, et~al., An adaptive gas cost mechanism for ethereum to defend against under-priced dos attacks, in: Proceedings of the International Conference on Information Security Practice and Experience, 2017, pp. 3--24.

\bibitem{DongEtAl2019}
C.~Dong, Y.~Li, L.~Tan, A new approach to prevent reentrant attack in solidity smart contracts, in: Proceedings of the CCF China Blockchain Conference, 2019, pp. 83--103.

\bibitem{El-DosukyEtAl2019}
M.~A. El-Dosuky, G.~H. Eladl, {DOORchain}: deep ontology-based operation research to detect malicious smart contracts, in: Proceedings of the World Conference on Information Systems and Technologies, 2019, pp. 538--545.

\bibitem{FeistEtAl2019}
J.~Feist, G.~Grieco, A.~Groce, {Slither}: A static analysis framework for smart contracts, in: Proceedings of the 2019 IEEE/ACM 2nd International Workshop on Emerging Trends in Software Engineering for Blockchain (WETSEB), 2019, pp. 8--15.

\bibitem{FuEtAl2019}
M.~Fu, L.~Wu, Z.~Hong, F.~Zhu, H.~Sun, W.~Feng, A critical-path-coverage-based vulnerability detection method for smart contracts, IEEE Access 7 (2019) 147327--147344.

\bibitem{GaoEtAl2019}
J.~Gao, H.~Liu, C.~Liu, Q.~Li, Z.~Guan, Z.~Chen, {Easyflow}: keep ethereum away from overflow, in: Proceedings of the 2019 IEEE/ACM 41st International Conference on Software Engineering: Companion Proceedings (ICSE-Companion), 2019, pp. 23--26.

\bibitem{GrechEtAl2018}
N.~Grech, M.~Kong, A.~Jurisevic, L.~Brent, B.~Scholz, Y.~Smaragdakis, {Madmax}: surviving out-of-gas conditions in ethereum smart contracts, Proc ACM Program Lang 2 (2018) 1--27.

\bibitem{GrossmanEtAl2017}
S.~Grossman, I.~Abraham, G.~Golan-Gueta, et~al., Online detection of effectively callback free objects with applications to smart contracts, Proc ACM Program Lang 2 (2017) 1--28.

\bibitem{HeEtAl2019}
J.~He, M.~Balunovic, N.~Ambroladze, P.~Tsankov, M.~Vechev, Learning to fuzz from symbolic execution with application to smart contracts, in: Proceedings of the 2019 ACM SIGSAC Conference on Computer and Communications Security, 2019, pp. 531--548.

\bibitem{HiraiEtAl2017}
Y.~Hirai, Defining the ethereum virtual machine for interactive theorem provers, in: International Conference on Financial Cryptography and Data Security, 2017, pp. 520--535.

\bibitem{HonigEtAl2019}
J.~J. Honig, M.~H. Everts, M.~Huisman, Practical mutation testing for smart contracts, in: P.-S. Cristina, N.-A. Guillermo, B.~Alex, G.-J. Joaquin (Eds.), Data Privacy Management, Cryptocurrencies and Blockchain Technology, Springer, 2019, pp. 289--303.

\bibitem{JiaoEtAl2020}
J.~Jiao, S.~W. Lin, J.~Sun, A generalized formal semantic framework for smart contracts, in: FASE, 2020, pp. 75--96.

\bibitem{LahbibEtAl2020}
A.~Lahbib, A.~A. Wakrime, A.~Laouiti, K.~Toumi, S.~Martin, An event-b based approach for formal modelling and verification of smart contracts, in: Proceedings of the International Conference on Advanced Information Networking and Applications, 2020, pp. 1303--1318.

\bibitem{LaiEtAl2020}
E.~Lai, W.~Luo, Static analysis of integer overflow of smart contracts in ethereum, in: Proceedings of the 2020 4th International Conference on Cryptography, Security and Privacy, 2020, pp. 110--115.

\bibitem{LiEtAl2019}
Z.~Li, H.~Wu, J.~Xu, X.~Wang, L.~Zhang, Z.~Chen, {MuSC}: a tool for mutation testing of ethereum smart contract, in: Proceedings of the 2019 34th IEEE/ACM International Conference on Automated Software Engineering (ASE), 2019, pp. 1198--1201.

\bibitem{MaEtAl2019}
F.~Ma, Y.~Fu, M.~Ren, et~al., {EVM*}: from offline detection to online reinforcement for ethereum virtual machine, in: Proceedings of the 2019 IEEE 26th International Conference on Software Analysis, Evolution and Reengineering (SANER), 2019, pp. 554--558.

\bibitem{MedeirosEtAl2019}
H.~Medeiros, P.~Vilain, J.~Mylopoulos, H.~A. Jacobsen, {SolUnit}: a framework for reducing execution time of smart contract unit tests, in: Proceedings of the 29th Annual International Conference on Computer Science and Software Engineering, 2019, pp. 264--273.

\bibitem{MomeniEtAl2019}
P.~Momeni, Y.~Wang, R.~Samavi, Machine learning model for smart contracts security analysis, in: Proceedings of the 2019 17th International Conference on Privacy, Security and Trust (PST), 2019, pp. 1--6.

\bibitem{MossbergEtAl2019}
M.~Mossberg, F.~Manzano, E.~Hennenfent, et~al., {Manticore}: a user-friendly symbolic execution framework for binaries and smart contracts, in: Proceedings of the 2019 34th IEEE/ACM International Conference on Automated Software Engineering (ASE), 2019, pp. 1186--1189.

\bibitem{NguyenEtAl2019}
Q.~B. Nguyen, A.~Q. Nguyen, V.~H. Nguyen, T.~Nguyen-Le, K.~Nguyen-An, Detect abnormal behaviours in ethereum smart contracts using attack vectors, in: Proceedings of the International Conference on Future Data and Security Engineering, 2019, pp. 485--505.

\bibitem{NikolicEtAl2018}
I.~Nikolic, A.~Kolluri, I.~Sergey, P.~Saxena, A.~Hobor, Finding the greedy, prodigal, and suicidal contracts at scale, 2018, pp. 653--663.

\bibitem{ParkEtAl2018}
D.~Park, Y.~Zhang, M.~Saxena, P.~Daian, G.~Rosu, A formal verification tool for ethereum vm bytecode, in: Proceedings of the 2018 26th ACM Joint Meeting on European Software Engineering Conference and Symposium on the Foundations of Software Engineering, 2018, pp. 912--915.

\bibitem{PengEtAl2019}
C.~Peng, S.~Akca, A.~Rajan, {SIF}: a framework for solidity contract instrumentation and analysis, in: Proceedings of the 2019 26th Asia-Pacific Software Engineering Conference (APSEC), 2019, pp. 466--473.

\bibitem{SergeyEtAl2019}
I.~Sergey, V.~Nagaraj, J.~Johannsen, A.~Kumar, A.~Trunov, K.~C.~G. Hao, Safer smart contract programming with {Scilla}, Proc ACM Program Lang 3 (2019) 1--30.

\bibitem{ShishkinEtAl2018}
E.~Shishkin, Debugging smart contract’s business logic using symbolic model checking, Program Comput Softw 45(8) (2019) 590--599.

\bibitem{TianEtAl2019}
Z.~Tian, Smart contract defect detection based on parallel symbolic execution, in: Proceedings of the 2019 3rd International Conference on Circuits, System and Simulation (ICCSS), 2019, pp. 127--132.

\bibitem{TorresEtAl2018}
C.~F. Torres, J.~Sch{\"u}tte, R.~State, {Osiris}: hunting for integer bugs in ethereum smart contracts, in: Proceedings of the 34th Annual Computer Security Applications Conference, 2018, pp. 664--676.

\bibitem{TsankovEtAl2018}
P.~Tsankov, A.~Dan, D.~Drachsler-Cohen, A.~Gervais, F.~Buenzli, M.~Vechev, {Securify}: practical security analysis of smart contracts, in: Proceedings of the 2018 ACM SIGSAC Conference on Computer and Communications Security, 2018, pp. 67--82.

\bibitem{WangEtAl2020}
Z.~Wang, W.~Dai, K.~K.~R. Choo, H.~Jin, D.~Zou, {FSFC}: an input filter-based secure framework for smart contract, J Netw Comput Appl 154 (2020) 102530.

\bibitem{YeEtAl2020}
J.~Ye, M.~Ma, T.~Peng, Y.~Xue, A software analysis based vulnerability detection system for smart contracts, in: J.~Stan, A.~P.-M. Aneta, M.~Lech (Eds.), Integrating Research and Practice in Software Engineering, Springer, 2020, pp. 69--81.

\bibitem{ZhangQEtAl2020}
Q.~Zhang, Y.~Wang, J.~Li, S.~Ma, {EthPloit}: from fuzzing to efficient exploit generation against smart contracts, in: Proceedings of the 2020 IEEE 27th International Conference on Software Analysis, Evolution and Reengineering (SANER), 2020, pp. 116--126.

\bibitem{ZhangYEtAl2020}
Y.~Zhang, S.~Ma, J.~Li, K.~Li, S.~Nepal, D.~Gu, {SMARTSHIELD}: automatic smart contract protection made easy, in: Proceedings of the 2020 IEEE 27th International Conference on Software Analysis, Evolution and Reengineering (SANER), 2020, pp. 23--34.

\bibitem{SongEtAl2019}
J.~Song, H.~He, Z.~Lv, C.~Su, G.~Xu, W.~Wang, An efficient vulnerability detection model for ethereum smart contracts, in: International Conference on Network and System Security, 2019, pp. 433--442.

\bibitem{GaoEtAl2021}
Z.~Gao, L.~Jiang, X.~Xia, D.~Lo, J.~Grundy, Checking smart contracts with structural code embedding, IEEE Transactions on Software Engineering 47~(12) (2021) 2874--2891.
\newblock \href {https://doi.org/10.1109/TSE.2020.2971482} {\path{doi:10.1109/TSE.2020.2971482}}.

\bibitem{TikhomirovEtAl2018}
S.~Tikhomirov, E.~Voskresenskaya, I.~Ivanitskiy, R.~Takhaviev, E.~Marchenko, Y.~Alexandrov, \href{https://doi.org/10.1145/3194113.3194115}{Smartcheck: static analysis of ethereum smart contracts}, in: Proceedings of the 1st International Workshop on Emerging Trends in Software Engineering for Blockchain, WETSEB '18, Association for Computing Machinery, New York, NY, USA, 2018, p. 9–16.
\newblock \href {https://doi.org/10.1145/3194113.3194115} {\path{doi:10.1145/3194113.3194115}}.
\newline\urlprefix\url{https://doi.org/10.1145/3194113.3194115}

\bibitem{WangEtAl2019}
S.~Wang, C.~Zhang, Z.~Su, \href{https://doi.org/10.1145/3360615}{Detecting nondeterministic payment bugs in ethereum smart contracts}, Proc. ACM Program. Lang. 3~(OOPSLA) (oct 2019).
\newblock \href {https://doi.org/10.1145/3360615} {\path{doi:10.1145/3360615}}.
\newline\urlprefix\url{https://doi.org/10.1145/3360615}

\bibitem{QianEtAl2020}
P.~Qian, Z.~Liu, Q.~He, R.~Zimmermann, X.~Wang, Towards automated reentrancy detection for smart contracts based on sequential models, IEEE Access 8 (2020) 19685--19695.
\newblock \href {https://doi.org/10.1109/ACCESS.2020.2969429} {\path{doi:10.1109/ACCESS.2020.2969429}}.

\bibitem{LiaoEtAl2019}
J.-W. Liao, T.-T. Tsai, C.-K. He, C.-W. Tien, Soliaudit: Smart contract vulnerability assessment based on machine learning and fuzz testing, in: 2019 Sixth International Conference on Internet of Things: Systems, Management and Security (IOTSMS), 2019, pp. 458--465.
\newblock \href {https://doi.org/10.1109/IOTSMS48152.2019.8939256} {\path{doi:10.1109/IOTSMS48152.2019.8939256}}.

\bibitem{FerreiraEtAl2020}
C.~Ferreira~Torres, M.~Baden, R.~Norvill, B.~B. Fiz~Pontiveros, H.~Jonker, S.~Mauw, \href{https://doi.org/10.1145/3320269.3384756}{\ae{}gis: Shielding vulnerable smart contracts against attacks}, in: Proceedings of the 15th ACM Asia Conference on Computer and Communications Security, ASIA CCS '20, Association for Computing Machinery, New York, NY, USA, 2020, p. 584–597.
\newblock \href {https://doi.org/10.1145/3320269.3384756} {\path{doi:10.1145/3320269.3384756}}.
\newline\urlprefix\url{https://doi.org/10.1145/3320269.3384756}

\bibitem{Hwang2020}
S.~R. Sungjae~Hwang, Gap between theory and practice : An empirical study of security patches in solidity, in: Proceedings of the 42nd International Conference on Software Engineering, 2020.

\bibitem{NguyenEtAl2020}
T.~D. Nguyen, L.~H. Pham, J.~Sun, Y.~Lin, Q.~T. Minh, \href{https://doi.org/10.1145/3377811.3380334}{sfuzz: an efficient adaptive fuzzer for solidity smart contracts}, in: Proceedings of the ACM/IEEE 42nd International Conference on Software Engineering, ICSE '20, Association for Computing Machinery, New York, NY, USA, 2020, p. 778–788.
\newblock \href {https://doi.org/10.1145/3377811.3380334} {\path{doi:10.1145/3377811.3380334}}.
\newline\urlprefix\url{https://doi.org/10.1145/3377811.3380334}

\bibitem{HaoEtAl2020}
X.~Hao, W.~Ren, W.~Zheng, T.~Zhu, Scscan: A svm-based scanning system for vulnerabilities in blockchain smart contracts, in: 2020 IEEE 19th International Conference on Trust, Security and Privacy in Computing and Communications (TrustCom), 2020, pp. 1598--1605.
\newblock \href {https://doi.org/10.1109/TrustCom50675.2020.00221} {\path{doi:10.1109/TrustCom50675.2020.00221}}.

\bibitem{XueEtAl2021}
Y.~Xue, M.~Ma, Y.~Lin, Y.~Sui, J.~Ye, T.~Peng, \href{https://doi.org/10.1145/3324884.3416553}{Cross-contract static analysis for detecting practical reentrancy vulnerabilities in smart contracts}, in: Proceedings of the 35th IEEE/ACM International Conference on Automated Software Engineering, ASE '20, Association for Computing Machinery, New York, NY, USA, 2021, p. 1029–1040.
\newblock \href {https://doi.org/10.1145/3324884.3416553} {\path{doi:10.1145/3324884.3416553}}.
\newline\urlprefix\url{https://doi.org/10.1145/3324884.3416553}

\bibitem{FerreiraEtAl2021}
J.~a.~F. Ferreira, P.~Cruz, T.~Durieux, R.~Abreu, \href{https://doi.org/10.1145/3324884.3415298}{Smartbugs: a framework to analyze solidity smart contracts}, in: Proceedings of the 35th IEEE/ACM International Conference on Automated Software Engineering, ASE '20, Association for Computing Machinery, New York, NY, USA, 2021, p. 1349–1352.
\newblock \href {https://doi.org/10.1145/3324884.3415298} {\path{doi:10.1145/3324884.3415298}}.
\newline\urlprefix\url{https://doi.org/10.1145/3324884.3415298}

\bibitem{AlbertEtAl2019}
E.~Albert, J.~Correas, P.~Gordillo, G.~Rom\'{a}n-D\'{\i}ez, A.~Rubio, \href{https://doi.org/10.1145/3293882.3338999}{Safevm: a safety verifier for ethereum smart contracts}, in: Proceedings of the 28th ACM SIGSOFT International Symposium on Software Testing and Analysis, ISSTA 2019, Association for Computing Machinery, New York, NY, USA, 2019, p. 386–389.
\newblock \href {https://doi.org/10.1145/3293882.3338999} {\path{doi:10.1145/3293882.3338999}}.
\newline\urlprefix\url{https://doi.org/10.1145/3293882.3338999}

\bibitem{ChenEtAl2019}
T.~Chen, Z.~Li, Y.~Zhang, X.~Luo, T.~Wang, T.~Hu, X.~Xiao, D.~Wang, J.~Huang, X.~Zhang, A large-scale empirical study on control flow identification of smart contracts, in: 2019 ACM/IEEE International Symposium on Empirical Software Engineering and Measurement (ESEM), 2019, pp. 1--11.
\newblock \href {https://doi.org/10.1109/ESEM.2019.8870156} {\path{doi:10.1109/ESEM.2019.8870156}}.

\bibitem{PrechtelEtAl2019}
D.~Prechtel, T.~Groß, T.~Müller, Evaluating spread of ‘gasless send’ in ethereum smart contracts, in: 2019 10th IFIP International Conference on New Technologies, Mobility and Security (NTMS), 2019, pp. 1--6.
\newblock \href {https://doi.org/10.1109/NTMS.2019.8763848} {\path{doi:10.1109/NTMS.2019.8763848}}.

\bibitem{EnceEtAl2018}
E.~Zhou, S.~Hua, B.~Pi, J.~Sun, Y.~Nomura, K.~Yamashita, H.~Kurihara, Security assurance for smart contract, in: 2018 9th IFIP International Conference on New Technologies, Mobility and Security (NTMS), 2018, pp. 1--5.
\newblock \href {https://doi.org/10.1109/NTMS.2018.8328743} {\path{doi:10.1109/NTMS.2018.8328743}}.

\bibitem{SchneidewindEtAl2020}
C.~Schneidewind, I.~Grishchenko, M.~Scherer, M.~Maffei, \href{https://doi.org/10.1145/3372297.3417250}{ethor: Practical and provably sound static analysis of ethereum smart contracts}, in: Proceedings of the 2020 ACM SIGSAC Conference on Computer and Communications Security, CCS '20, Association for Computing Machinery, New York, NY, USA, 2020, p. 621–640.
\newblock \href {https://doi.org/10.1145/3372297.3417250} {\path{doi:10.1145/3372297.3417250}}.
\newline\urlprefix\url{https://doi.org/10.1145/3372297.3417250}

\bibitem{ZhouEtAl2020}
S.~Zhou, Z.~Yang, J.~Xiang, Y.~Cao, Z.~Yang, Y.~Zhang, \href{https://www.usenix.org/conference/usenixsecurity20/presentation/zhou-shunfan}{An ever-evolving game: Evaluation of real-world attacks and defenses in ethereum ecosystem}, in: 29th USENIX Security Symposium (USENIX Security 20), USENIX Association, 2020, pp. 2793--2810.
\newline\urlprefix\url{https://www.usenix.org/conference/usenixsecurity20/presentation/zhou-shunfan}

\bibitem{btcrelay}
Btc relay, \url{http://btcrelay.org/} (2016).

\bibitem{he2020joint}
H.~He, Z.~Luo, Q.~Wang, M.~Chen, H.~He, L.~Gao, H.~Zhang, Joint operation mechanism of distributed photovoltaic power generation market and carbon market based on cross-chain trading technology, IEEE Access 8 (2020) 66116--66130.

\bibitem{notiondefihacks}
{Web3 Security Community}, Defi hacks analysis - root cause, \url{https://wooded-meter-1d8.notion.site/0e85e02c5ed34df3855ea9f3ca40f53b?v=22e5e2c506ef4caeb40b4f78e23517ee}, accessed: 2025-04-12 (2025).

\bibitem{frontrunning2023}
Frontrunning, \url{https://en.wikipedia.org/wiki/Front-running} (2023).

\bibitem{crossmarket2023}
Cross-market manipulation, \url{https://en.wikipedia.org/wiki/Market-manipulation\#Cross-market-manipulation} (2023).

\bibitem{DurieuxEtAl2020ICSE}
T.~Durieux, J.~F. Ferreira, R.~Abreu, P.~Cruz, Empirical review of automated analysis tools on 47,587 {Ethereum} smart contracts, in: Proceedings of the ACM/IEEE 42nd International conference on software engineering, 2020, pp. 530--541.

\bibitem{rekt}
{Rekt}, Rekt, \url{https://rekt.news/}, accessed: 2025-04-12 (2025).

\bibitem{slowmist}
{SlowMist}, Slowmist, \url{https://hacked.slowmist.io/}, accessed: 2025-04-12 (2025).

\bibitem{certik}
{CertiK}, Certik, \url{https://www.certik.com/}, accessed: 2025-04-12 (2025).

\bibitem{blocksec}
{BlockSec Team}, Blocksec, \url{https://blocksecteam.medium.com/}, accessed: 2025-04-12 (2025).

\bibitem{Code4rena}
{Code4rena}, Code4rena, \url{https://code4rena.com/}, accessed: 2025-04-12 (2025).

\bibitem{DISORBO2022111193}
A.~{Di Sorbo}, S.~Laudanna, A.~Vacca, C.~A. Visaggio, G.~Canfora, \href{https://www.sciencedirect.com/science/article/pii/S0164121221002697}{Profiling gas consumption in solidity smart contracts}, Journal of Systems and Software 186 (2022) 111193.
\newblock \href {https://doi.org/https://doi.org/10.1016/j.jss.2021.111193} {\path{doi:https://doi.org/10.1016/j.jss.2021.111193}}.
\newline\urlprefix\url{https://www.sciencedirect.com/science/article/pii/S0164121221002697}

\bibitem{marchesi2020design}
L.~Marchesi, M.~Marchesi, G.~Destefanis, G.~Barabino, D.~Tigano, Design patterns for gas optimization in ethereum, in: 2020 IEEE International Workshop on Blockchain Oriented Software Engineering (IWBOSE), IEEE, 2020, pp. 9--15.

\bibitem{Jiang}
J.~Jiang, Z.~Li, H.~Qin, M.~Jiang, X.~Luo, X.~Wu, H.~Wang, Y.~Tang, C.~Qian, T.~Chen, Unearthing gas-wasting code smells in smart contracts with large language models, IEEE Transactions on Software Engineering 51~(4) (2025) 879--903.
\newblock \href {https://doi.org/10.1109/TSE.2024.3491578} {\path{doi:10.1109/TSE.2024.3491578}}.

\bibitem{solidifi}
A.~Ghaleb, K.~Pattabiraman, How effective are smart contract analysis tools? evaluating smart contract static analysis tools using bug injection, in: Proceedings of the 29th ACM SIGSOFT international symposium on software testing and analysis, 2020, pp. 415--427.

\bibitem{scanogenerator}
P.~Zhang, B.~Wang, X.~Luo, H.~Dong, Scanogenerator: Automatic anomaly injection for ethereum smart contracts, IEEE Transactions on Software Engineering 50~(11) (2024) 2983--3006.
\newblock \href {https://doi.org/10.1109/TSE.2024.3464539} {\path{doi:10.1109/TSE.2024.3464539}}.

\bibitem{gupta2021realworld}
S.~Gupta, 10 real world use cases for ethereum, \url{https://medium.com/blockchain-vidhya/11-real-world-use-cases-for-ethereum-352ad4509fa1}, accessed: 18-Nov-2025.

\bibitem{threats}
X.~Zhou, Y.~Jin, H.~Zhang, S.~Li, X.~Huang, A map of threats to validity of systematic literature reviews in software engineering, in: 2016 23rd Asia-Pacific Software Engineering Conference (APSEC), 2016, pp. 153--160.
\newblock \href {https://doi.org/10.1109/APSEC.2016.031} {\path{doi:10.1109/APSEC.2016.031}}.

\bibitem{VIDAL2024112160}
F.~R. Vidal, N.~Ivaki, N.~Laranjeiro, \href{https://www.sciencedirect.com/science/article/pii/S016412122400205X}{Vulnerability detection techniques for smart contracts: A systematic literature review}, Journal of Systems and Software 217 (2024) 112160.
\newblock \href {https://doi.org/https://doi.org/10.1016/j.jss.2024.112160} {\path{doi:https://doi.org/10.1016/j.jss.2024.112160}}.
\newline\urlprefix\url{https://www.sciencedirect.com/science/article/pii/S016412122400205X}

\end{thebibliography}
